\documentclass[11pt]{article}

\usepackage{a4wide}
\usepackage{changebar}
\usepackage{multirow}

\usepackage[utf8]{inputenc} % allow utf-8 input
\usepackage[T1]{fontenc}    % use 8-bit T1 fonts
\usepackage{xurl}          % simple URL typesetting
\usepackage{hyperref}       % hyperlinks
\usepackage{booktabs}       % professional-quality tables
\usepackage{amsfonts}       % blackboard math symbols
\usepackage{nicefrac}       % compact symbols for 1/2, etc.
\usepackage{microtype}      % microtypography
\usepackage{lipsum}
\usepackage{graphicx}
\graphicspath{ {./images/} }

\usepackage{bbm}
\usepackage{enumitem}

\usepackage{crop}
\usepackage{caption}
\usepackage{amsmath}
\usepackage{mathrsfs}
\usepackage{array}
\usepackage{color}
\usepackage{xcolor}
\usepackage{amssymb}
\usepackage{flushend}
\usepackage{stfloats}
\usepackage[figuresright]{rotating}
\usepackage{chngpage}
\usepackage{totcount}
\usepackage{fix-cm}
\usepackage{amsthm}
\usepackage{centernot}
\usepackage{ulem}

\title{Forecasting UK Consumer Price Inflation with RaGNAR: Random Generalised Network Autoregressive Processes}

\newtheorem{proposition}{Proposition}

\newtheorem{remark}{Remark}
\theoremstyle{definition}
\newtheorem{definition}{Definition}

\usepackage{natbib}

\author{
  Guy P. Nason and Henry Antonio Palasciano\thanks{
    \texttt{g.nason@imperial.ac.uk}, \texttt{henry.palasciano17@imperial.ac.uk}\\
    Department of Mathematics, Imperial College\\
    SW7 2AZ, London, UK}}

\begin{document}

\maketitle

\begin{abstract}
    This article forecasts CPI inflation in the United Kingdom using Random Generalised Network Autoregressive (RaGNAR) Processes. More specifically, we fit Generalised Network Autoregressive (GNAR) Processes to a large set of random networks generated according to the Erd\H{o}s–R\'{e}nyi–Gilbert model and select the best-performing networks each month to compute out-of-sample forecasts. RaGNAR significantly outperforms traditional benchmark models across all horizons.
    Remarkably, RaGNAR also delivers materially more accurate predictions than the Bank of England’s four to six month inflation rate forecasts published in their quarterly Monetary Policy Reports. Our results are remarkable not only
    for their accuracy, but also because of
    their speed, efficiency and simplicity compared
    to the Bank's current forecasting processes.
    RaGNAR's performance improvements manifest both in terms of their
    root mean squared error and mean absolute percentage error, which measure
    different, but crucial, aspects of the methods' performance.
    GNAR processes demonstrably predict future changes to
    CPI inflation more accurately and quickly
    than the benchmark models, especially
    at medium- to long-term forecast
    horizons, which is of great importance
    to policymakers charged with setting interest rates. We find that the most robust forecasts are those which combine the predictions from multiple GNAR processes via the use of various model averaging techniques. By analysing the structure of the best-performing graphs, we are also able to identify the key components that influence inflation rates during different periods.
\end{abstract}


\section{Introduction}

Forecasting inflation is crucial for economic policymakers, such as the Bank of England, and plays a significant role in the broader economy by aiding businesses and firms in their strategic planning. Precise inflation forecasts are fundamental to shaping monetary policy, with the primary objectives of maintaining price stability and fostering economic growth, while also serving as tools to communicate the rationale behind policy decisions to financial institutions and the general public~\citep{bernanke1997inflation, batini1999monetary, central_banks, boe2024}.
This has recently been highlighted by the UK's (and the global)
struggle with high inflation, which  presented considerable challenges for policymakers. Following a sustained period of elevated inflation, the rate has only now begun to approach the Bank of England's target, highlighting the need for robust and reliable forecasting models to enable better decision-making earlier on. This need is further emphasised by the fact that inflation forecasting accuracy has deteriorated in recent years, both among central banks and external forecasters~\citep{central_banks, boe2024}.

For many years, univariate time series forecasting has been a fundamental approach for predicting inflation. \citet{inflationforecasting2013} provide a comprehensive review of various forecasting techniques, including autoregressive models, direct forecasting methods, random walks, and the Phillips curve. The authors find that the most effective approaches tend to be simple methods requiring few or no parameters. The effectiveness of the Phillips curve, a method which essentially extends an autoregressive model by incorporating the unemployment rate as an additional predictor, has been the subject of much debate~\citep{stock1999forecasting, stock2009phillips, atkeson2001phillips, inflationforecasting2013}. For example,~\citet{atkeson2001phillips} show that the Phillips curve struggles to outperform even simple naive forecasts and advocate for the use of a random walk of order four instead. On the other hand,~\citet{stock2009phillips} support the Phillips curve over more complex multivariate forecasting techniques, although its performance relative to other univariate methods is inconsistent and highly dependent on prevailing macroeconomic conditions. Furthermore,~\citet{stock2007has} argue that univariate models with time varying parameters are the best models for forecasting inflation. In general, much of the literature seems to suggest that simple univariate time series are remarkably difficult to surpass when it comes to forecasting inflation~\citep{atkeson2001phillips, stock2003forecasting, stock2007has, inflationforecasting2013}.

There have been several attempts to improve upon linear univariate time series inflation rate forecasts using more advanced techniques. For instance, \citet{uk_2024} explore a range of machine learning and shrinkage methods to forecast UK inflation, using the individual item series as predictors. They find that only shrinkage methods significantly outperform the benchmarks, and even then, only during specific sub-periods. \citet{barkan2023forecasting, boaretto2023forecasting, medeiros2021forecasting} have outperformed  classical benchmarks with machine learning methods, including recurrent neural networks and long short-term memory models, when applied to the disaggregate item series in the US and Brazil.  In contrast, \citet{almosova2023nonlinear} find that such non-linear forecasting techniques do not yield significant improvements in performance over autoregressive models when applied to the US inflation rate time series alone. \citet{makridakis2018statistical} argue that standard statistical techniques are more accurate than machine learning methods when applied to a large range of economic and business data series, although such a conclusion was reached on average, and in certain areas, machine learning methods might still be better suited. Alternative data sources have also been explored for the purpose of forecasting inflation. For example \citet{aparicio2020forecasting} demonstrated success using online prices to forecast inflation rates across multiple countries, exploiting the fact that such information is available at a higher frequency. More recently, \citet{eugster2024forecasting} applied sentiment analysis to a large corpus of news articles, producing US inflation forecasts that outperform a naive random walk model. Hence, although improvements in performance can be obtained, univariate time series models remain highly robust inflation rate forecasters, especially when one takes into account the simplicity of such methods.

Generally speaking, the best kinds of forecasts are those that involve the use of surveys or a degree of expert judgement~\citep{ang2007macro,inflationforecasting2013, central_banks}, such as the Bloomberg, U.S. and European Central Bank Surveys of Professional Forecasters and the Bank of England Monetary Policy Reports and Survey of External Forecasters~\citep{boe_survey, central_banks, clements2023, clements2024}. These methods integrate expert insights and blend forecasts from a wide array of models. A simple example is the AR-GAP model discussed in~\citet{inflationforecasting2013}, which adjusts the inflation rate series by removing the five-to-ten-year-ahead inflation forecast from Blue Chip, treating it as a time-varying local mean before fitting an AR process. Alternatively, the Bank of England's forecasting model~\citep{boe2013, boe2024}, typically makes use of a suite of at least forty models, including its central model, COMPASS, when generating a forecast. Further, a single forecast involves several hundred runs of the COMPASS algorithm and at least five Monetary Policy Committee meetings, where judgemental adjustments are made to both the forecasts and the models used. Given the complexity and thoroughness of such processes, it is challenging for other methods to compete. For example,~\citet{uk_var} find that a Bayesian VAR is unable to outperform the Bank of England's inflation rate forecasts. This is why simple univariate time series models are often used as benchmarks ---  not only are they straightforward, but also among the most effective when relying on historical data alone.

Graphs are  popular tools that have been extensively used in many fields, such as Bayesian statistics \citep{heckerman2008tutorial, koski2011bayesian}, time series analysis
\citep{GNARorig, zhu2017network, GNAR, kang2021dynamic, lucchese2023estimation}, feature extraction and classification
\citep{peach2019semi, peach2021hcga}, deep learning \citep{scarselli2008graph, wu2020comprehensive, zhou2020graph}, causal inference \citep{pearl2009causal, pearl2009causality, yao2021survey}, and neuroscience \citep{bullmore2009complex, zhou2020toolbox}, to name but a few. Excellent and comprehensive introductions to statistical network analyses
can be found in \citet{Kolaczyk09} and \citet{KolaczykCsardi20}.

In this article we forecast monthly inflation in the United Kingdom using Random Generalised Network Autoregressive (RaGNAR) Processes. More specifically, we generate a set of random graphs according to the Erd\H{o}s–R\'{e}nyi–Gilbert model~\citep{random_graphs} and fit the Generalised Network Autoregressive (GNAR) Processes introduced by~\citet{GNAR}. GNAR models are a class of autoregressive models which use information from neighbouring nodes (vertices that represent individual time series) as additional predictors. We then select the ones which perform best at each time step and use these to compute forecasts for horizons up to 12 months ahead. We find that our best forecasts are those which average the predictions of multiple graphs and GNAR processes. RaGNAR consistently outperforms classic univariate benchmarks, particularly at longer horizons. A welcome and surprising finding is that our forecasts are also able to beat the Bank of England's four- to six-month ahead forecasts, further underlining the strength of our approach. Moreover, by analysing the structure of the best performing networks, we are also able to identify the set of specific components which  most influence the inflation rate forecast each month.

Significant research has been conducted on studying and forecasting inflation using disaggregated item series. For instance,~\citet{stock2020slack} examine the components of inflation in the US with a focus on cyclical variation, while~\citet{hubrich2005forecasting, hendry2011combining, chu2018evolution} explore the effectiveness of using aggregate data for forecasting in the US, the Euro area, and the UK, respectively. As previously mentioned, studies by~\citet{medeiros2021forecasting, barkan2023forecasting, boaretto2023forecasting, uk_2024} have also attempted to forecast inflation using various shrinkage and machine learning techniques, yielding mixed results. The literature generally indicates that individual component series offer valuable insights for forecasting inflation, which networks and GNAR processes are particularly well-suited to exploit.

This article is organised as follows. Section~\ref{sec:GNAR_review} introduces GNAR processes. Section~\ref{sec:CPI} provides a brief review of CPI in the UK and describes the data used in this work. Section~\ref{sec:methodology} describes our approach in more detail. Section~\ref{sec:results} compares our approach to various benchmark models, as well as to the Bank of England forecasts. Finally,~\ref{sec:structure} analyses the structure of the most successful networks in order to determine the most influential components.

\section{Generalised Network Autoregressive Processes} \label{sec:GNAR_review}
Generalised Network Autoregressive (GNAR) Processes~\cite{GNARorig, GNAR2, GNAR} describe the behaviour of multivariate time series linked according to an associated graph.
They have found successful application in economics~\cite{GNAR_covid, GNAR_GDP}, medicine~\cite{GNAR_covid2}, finance~\cite{GNAR_sm} and computational politics~\cite{us_politics}, to name a few.
For many multivariate time series it has been shown, by several of the previous papers mentioned, e.g., that GNAR models provide
unusually parsimonious models that are extremely rapid to compute and store and exhibit excellent forecasting properties.
We briefly review GNAR processes, following~\cite{GNAR}. 

A graph $\mathcal{G} = (\mathcal{K}, \mathcal{E})$ consists of a set of nodes $\mathcal{K}$ and a set of edges $\mathcal{E}$. Edges between nodes $i$ and $j$ can be either directed, denoted by $i \rightsquigarrow j$, or undirected, denoted by $i \leftrightsquigarrow j$. For brevity, we define everything for undirected graphs; the definitions for directed graphs are analogous by replacing $\leftrightsquigarrow$ with $\rightsquigarrow$.

The edge set of a network is defined as $\mathcal{E} = \{(i,j):i \leftrightsquigarrow j, \, i,j\in\mathcal{K}\}$, while the neighbour set of a set of nodes $A\subset \mathcal{K}$ is defined as $\mathcal{N}(A)= \{j\in\mathcal{K} : i \leftrightsquigarrow j\text{ for some } i\in A\}$. The stage $r$ neighbour set can then be defined as `neighbours of the $(r-1)$ stage neighbours,
not including any previous neighbour already seen', which, written mathematically is
\begin{equation}
    \mathcal{N}^{(r)}(A) = \mathcal{N}\{\mathcal{N}^{(r-1)}(A)\}/[\{\cup_{q=1}^{r-1}\mathcal{N}^{(q)}(A)\}\cup A],
\end{equation}
which represents the set of nodes whose shortest path to $A$ is of length $r$ edges. For simplicity, we write $\mathcal{N}_i^{(r)} = \mathcal{N}^{(r)}(\{i\})$ when referring to the stage-$r$ neighbour set of a single node. Finally, the connection weight between nodes $i$ and $j$, given that $j\in\mathcal{N}_i^{(r)}$, is defined as 
\begin{equation}
    \omega_{i,j} = |\mathcal{N}_i^{(r)}|^{-1},
\end{equation}
where $|A|$ denotes the cardinality of a set $A$.  If the network edges are weighted, the stage-$r$ connection weights are determined by the combined weights of the shortest path between node $i$ and its stage-$r$ neighbour, normalised so that the sum of all stage-$r$ connection weights to that node equals one.

We can now define GNAR processes as follows.

\begin{definition}\textbf{(Generalised Network Autoregressive Process).} \cite{GNAR}. Let $\mathcal{G} = (\mathcal{K}, \mathcal{E})$ be a graph and $\boldsymbol{X}_t = (X_{1,t},\ldots,X_{N, t})^T$ be an \mbox{$N$-dimensional}
vector time series, i.e. $X_{i,t}$ corresponds to the time series at the $i$-th node/vertex on the graph $\mathcal{G}$.
Then the generalised autoregressive process of order $(p, \boldsymbol{s})\in\mathbb{N}\times \mathbb{N}_0^p$ for $\boldsymbol{X}_t$, denoted GNAR$(p, \boldsymbol{s})$, is given by 
\begin{equation}\label{eq:GNAR}
    X_{i,t} = \sum_{j=1}^p\Biggl(\alpha_{i, j}X_{i, t-j} + \sum_{r=1}^{s_j}\beta_{i,j,r}\sum_{q\in \mathcal{N}_{i, t}^{(r)}}\omega_{i,q}^{(t)}X_{q, t-j}\Biggr) + u_{i,t},
\end{equation}
for $i \in \mathcal{K}$ and $t \in \{ 1, \ldots, T\} = \mathcal{T}$ and
where $p$ is the maximum time lag, $\boldsymbol{s} = (s_1,\ldots, s_p)$ and $s_j\in\mathbb{N}_0$ the maximum stage of neighbour dependence for lag $j$, with $\mathbb{N}_0 = \mathbb{N}\cup\{0\}$, $\mathcal{N}_{i, t}^{(r)}$ is the $r$-th stage neighbour set of node $i$ at time $t$ and $\omega_{i,q}^{(t)}\in[0,1]$ is the connection weight between node $i$ and $q$ at time
$t$. 
The $\{ u_{i, t}\}_{i\in \mathcal{K}, t\in\mathcal{T}}$ is assumed to be an independent and identically distributed set of random variables each with zero mean and variance of $\sigma_i^2$. The $\alpha_{i,j}\in\mathbb{R}$ represent the `standard' autoregressive coefficients for $i\in\mathcal{K}$ and $j=1, \ldots, p$,
whereas the $\beta_{i,j,r}\in\mathbb R$ capture the influence of $r$-th stage neighbours at lag $j$ on node $i$. 
Here $r=1, \ldots, s_j$, where $s_j \geq 0$ are integers for $j=1, \ldots, p$.
\end{definition}

In the standard GNAR model formulation due to~\cite{GNAR}, the parameters $\beta_{i,j,r}$ are global, meaning that $\beta_{i,j,r} = \beta_{j,r}$. Hence, we refer to the model in which the $\beta_{i,j,r}$ are allowed to vary from node to node as a local-$\alpha\beta$ GNAR$(p, \boldsymbol{s})$ process, distinguishing it from the original model.
Additionally,~\cite{GNAR} define a global-$\alpha$ GNAR$(p, \boldsymbol{s})$ model where the parameters $\alpha_{i,j}$ are shared among nodes, or more formally $\alpha_{i,j} = \alpha_j$. If both
the $\alpha, \beta$ parameters are both local,
as we use here, we refer to the model as
\mbox{local-$\alpha\beta$}.

%\begin{remark}
%    In equation~\eqref{eq:GNAR} we have not included the sum due to the covariates of the model as in \cite{GNAR}, since in this work we will only be dealing with a single covariate at each node. 
%\end{remark}

%For a more comprehensive review of networks see \cite{networks} and of GNAR processes see \cite{GNAR}.

\section{Consumer Price Index Data} \label{sec:CPI}

The Consumer Price Index (CPI) is a key indicator used to measure inflation, which tracks changes in the cost of a basket of approximately 700 goods and services consumed by households in the United Kingdom. This basket is carefully designed to represent typical consumer spending patterns, with the weight assigned to each item reflecting its relative importance in the average household’s expenditure. Both the composition of the basket and the weights assigned to each item are reviewed and adjusted annually to ensure they remain relevant to current trends in the economy. Each month, the Office for National Statistics (ONS) gathers price data from a wide range of locations across the country, including physical retail outlets and online stores and service providers. These prices are first combined to produce item-level indices, which are then systematically aggregated to create broader indices that track prices at the class, group, and division levels. Finally, these division-level indices are combined to produce the overall CPI, which is used to calculate the inflation rate. \cite{cpi_bg, cpi_manual}. An excellent introduction to price indices and CPI can be found in \cite{Ralph15}.
The twelve divisions are food and non-alcoholic beverages, alcoholic beverages and tobacco, clothing and footwear,
housing, water, electricity, gas and other fuels, furniture, household equipment and maintenance, health, transport, communication, recreation and culture, education, restaurants and hotels and miscellaneous goods and services.

The data used here consists of monthly observations from January 1988 to the present, sourced from the Office for National Statistics (ONS)~\cite{cpi_data}. As of the latest release, the twelve main divisions
are further broken down into 40 groups, 71 classes, and 192 item-level series. Our study considers time series up to and including the class level, excluding any series consisting of annual observations. In total we consider
$N=114$ individual time series, including the overall CPI. These series are transformed into year-on-year percentage changes to mitigate
for trends and seasonality, and to help stabilise the variance.

\begin{figure}
    \centering
    \resizebox{\textwidth}{!}{\includegraphics{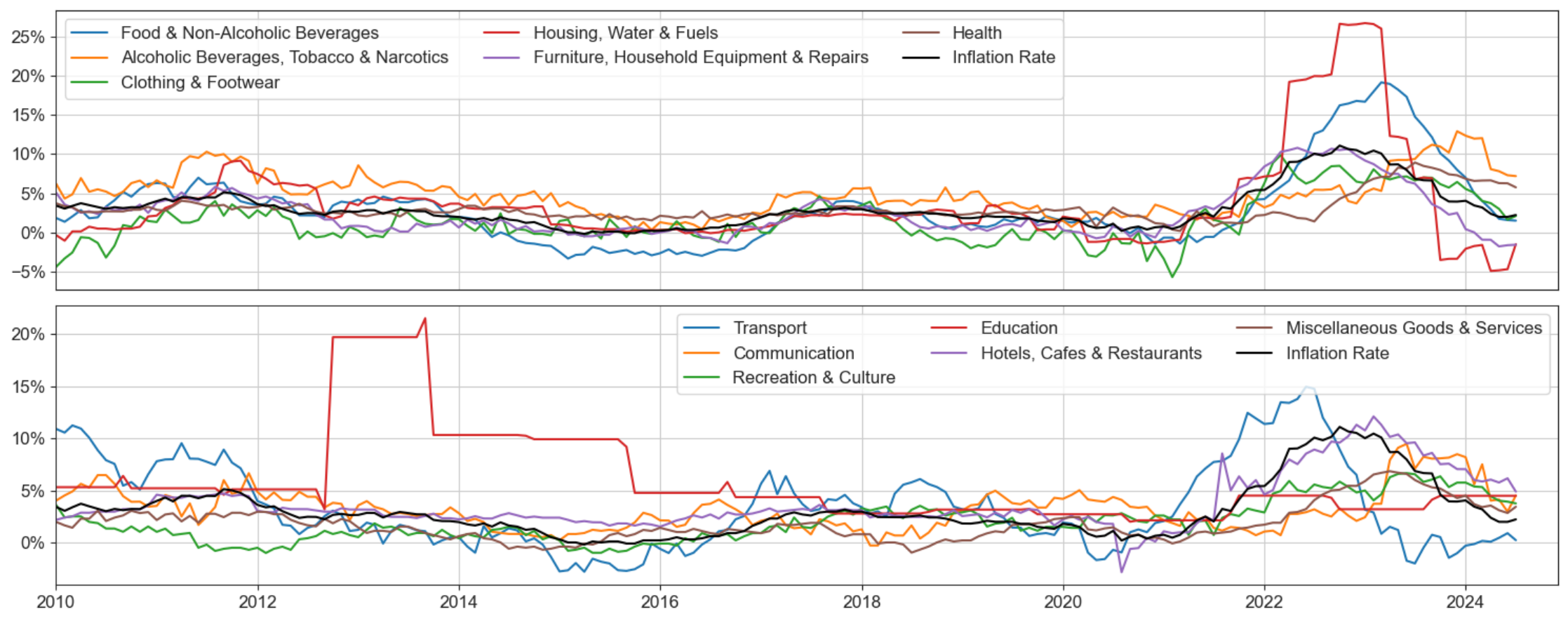}}
    \caption{Year-on-year percentage change time series of the Consumer Price Index divisions.}
    \label{fig:lvl1}
\end{figure}

For visualisation purposes, Figure~\ref{fig:lvl1} presents the year-on-year percentage changes for the
twelve CPI division-level series and the inflation rate.  Here, we clearly see that the Education division is of a yearly nature, and in fact its sub-components are excluded from our analysis.  Disaggregated CPI  component series can contain valuable information, revealing dynamics that may not be immediately visible in the aggregated CPI. For example, from 2020 onwards, the Transport division time series appears to anticipate movements in the overall inflation rate, illustrating how certain sub-components might act as early signals. For a more in-depth review of the CPI disaggregated item series, including descriptive statistics, see Section 2 of~\cite{uk_2024}.

\section{Modelling and Forecasting Methodology} \label{sec:methodology}

\subsection{Network Construction}
\label{sec:networkconstruction}

Our approach involves selecting the best networks each month from a set of possible networks. With $N=114$ nodes
(separate time series) in the dataset, this corresponds to a possible $2^{6441}$ distinct networks, making it computationally infeasible to iterate through all of them to see how well each one can be used to forecast. Instead, we propose selecting an `optimal' (or very good) network(s) from a set of $G$ undirected random networks generated according to the Erd\H{o}s–R\'{e}nyi–Gilbert model~\cite{random_graphs}, where each edge exists with probability $\pi$. The next two propositions provide some insight into the behaviour of the neighbour sets as a function of $\pi$.

\begin{proposition}~\label{prop:dist}
    Let $(\mathcal{G}, \mathcal{E})$ be a graph generated according to the Erd\H{o}s–R\'{e}nyi–Gilbert model~\cite{random_graphs}, where edges exist with probability $\pi$ and $|\mathcal{G}| = N$. Then, for any $i\in\mathcal{G}$, 
    \begin{equation}\label{eq:prop1}
        \mathbb{P}\left(\bigl|\mathcal{N}_i^{(r)}\bigr| = n_r\right) = \sum_{\boldsymbol{n}\in\Lambda_{r-1}(N)}\prod_{s=1}^{r}\mathbb{P}(Z_{n_1:n_{s-1}}(N) = n_s),
    \end{equation}
    for $n_r\in\{0,1,\ldots, N - r\}$, where 
    \begin{equation}
        \Lambda_{r-1}(N)=\left\{\boldsymbol{n}=(n_1,\ldots,n_{r-1})\in \mathbb{N}^{r-1} : \sum_{s=0}^{r-1} n_s \leq N\right\}
    \end{equation}
    is the set of all possible combinations of neighbour set sizes up to stage $r-1$, 
    \begin{equation}
        Z_{n_1:n_{s-1}}(N)\sim \text{Bin}\left(N - \sum_{k=0}^{s-1} n_k,  1 - (1 - \pi)^{n_{s-1}}\right)
    \end{equation}
    and $n_0=1$.
\end{proposition}

\begin{proposition}\label{prop:element}
    Let $(\mathcal{G}, \mathcal{E})$ be a graph generated according to the Erd\H{o}s–R\'{e}nyi–Gilbert model~\cite{random_graphs}, where edges exist with probability $\pi$ and $|\mathcal{G}| = N$. Then, for any distinct $i,j\in\mathcal{G}$,
    \begin{align}\nonumber
        \pi_r & = \mathbb{P}\left(j\in\mathcal{N}_i^{(r)}\right)\\ & = \left(1 - \sum_{s=1}^{r-1} \pi_s\right)\left\{1 - \sum_{n=0}^{N - r}(1 - \pi)^n\sum_{\boldsymbol{n}\in\Lambda_{r-2}(N-1)} \prod_{s=1}^{r-1}\mathbb{P}\left(Z_{n_1:n_{s-1}}(N-1) = n \right)\right\}
    \end{align}
    where $\pi_1 = \pi$, and $\Lambda_{r-2}(N-1)$ and $Z_{n_1:n_{s-1}}(N-1)$ are as defined in Proposition~\ref{prop:dist}.
\end{proposition}

\begin{remark}
    For the $r=2$ case the result in Proposition~\ref{prop:element} simplifies to
    \begin{equation}
        \pi_2 = (1 - \pi)\left\{1 - \left(1 - \pi^2\right)^{N-2}\right\},
    \end{equation}
    since there is only the binomial $Z_{n_1}(N-1) = \bigl|\mathcal{N}_i^{(1)}\bigr| \Big | j\notin \mathcal{N}_i^{(1)} \sim \text{Bin}(N-2, \pi)$ to consider.
\end{remark}

\begin{figure}
    \centering
    \resizebox{\textwidth}{!}{\includegraphics{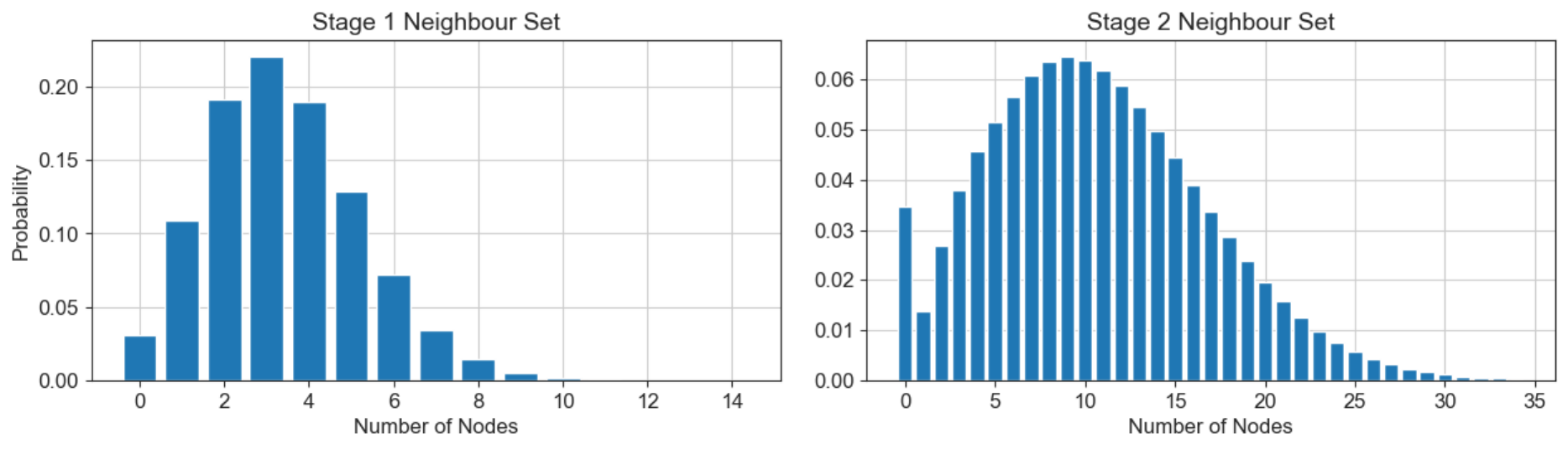}}
    \caption{Probability distribution of the number of nodes in each neighbour set when $N=114$ and  $\pi=0.03$.}
    \label{fig:ns_cardinality}
\end{figure}

Proposition 1 provides insight into the behaviour of neighbour set sizes as a function of $\pi$. We find that dense networks perform poorly in GNAR models, as large neighbour sets lead to information loss due to excessive averaging. In CPI forecasting, averaging many item series produces a time series that closely resembles aggregate CPI, reducing its usefulness. Generating such networks can lead to: 1. poor forecasts due to linear dependencies among predictors and 2. redundant network rankings, as different graphs may yield similar forecasting performance, given that large sets tend to have similar averages. Since higher-stage neighbour sets naturally contain more nodes, large sets can still be explored while ensuring that smaller sets are also considered. In our case, we find that GNAR models using only stage-1 neighbour sets often perform just as well as those including stage-2, suggesting that most of the predictive power comes indeed from the small stage-1 neighbour sets. Hence, Proposition 1 helps guide the selection of $\pi$, ensuring that networks with appropriate neighbour set sizes are explored. While the choice of $\pi$ remains problem-specific, in general, larger networks require a lower $\pi$.
    
Proposition 2, which provides the probability of a node appearing in another’s neighbour set, is less central to our approach but can help assess node frequency. If this probability is too high, networks become more similar, limiting the effectiveness of our approach.

Figure~\ref{fig:ns_cardinality} shows the distribution of stage-1 and stage-2 neighbour set sizes for networks sampled using the Erdős–Rényi–Gilbert model~\cite{random_graphs} with $\pi=0.03$. This setting allows us to explore a wide range of small stage-1 neighbour sets, ensuring that only the most influential components are selected as connections to the CPI node each month. In practice, the best-performing networks typically have three to four nodes in their stage-1 neighbour sets, though this varies. The increased probability of a stage-2 neighbour set size of zero arises because if a stage-1 neighbour set is empty, stage-2 must also be empty.

Empirically, we find that generating $G=10000$ random graphs is sufficient for identifying some networks that perform very well. We also tested $G=1000$ and $G=100000$, finding that increasing from $G=1000$ to $G=10000$ improved performance, but further increasing to $G=100000$ provided no meaningful gains while adding computational cost. Similarly, we tested alternative values of $\pi=0.05$ and $\pi=0.07$, but there was no discernible performance improvement. The lower chosen value of $\pi= 0.03$ also means fewer neighbours on average, which makes it easier for our algorithm to discern more dominant nodes (time series).

Throughout this article, the networks we consider are constructed from unweighted edges, these being sampled from a Bernoulli distribution with probability of success $\pi$. Hence, the GNAR($p,\,\boldsymbol{s})$ process~\eqref{eq:GNAR} can be written as 
\begin{equation}\label{eq:GNAR_unweighted}
    X_{i,t} = \sum_{j=1}^p\left(\alpha_{i, j}X_{i, t-j} + \sum_{r=1}^{s_j}\beta_{i,j,r}\overline{\mathcal{N}}_{i, t-j}^{(r)}\right) + u_{i,t},
\end{equation}
where $\overline{\mathcal{N}}_{i, t-j}^{(r)}$ denotes the average over the stage $r$ neighbour set of node $i$ at time $t-j$. Having fitted the model, we can then obtain the one-month ahead forecast as 
\begin{equation}
    \hat{X}_{i,t+1|t} = \sum_{j=1}^p\left(\hat{\alpha}_{i, j}X_{i, t-j+1} + \sum_{r=1}^{s_j}\hat{\beta}_{i,j,r}\overline{\mathcal{N}}_{i, t-j+1}^{(r)}\right),
\end{equation}
iterating this procedure in order to compute the $h$-step ahead forecast $\hat{X}_{i,t+h|t}$.

In science it is important to discuss ideas that did {\em not} work so well as well as those that did. We
briefly mention an early idea that we tried that did not succeed.
The construction of the Consumer Price Index itself closely resembles a graph: individual item series can be
progressively combined into larger components in a tree-like network, using basket weights as connection weights,
culminating in the final CPI product. This structure suggests a natural fit for applying GNAR processes to the graph, with the aim of generating inflation rate forecasts. However, the inherent properties of GNAR models pose challenges in this context. Specifically, the CPI neighbour sums in such a network then often become redundant relative to the CPI time series itself, leading to parameter estimation instability. Additionally, this method inherently includes edges between CPI components, which may not always contribute meaningfully to forecasting accuracy.

\subsection{Model Fitting and Selection}

We consider GNAR processes in which the maximum neighbour stage dependence $s$ is kept the same at all lags, and hence write GNAR($p,\boldsymbol{s}$) where $\boldsymbol{s}$ denotes a vector of length $p$ with all entries equal to $s$. Having generated $G=10000$ undirected graphs, we fit the GNAR models on a rolling basis using $n_{train}=150$ observations, equivalent to 12.5 years of data, and use these to compute an out-of-sample one-month ahead inflation forecast. The models are then ranked each month based on the root mean squared error (RMSE) of the previous $n_{val} = 30$ one-step ahead inflation rate forecasts. We experimented with alternative window sizes, including 12 and 60 months, as well as an exponentially weighted moving average that assigns greater weight to recent months. While none of these alternatives led to substantial changes in performance, $n_{val}=30$ provided slightly better results in terms of forecast accuracy, offering a balance between robustness and representativeness of recent performance. Formally, the RMSE for the GNAR($p,\boldsymbol{s}$) model fit to the $k$-th graph at time $t$ is given by
\begin{equation}\label{eq:rmse_def}
    \operatorname{RMSE}_{t}^{(k, p,\boldsymbol{s})} = \left\{ n^{-1}_{val} \sum_{i=0}^{n_{val}-1} \left(X_{t-i} - \hat{X}_{t-i|t-i-1}^{(k, p,\boldsymbol{s})}\right)^2  \right\}^{1/2},
\end{equation}
where $\hat{X}_{t|t-1}^{(k, p,\boldsymbol{s})}$ denotes the corresponding forecast. The best performing models are then re-fitted with the data up to time $t$ and used to compute inflation rate forecasts for times $t+1$ up to $t+12$.

In selecting the best network, we consider the one-step ahead RMSE at (only) the node corresponding to the CPI time series as a metric for performance. Hence, we only fit the GNAR model to that node, speeding the computation up significantly. Once a graph, or set of graphs, is chosen for a given month, we then fit the entire model and forecast CPI inflation multiple months ahead. A potential drawback is that, by selecting networks like so, edges between nodes excluding the CPI one are essentially chosen at random. However, since these other nodes only affect CPI through second order effects and, even so, these effects consist of averages over sets of nodes, this randomness should have a minimal impact on the effectiveness of our method, and may even be beneficial in preventing our models from over-fitting. Moreover, we will also compute forecasts by averaging the inflation rate predictions of several GNAR processes of the same order fit to multiple graphs, further diminishing the importance of the specific network structure outside that of the CPI node, while allowing information to flow between the disaggregated item series.

We select the appropriate model order each month using the Bayesian Information Criterion (BIC)~\citet{brd, GNAR}. To accurately assess the forecasting performance of a GNAR model for a given set of parameters, we use the RMSE, as defined in equation~\eqref{eq:rmse_def}, averaged across the top 25\% of networks for that month. Denoting the set of $K$ best-performing networks ($K=2500$) at time $t$ by $\mathcal{G}_t^{(K, p,\boldsymbol{s})}$, the BIC in our setting is
\begin{equation}
    \operatorname{BIC}_t^{(K, p,\boldsymbol{s})} = K^{-1}\sum_{k\in\mathcal{G}_t^{(K, p,\boldsymbol{s})}} 2 \log\left(\operatorname{RMSE}_t^{(k, p,\boldsymbol{s})}\right) + n_{val}^{-1} \,p (s+1)\log(n_{val}),
\end{equation}
which can be interpreted as the average BIC of the top $K$ networks at the CPI node. We avoid averaging over a small subset of top-performing networks to reduce the risk of overfitting. Since we have already ranked our set of networks and our goal is now to select the model parameters, choosing a larger $K$ provides a more representative measure of out-of-sample performance for the GNAR($p,\boldsymbol{s}$) class at each time step.

\begin{figure}
        \centering
        \includegraphics[width=1\linewidth]{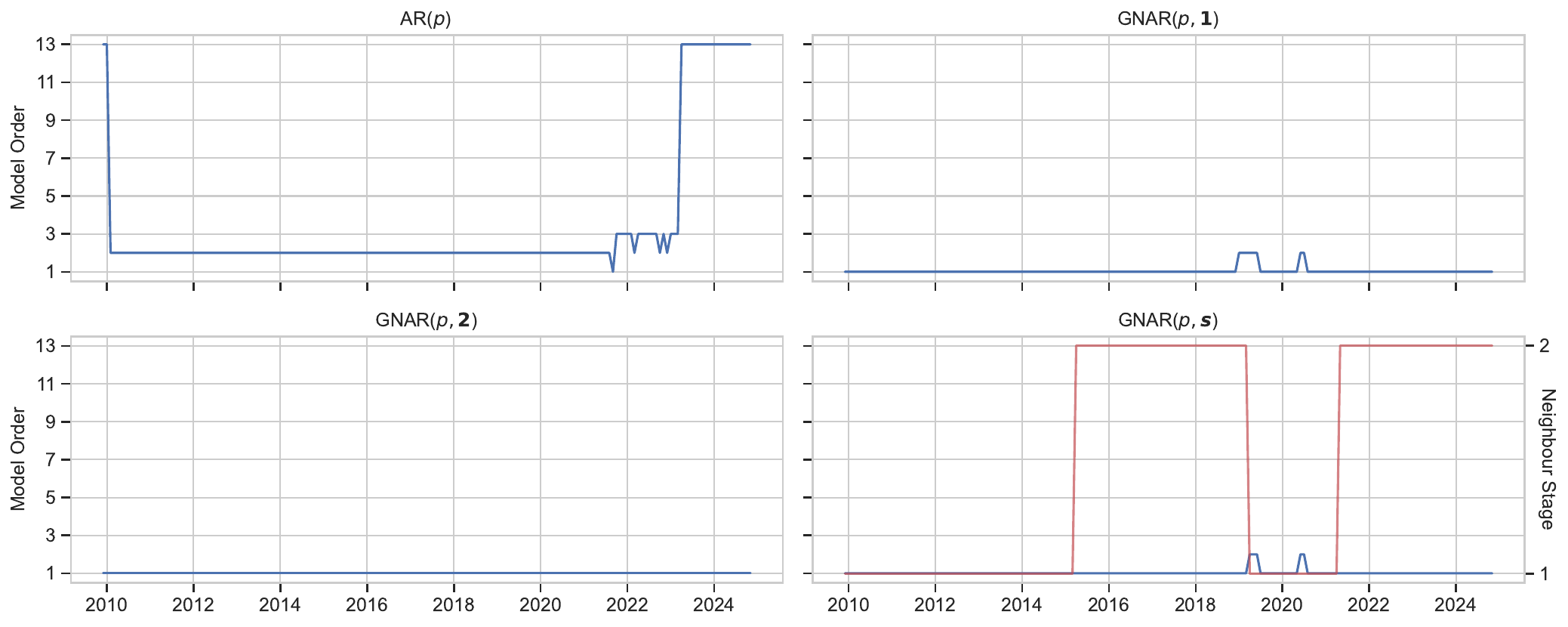}
        \caption{Model orders selected by the Bayesian Information Criterion (BIC) over time. The autoregressive order $p$ is shown in blue, while the maximum neighbour stage $s$ is shown in red.}
        \label{fig:bic_parameters}
    \end{figure}

We consider GNAR processes where $s$ is fixed at either 1 or 2, selecting only the autoregressive order $p$, as well as a more flexible version where both $p$ and $s$ are chosen each month.
We find no benefit in including higher neighbour stages, so the BIC models we consider are the global-$\alpha$, standard, and local-$\alpha\beta$ GNAR($p,\boldsymbol{1}$), GNAR($p,\boldsymbol{2}$), and GNAR($p,\boldsymbol{s}$) processes. Figure~\ref{fig:bic_parameters} displays the selected parameters over time. In agreement with the existing literature on forecasting inflation (see for example~\citet{uk_2024}), low-order models are preferred. This is because GNAR models effectively leverage disaggregated item series, reducing the need for longer orders, while the greater number of parameters in high-order models is penalized more heavily. However, \citet{long_ar1, long_ar2} suggest that higher-order models are more appropriate for inflation modelling, and indeed, we find that higher-order GNAR processes perform better at shorter forecasting horizons (see Appendix C). In the GNAR($p,\boldsymbol{s}$) process, the selected neighbour stage $s$ is often 1, suggesting that a lot of the additional predictive power does indeed come from stage-1 neighbour sets. 

We compare fixed-order GNAR and AR processes in Appendix C, where we find that low-order GNAR models remain competitive against higher-order alternatives, often outperforming them - especially at longer horizons. This contrasts sharply with AR processes, where higher-order models are overall superior, due to their ability to capture inflation dynamics more effectively in recent periods. This reinforces the idea that disaggregated item series add significant value to our forecasts, as low-order GNAR models frequently match or even surpass their higher-order counterparts.

We also repeated our analyses by model selection using the one-step ahead forecast error
criterion, but the results were similar in this case.

In addition to selecting models using the BIC, we explore the use of model averaging, a technique dating back to~\citet{model_averaging_1} that has been widely applied in inflation forecasting. For example, \citet{stock2003forecasting, stock_avg, inflationforecasting2013} forecast inflation using equally weighted averages of various models, each incorporating additional predictors within an autoregressive framework. Such forecasts have been praised in the literature as being among the strongest methods relying solely on historical data~\citet{stock2003forecasting, inflationforecasting2013}. Alternatively, \citet{boe_avg, wright_avg, inf_avg_1, inflationforecasting2013} have combined forecasts using approaches such as Bayesian and information-theoretic model averaging. Moreover, averaging AR model forecasts is a common practice beyond the context of inflation forecasting.

We implement two distinct forms of model averaging. First, we average forecasts across different networks while keeping the model fixed, similar to the equally weighted forecasting techniques of~\citet{stock2003forecasting, stock_avg, inflationforecasting2013}. Second, we compute averages across multiple models within the same class, a standard approach in autoregressive modelling, applied here to GNAR. We write AvGNAR($\{p_1, \ldots, p_\ell\}, \{\boldsymbol{s}_1, \ldots, \boldsymbol{s}_m\}$), for the average over the set of models ${\text{GNAR}(p_i, \boldsymbol{s}_j), i=1, \ldots, \ell, j=1, \ldots, m}$. Whenever forecasts are averaged across multiple networks, this is explicitly stated in the text. Averaging across models of different orders allows us to incorporate both long- and short-term information from CPI and its disaggregate item series into a single prediction. Moreover, model averaging produces more robust forecasts by reducing reliance on any single model, mitigating the risk of overfitting. The Appendix provides the rationale behind our selection of model orders for averaging. All averages are equally weighted; alternative weighting schemes, such as the Bayesian model averaging techniques used in~\citet{boe_avg, wright_avg}, are left for future work.

The process of trying each random network
can be carried out quite independently for each
network. Hence, our method would be simple
to parallelize. However, the results below,
fitting a new model each month from 2010, selecting the best network(s) and then fitting GNAR processes of different classes and sets of parameters to these can already
be carried out in just a few hours on a
regular single processor.

The basic idea of using random networks in GNAR modelling and prediction was introduced by \cite{GNAR2,GNAR}, but our work has
tailored the idea for CPI forecasting  and there are several key differences. The original \cite{GNAR2} work used random networks to fit GNAR models to a dataset consisting of 35 GDP time series from various countries, selecting the graph which minimised the prediction error at all nodes, with the interest lying in finding `the' `correct' network, which may or may not exist (and in many real cases probably does not).
Here, we only use different random networks
as a means to an end and we do not believe any one network is the `correct' network, or at least not for 
any sustained period, and in fact find that averaging the forecasts of multiple graphs aids prediction. Having said that, we do note the components that enter the best set of networks to get an idea
of which CPI components are important for forecasting, see Section~\ref{sec:structure}. 
In contrast with~\cite{GNAR2},  our method also does not 
use connection weights --- the averages in~\eqref{eq:GNAR_unweighted} are ordinary and not weighted averages, in spite
of having some idea of what connection weights might be by considering functions of the CPI basket weights and
sales volumes. A key innovation is that prediction accuracy in \cite{GNAR2} was an aggregate prediction error taken
over all time series, whereas here we focus intensely on the single CPI variable (as mentioned in the previous but one paragraph),
which confers considerable speed and prediction accuracy advantages. The other major innovation is that we
consider both local-$\alpha$ {\em and} local-$\alpha\beta$ models.

\subsection{Benchmark Models}

As suggested in the introduction and highly-respected articles such as \cite{uk_2024},
we use random walks and autoregressive processes as benchmark models, which have been consistently found to be difficult to beat and among the best forecasting techniques that rely solely on the historical data. Table~\ref{tab:benchmarks} displays the RMSEs for forecasts from 2010 onwards for various benchmark models discussed next.

\textbf{Random Walk (RW) Models} \citep{atkeson2001phillips, stock2007has, inflationforecasting2013}.  The order $n$ RW  forecast for all horizons $h$ is defined as the average of the past $n$ observations
\begin{equation}
    \hat{y}_{t+h|t} = \frac{1}{n}\sum_{i=0}^{n-1} y_{t-i},
\end{equation}
which we denote RW($n$). We consider both a naive RW(1) forecaster, and a longer order RW(4) process as proposed by~\citet{atkeson2001phillips}.

\textbf{Autoregressive (AR) Processes} \citep{brd, stock2003forecasting, inflationforecasting2013}. An AR process of order $p$, which we denote AR$(p)$, is defined as 
\begin{equation}
    y_t = \phi_1 y_{t-1} + \cdots + \phi_p y_{t-p} + u_t,
\end{equation}
where $\phi_i \in\mathbb R$ are the coefficients of the model in the zone of stationarity
with $\phi_p\neq 0$ and the `noise' $ \{ u_t \}$ is a set of zero mean constant variance independent and identically
random variables. Recently, \citet{uk_2024} compared their models to a fixed AR(2) process. However, \citet{long_ar1, long_ar2} argue that, since inflation is calculated using year-on-year percentage changes, longer-order AR processes may be more suitable for capturing these effects. To account for this, we select the AR order each month based on the BIC. Figure~\ref{fig:bic_parameters} displays the optimal order over time. For most periods, an AR(2) model is deemed the most appropriate. However, in the recent period of heightened inflation, the BIC selects an AR(13) process, suggesting that longer memory effects have become more relevant.

\textbf{Autoregressive Model Averaging.} In addition to forecasting with a single AR model, where the order is dynamically selected each month, we also combine forecasts from AR processes of different orders. As discussed earlier, model averaging is a standard forecasting technique beyond the context of inflation. We denote the average of a set of AR models ${\text{AR}(p_i): i=1, \ldots, \ell}$ by AvAR($\{p_1, \ldots, p_\ell\}$). The two sets of model orders we average over are $\mathcal{P}_1 = \{1, 13, 25\}$ and $\mathcal{P}_2 = \{2, 13, 25\}$, with the rationale behind these choices being detailed in the Appendix.

The AvAR models consistently achieve the lowest forecasting error overall, with AvAR($\mathcal{P}_2$) performing best, making it our main benchmark throughout this article. To our knowledge, no prior work has explicitly averaged multiple AR processes for inflation forecasting, and more specifically in the UK, where recent studies—such as~\citet{boe2024}—have struggled to outperform a simple AR(2) benchmark. Thus, averaging across AR processes for inflation forecasting represents a novel, if somewhat minor, contribution of our work.

\textbf{Chronos} \citep{chronos}. Chronos is a collection of pre-trained probabilistic time series forecasting models developed by Amazon Science, leveraging transformer-based language model architectures. Its core architecture is T5~\citep{t5}, with models ranging from 8 million to 710 million parameters, trained on both publicly available and synthetic datasets. A key objective of Chronos is zero-shot forecasting—the ability to generate forecasts for unseen datasets. We tested various models and report results for the tiny (8 million parameters), base (200 million parameters), and large (710 million parameters) models. Recently, the family of Chronos-Bolt models were introduced promising improved accuracy, speed and memory efficiency. However, we found that these performed similarly to the original Chronos models and so are not reported here.

\begin{table}
    \centering
    \begin{tabular}{l|lllll}
    \toprule
     &  \multicolumn{5}{c}{Forecast Horizon (months)} \\
    Model & 1 & 3 & 6 & 9 & 12 \\
    \midrule
    RW(1) & 0.43 & 0.88 & 1.56 & 2.22 & 2.79 \\
    RW(4) & 0.73 & 1.19 & 1.87 & 2.48 & 3.00 \\
    AR($p$) & 0.41 & 0.79 & 1.43 & 2.13 & 2.73 \\
    AvAR($\mathcal{P}_1$) & 0.37 & 0.74 & 1.32 & 1.96 & 2.54 \\
    AvAR($\mathcal{P}_2$) & 0.37 & 0.74 & 1.32 & 1.95 & 2.51 \\
    Chronos (Tiny) & 0.41 & 0.83 & 1.53 & 2.23 & 2.85 \\
    Chronos (Base) & 0.43 & 0.84 & 1.49 & 2.12 & 2.66 \\
    Chronos (Large) & 0.43 & 0.84 & 1.49 & 2.13 & 2.62 \\
    \bottomrule
    \end{tabular}
    \caption{RMSEs for benchmark model CPI inflation forecasts from 2010 onwards.}
    \label{tab:benchmarks}
\end{table}

Several related works use the Phillips curve as a benchmark as well. Broadly speaking, the Phillips curve can be viewed as
equivalent to an AR$(p)$ process with an additional measure of economic activity as an extra predictor, often the unemployment rate (see~\cite{stock1999forecasting, atkeson2001phillips, stock2009phillips, inflationforecasting2013} for more details)
However, the literature seems to suggest that the Phillips curve does not provide significant improvements over other univariate time series models, and, in fact,~\cite{atkeson2001phillips} show that a naive RW forecaster generally outperforms it. Other benchmarks often also include the direct forecasting method~\cite{inflationforecasting2013}, which is shown to be outperformed by the iterative forecasts of an autoregressive model~\cite{marcellino2006comparison}, and hence we do not include these here either.

\section{Results}\label{sec:results}
This section presents the results for the various GNAR models discussed earlier, including a comparison of our forecasts with those released by the Bank of England. We generate 100 sets of $G=10000$ random networks using the Erdős–Rényi–Gilbert model~\cite{random_graphs} and compute inflation rate forecasts based on each previously described method. This approach allows us to estimate the average RMSE across the 100 sets of random graphs, as well as the standard deviation in performance. All RMSEs are calculated for forecasts spanning the 1\textsuperscript{st} January 2010 to the 31\textsuperscript{st} December 2024. In order to keep the notation compact, we define the model order sets $\mathcal{P}_1 = \{1,13,25\}$ and $\mathcal{P}_2 = \{2,13,25\}$ as for the AR processes above, as well as the neighbour sum sets $\mathcal{S}_1 = \{\boldsymbol{1}\}$, $\mathcal{S}_2 = \{\boldsymbol{2}\}$ and $\mathcal{S}_3 = \{\boldsymbol{1}, \boldsymbol{2}\}$. The results are detailed in the tables throughout this section.

\subsection{Best Performing Graph} \label{sec:best_graph}

We first present results obtained by
selecting the best network at each step and using it to forecast CPI inflation for the next twelve months. Tables~\ref{tab:best_net} and ~\ref{tab:best_net_mavg} display the mean
relative-to-benchmark RMSE over 100 runs for different GNAR processes. The models selected using BIC (shown in Table~\ref{tab:best_net}) generally struggle to outperform the benchmark, though standard and local-$\alpha\beta$ processes often achieve better results. However, all these models consistently outperform the corresponding BIC-selected AR($p$) forecasts, which has been argued by others  as a more traditional benchmark, as mentioned above. By contrast, AvGNAR models (shown in Table~\ref{tab:best_net_mavg}) consistently outperform the benchmark on average, often by a substantial margin. This is particularly notable given the number of parameters involved: after removing the mean from each time series, the largest global-$\alpha$ model, GNAR($25,\boldsymbol{2}$), has only 75 parameters.

By contrast a standard vector autoregressive model of order one, VAR$(1)$, contains $114^2 = 12996$ possible parameters.
Our CPI data consists of twenty years of monthly data for $N=114$ time series, which equates to 27360 observations in total.
The raw data size to
parameter ratio for GNAR is dramatically better than for standard VAR, which goes some way to explaining its attractive predictive abilities; GNAR models are rarely in danger of overfitting. Of course, the parameter explosion
in VAR models is well known and so modellers often use sparse or restricted VAR models, which cut down the number of
parameters fitted, but even here GNAR typically outpredicts them, see \cite{GNAR_covid, GNAR_covid2, us_politics}, for example. While GNAR can indeed be viewed as a form of restricted VAR, its key distinction lies in how the restrictions are imposed and their implications for estimation. In GNAR, the constraints are dictated by the network structure, effectively treating many coefficients as identical. In contrast, restricted VAR models typically retain a larger number of parameters, which can increase estimation error, particularly in small samples.
Indeed, we also tried VAR models (not reported) but they were not competitive.

\begin{table}
    \centering
    \begin{tabular}{ccr|lllll}
    \toprule
        \multicolumn{2}{c}{RaGNAR} & Number of &  \multicolumn{5}{c}{Forecast Horizon (months)} \\
    Class & Model & Parameters & 1 & 3 & 6 & 9 & 12 \\
    \midrule
    \multirow{3}{*}{\mbox{global-$\alpha$}} 
    & GNAR($p, \boldsymbol{1}$) & 2 - 4 & 1.08\tiny{$\pm$0.05} & 1.08\tiny{$\pm$0.07} & 1.04\tiny{$\pm$0.06} & 0.99\tiny{$\pm$0.04} & 0.95\tiny{$\pm$0.02} \\
    & GNAR($p, \boldsymbol{2}$) & 3 - 6 & 1.06\tiny{$\pm$0.04} & 1.06\tiny{$\pm$0.06} & 1.05\tiny{$\pm$0.06} & 1.04\tiny{$\pm$0.05} & 1.02\tiny{$\pm$0.05} \\
    & GNAR($p, \boldsymbol{s}$) & 2 - 6 & 1.07\tiny{$\pm$0.04} & 1.07\tiny{$\pm$0.06} & 1.06\tiny{$\pm$0.06} & 1.04\tiny{$\pm$0.05} & 1.02\tiny{$\pm$0.05} \\
    \midrule
    \multirow{3}*{\parbox{1.5cm}{standard\\\mbox{(local-$\alpha$)}}} 
    & GNAR($p, \boldsymbol{1}$) & 115 - 230 & 1.08\tiny{$\pm$0.04} & 1.03\tiny{$\pm$0.05} & 0.99\tiny{$\pm$0.06} & 0.99\tiny{$\pm$0.06} & 0.98\tiny{$\pm$0.06} \\
    & GNAR($p, \boldsymbol{2}$) & 116 - 232 & 1.04\tiny{$\pm$0.04} & 1.00\tiny{$\pm$0.05} & 0.97\tiny{$\pm$0.07} & 0.97\tiny{$\pm$0.07} & 0.97\tiny{$\pm$0.07} \\
    & GNAR($p, \boldsymbol{s}$) & 115 - 232 & 1.05\tiny{$\pm$0.04} & 1.01\tiny{$\pm$0.05} & 0.98\tiny{$\pm$0.06} & 0.98\tiny{$\pm$0.06} & 0.98\tiny{$\pm$0.07} \\
    \midrule
    \multirow{3}{*}{\mbox{local-$\alpha\beta$}} 
    & GNAR($p, \boldsymbol{1}$) & 228 - 456 & 1.04\tiny{$\pm$0.02} & 0.96\tiny{$\pm$0.03} & 0.88\tiny{$\pm$0.03} & 0.88\tiny{$\pm$0.03} & 0.87\tiny{$\pm$0.03} \\
    & GNAR($p, \boldsymbol{2}$) & 342 - 684 & 1.00\tiny{$\pm$0.02} & 0.93\tiny{$\pm$0.03} & 0.87\tiny{$\pm$0.04} & 0.87\tiny{$\pm$0.04} & 0.88\tiny{$\pm$0.04} \\
    & GNAR($p, \boldsymbol{s}$) & 228 - 684 & 1.01\tiny{$\pm$0.02} & 0.94\tiny{$\pm$0.03} & 0.87\tiny{$\pm$0.04} & 0.87\tiny{$\pm$0.04} & 0.87\tiny{$\pm$0.05} \\
    \bottomrule
    \end{tabular}
    \caption{Relative to benchmark RMSEs for inflation rate forecasts using GNAR models with parameters selected via BIC and fitted to the best-performing network each month. All RMSEs are reported relative to the AR($\mathcal{P}_2$) benchmark and averaged over 100 runs of the algorithm, with $\pm1$ standard deviation displayed in a smaller font.}
    \label{tab:best_net}
\end{table}

\begin{table}
    \centering
    \begin{tabular}{cc|lllll}
    \toprule
        \multicolumn{2}{c|}{RaGNAR} &  \multicolumn{5}{c}{Forecast Horizon (months)} \\
    Class & Model & 1 & 3 & 6 & 9 & 12 \\
    \midrule
    \multirow{6}{*}{\mbox{global-$\alpha$}} 
    & AvGNAR($\mathcal{P}_1,\mathcal{S}_1$) & 0.97\tiny{$\pm$0.02} & 0.98\tiny{$\pm$0.03} & 0.93\tiny{$\pm$0.03} & 0.92\tiny{$\pm$0.02} & 0.92\tiny{$\pm$0.02} \\
    & AvGNAR($\mathcal{P}_2,\mathcal{S}_1$) & 0.96\tiny{$\pm$0.02} & 0.96\tiny{$\pm$0.03} & 0.91\tiny{$\pm$0.02} & 0.91\tiny{$\pm$0.02} & 0.90\tiny{$\pm$0.02} \\
    & AvGNAR($\mathcal{P}_1,\mathcal{S}_2$) & 0.94\tiny{$\pm$0.04} & 0.92\tiny{$\pm$0.05} & 0.87\tiny{$\pm$0.04} & 0.88\tiny{$\pm$0.03} & 0.89\tiny{$\pm$0.02} \\
    & AvGNAR($\mathcal{P}_2,\mathcal{S}_2$) & 0.94\tiny{$\pm$0.06} & 0.93\tiny{$\pm$0.08} & 0.88\tiny{$\pm$0.06} & 0.89\tiny{$\pm$0.06} & 0.90\tiny{$\pm$0.04} \\
    & AvGNAR($\mathcal{P}_1,\mathcal{S}_3$) & 0.94\tiny{$\pm$0.02} & 0.93\tiny{$\pm$0.02} & 0.89\tiny{$\pm$0.02} & 0.89\tiny{$\pm$0.02} & 0.89\tiny{$\pm$0.01} \\
    & AvGNAR($\mathcal{P}_2,\mathcal{S}_3$) & 0.93\tiny{$\pm$0.03} & 0.92\tiny{$\pm$0.03} & 0.88\tiny{$\pm$0.02} & 0.88\tiny{$\pm$0.02} & 0.89\tiny{$\pm$0.02} \\
    \midrule
    \multirow{6}*{\parbox{1.5cm}{standard\\\mbox{(local-$\alpha$)}}} 
    & AvGNAR($\mathcal{P}_1,\mathcal{S}_1$) & 0.96\tiny{$\pm$0.02} & 0.92\tiny{$\pm$0.02} & 0.89\tiny{$\pm$0.02} & 0.88\tiny{$\pm$0.02} & 0.89\tiny{$\pm$0.02} \\
    & AvGNAR($\mathcal{P}_2,\mathcal{S}_1$) & 0.96\tiny{$\pm$0.02} & 0.92\tiny{$\pm$0.03} & 0.88\tiny{$\pm$0.02} & 0.88\tiny{$\pm$0.02} & 0.89\tiny{$\pm$0.01} \\
    & AvGNAR($\mathcal{P}_1,\mathcal{S}_2$) & 0.95\tiny{$\pm$0.04} & 0.91\tiny{$\pm$0.03} & 0.88\tiny{$\pm$0.02} & 0.89\tiny{$\pm$0.02} & 0.90\tiny{$\pm$0.02} \\
    & AvGNAR($\mathcal{P}_2,\mathcal{S}_2$) & 0.97\tiny{$\pm$0.07} & 0.92\tiny{$\pm$0.04} & 0.89\tiny{$\pm$0.03} & 0.90\tiny{$\pm$0.07} & 0.93\tiny{$\pm$0.32} \\
    & AvGNAR($\mathcal{P}_1,\mathcal{S}_3$) & 0.94\tiny{$\pm$0.02} & 0.90\tiny{$\pm$0.02} & 0.88\tiny{$\pm$0.02} & 0.88\tiny{$\pm$0.01} & 0.89\tiny{$\pm$0.01} \\
    & AvGNAR($\mathcal{P}_2,\mathcal{S}_3$) & 0.95\tiny{$\pm$0.03} & 0.91\tiny{$\pm$0.02} & 0.88\tiny{$\pm$0.02} & 0.88\tiny{$\pm$0.03} & 0.90\tiny{$\pm$0.13} \\
    \midrule
    \multirow{6}{*}{\mbox{local-$\alpha\beta$}}
    & AvGNAR($\mathcal{P}_1,\mathcal{S}_1$) & 0.90\tiny{$\pm$0.02} & 0.89\tiny{$\pm$0.02} & 0.86\tiny{$\pm$0.02} & 0.86\tiny{$\pm$0.03} & 0.88\tiny{$\pm$0.03} \\
    & AvGNAR($\mathcal{P}_2,\mathcal{S}_1$) & 0.89\tiny{$\pm$0.02} & 0.90\tiny{$\pm$0.02} & 0.87\tiny{$\pm$0.02} & 0.87\tiny{$\pm$0.02} & 0.89\tiny{$\pm$0.02} \\
    & AvGNAR($\mathcal{P}_1,\mathcal{S}_2$) & 0.90\tiny{$\pm$0.02} & 0.90\tiny{$\pm$0.03} & 0.88\tiny{$\pm$0.03} & 0.89\tiny{$\pm$0.08} & 0.96\tiny{$\pm$0.53} \\
    & AvGNAR($\mathcal{P}_2,\mathcal{S}_2$) & 0.90\tiny{$\pm$0.03} & 0.91\tiny{$\pm$0.03} & 0.89\tiny{$\pm$0.03} & 0.91\tiny{$\pm$0.08} & 0.98\tiny{$\pm$0.53} \\
    & AvGNAR($\mathcal{P}_1,\mathcal{S}_3$) & 0.88\tiny{$\pm$0.02} & 0.88\tiny{$\pm$0.02} & 0.85\tiny{$\pm$0.02} & 0.85\tiny{$\pm$0.03} & 0.90\tiny{$\pm$0.24} \\
    & AvGNAR($\mathcal{P}_2,\mathcal{S}_3$) & 0.87\tiny{$\pm$0.02} & 0.89\tiny{$\pm$0.02} & 0.86\tiny{$\pm$0.02} & 0.87\tiny{$\pm$0.03} & 0.92\tiny{$\pm$0.23} \\
    \bottomrule
    \end{tabular}
    \caption{Relative to benchmark RMSEs for inflation rate forecasts computed by averaging different order GNAR models fitted to the best-performing network each month. Here $\mathcal{P}_1 = \{1,13,25\}$, $\mathcal{P}_2 = \{2,13,25\}$, $\mathcal{S}_1 = \{\boldsymbol{1}\}$, $\mathcal{S}_2 = \{\boldsymbol{2}\}$, and $\mathcal{S}_3 = \{\boldsymbol{1}, \boldsymbol{2}\}$. All RMSEs are reported relative to the AvAR($\mathcal{P}_2$) benchmark and averaged over 100 runs of the algorithm, with $\pm1$ standard deviation displayed in a smaller font.}
    \label{tab:best_net_mavg}
\end{table}

\subsection{Model Averaging}

We now consider averaging the predictions of the GNAR processes fitted to the top $n$ networks. Table~\ref{tab:averages5} compares our forecasts for $n=2$ and $n=5$ against those of the AR($\mathcal{P}_2$) benchmark. Our results improve as $n$ increases (the results for $n=2, 3$ and $4$ are not presented for brevity), while displaying the same patterns as for the $n=1$ case discussed in the previous sub-section. More importantly, as $n$ increases, the RMSE standard deviations decrease significantly, indicating that averaging predictions across multiple networks yields more robust and better forecasts. This effect is more pronounced at longer horizons, since these are the ones which rely on the network structure outside of the CPI node neighbour sets.

\begin{figure}
    \centering
    \resizebox{\textwidth}{!}{\includegraphics{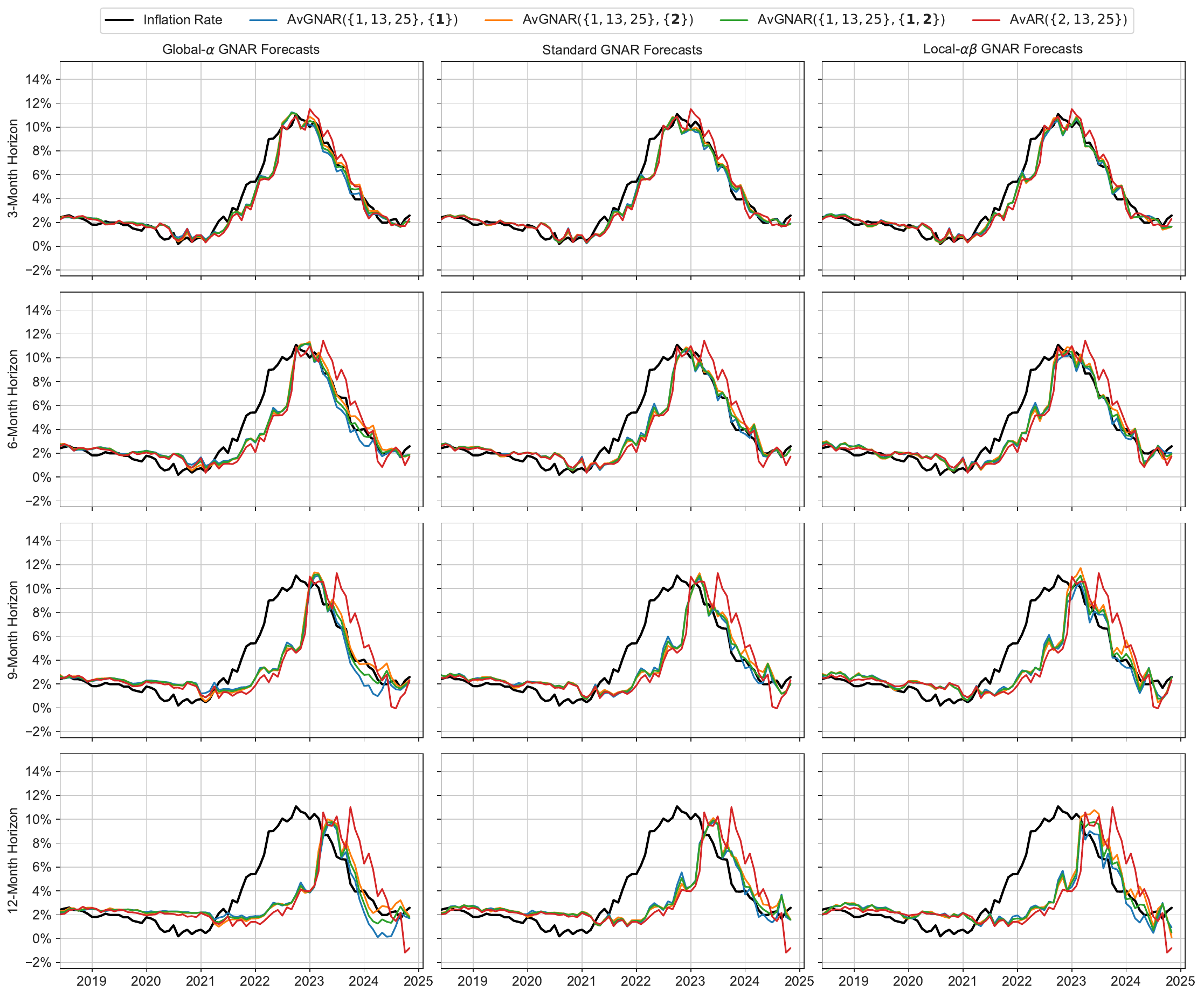}}
    \caption{Comparison of some of our inflation forecasts computed by averaging the predictions of GNAR models fitted to the $n=5$ best-performing networks each month against the AvAR($\mathcal{P}_2$) benchmark model.The better performance
    of the AvGNAR models against AvAR during the CPI stabilisation phase
    can be clearly seen, as the dark red AvAR line clearly lags the others and CPI.}
    \label{fig:avg5}
\end{figure}

Figure~\ref{fig:avg5} compares some of our forecasts for the $n=5$ case with the AR($\mathcal{P}_2$) benchmark from 2018 onwards, a period marked by high inflation rates, peaking at 11\% in October 2022 before returning to the Bank of England's 2\% target in the second quarter of 2024. As seen in the plots, RaGNAR slightly outperforms other models in forecasting the rise in inflation, though all models struggle during this period. However, RaGNAR’s real strength lies when predicting inflation rates during the period of stabilisation, where it significantly outperforms the benchmark. While the AR($\mathcal{P}_2$) model overshoots the inflation peak and overcorrects during the decline, RaGNAR more accurately anticipates the return to stable inflation rates. This superior performance is due to RaGNAR’s ability to exploit the information contained in the disaggregated item series, demonstrating a clear and practical advantage of our approach. During calmer periods, however, these item series contribute less additional value, making RaGNAR’s forecasts more similar to those of the benchmark.

Figure~\ref{fig:avg5} also highlights a lag in how information from disaggregated item series influences forecasts, with this delay increasing as the forecast horizon extends. Specifically, all models fail to anticipate the rise in inflation around mid-2021. At this stage, RaGNAR has yet to identify which item series are most useful, and only after some time do these components appear in the neighbour sets, leading to significant forecast improvements—a trend that is even more pronounced at longer horizons.

\begin{table}
    \centering
    \begin{tabular}{cl|lllll}
    \toprule
        \multicolumn{2}{c|}{RaGNAR} &  \multicolumn{5}{c}{Forecast Horizon (months)} \\
    Class & \multicolumn{1}{c|}{Model} & 1 & 3 & 6 & 9 & 12 \\
    \midrule
    \multirow{9}{*}{\mbox{global-$\alpha$}}
    & GNAR($p, \boldsymbol{1}$) & 1.04\tiny{$\pm$0.02} & 1.03\tiny{$\pm$0.04} & 1.00\tiny{$\pm$0.03} & 0.97\tiny{$\pm$0.02} & 0.93\tiny{$\pm$0.01} \\
    & GNAR($p, \boldsymbol{2}$) & 1.03\tiny{$\pm$0.02} & 1.01\tiny{$\pm$0.03} & 1.02\tiny{$\pm$0.03} & 1.02\tiny{$\pm$0.02} & 1.00\tiny{$\pm$0.02} \\
    & GNAR($p, \boldsymbol{s}$) & 1.04\tiny{$\pm$0.02} & 1.03\tiny{$\pm$0.03} & 1.03\tiny{$\pm$0.03} & 1.03\tiny{$\pm$0.02} & 1.01\tiny{$\pm$0.02} \\
    & AvGNAR($\mathcal{P}_1,\mathcal{S}_1$) & 0.96\tiny{$\pm$0.01} & 0.96\tiny{$\pm$0.02} & 0.92\tiny{$\pm$0.01} & 0.91\tiny{$\pm$0.01} & 0.91\tiny{$\pm$0.01} \\
    & AvGNAR($\mathcal{P}_2,\mathcal{S}_1$) & 0.94\tiny{$\pm$0.01} & 0.94\tiny{$\pm$0.01} & 0.90\tiny{$\pm$0.01} & 0.90\tiny{$\pm$0.01} & 0.90\tiny{$\pm$0.01} \\
    & AvGNAR($\mathcal{P}_1,\mathcal{S}_2$) & 0.90\tiny{$\pm$0.01} & 0.89\tiny{$\pm$0.02} & 0.85\tiny{$\pm$0.01} & 0.86\tiny{$\pm$0.01} & 0.88\tiny{$\pm$0.01} \\
    & AvGNAR($\mathcal{P}_2,\mathcal{S}_2$) & 0.90\tiny{$\pm$0.01} & 0.89\tiny{$\pm$0.02} & 0.85\tiny{$\pm$0.01} & 0.86\tiny{$\pm$0.01} & 0.88\tiny{$\pm$0.01} \\
    & AvGNAR($\mathcal{P}_1,\mathcal{S}_3$) & 0.92\tiny{$\pm$0.01} & 0.91\tiny{$\pm$0.01} & 0.87\tiny{$\pm$0.01} & 0.88\tiny{$\pm$0.01} & 0.89\tiny{$\pm$0.01} \\
    & AvGNAR($\mathcal{P}_2,\mathcal{S}_3$) & 0.92\tiny{$\pm$0.01} & 0.91\tiny{$\pm$0.01} & 0.86\tiny{$\pm$0.01} & 0.87\tiny{$\pm$0.01} & 0.88\tiny{$\pm$0.01} \\
    \midrule
    \multirow{9}*{\parbox{1.5cm}{standard\\\mbox{(local-$\alpha$)}}}
    & GNAR($p, \boldsymbol{1}$) & 1.06\tiny{$\pm$0.02} & 1.00\tiny{$\pm$0.02} & 0.96\tiny{$\pm$0.03} & 0.97\tiny{$\pm$0.04} & 0.96\tiny{$\pm$0.04} \\
    & GNAR($p, \boldsymbol{2}$) & 1.02\tiny{$\pm$0.02} & 0.97\tiny{$\pm$0.03} & 0.95\tiny{$\pm$0.04} & 0.96\tiny{$\pm$0.03} & 0.96\tiny{$\pm$0.03} \\
    & GNAR($p, \boldsymbol{s}$) & 1.03\tiny{$\pm$0.02} & 0.97\tiny{$\pm$0.03} & 0.95\tiny{$\pm$0.04} & 0.96\tiny{$\pm$0.03} & 0.97\tiny{$\pm$0.03} \\
    & AvGNAR($\mathcal{P}_1,\mathcal{S}_1$) & 0.95\tiny{$\pm$0.01} & 0.92\tiny{$\pm$0.01} & 0.87\tiny{$\pm$0.01} & 0.88\tiny{$\pm$0.01} & 0.89\tiny{$\pm$0.01} \\
    & AvGNAR($\mathcal{P}_2,\mathcal{S}_1$) & 0.96\tiny{$\pm$0.01} & 0.93\tiny{$\pm$0.01} & 0.88\tiny{$\pm$0.01} & 0.88\tiny{$\pm$0.01} & 0.88\tiny{$\pm$0.01} \\
    & AvGNAR($\mathcal{P}_1,\mathcal{S}_2$) & 0.93\tiny{$\pm$0.02} & 0.89\tiny{$\pm$0.01} & 0.87\tiny{$\pm$0.01} & 0.88\tiny{$\pm$0.01} & 0.89\tiny{$\pm$0.01} \\
    & AvGNAR($\mathcal{P}_2,\mathcal{S}_2$) & 0.93\tiny{$\pm$0.02} & 0.89\tiny{$\pm$0.01} & 0.87\tiny{$\pm$0.01} & 0.88\tiny{$\pm$0.01} & 0.90\tiny{$\pm$0.03} \\
    & AvGNAR($\mathcal{P}_1,\mathcal{S}_3$) & 0.94\tiny{$\pm$0.01} & 0.90\tiny{$\pm$0.01} & 0.87\tiny{$\pm$0.01} & 0.87\tiny{$\pm$0.01} & 0.89\tiny{$\pm$0.01} \\
    & AvGNAR($\mathcal{P}_2,\mathcal{S}_3$) & 0.94\tiny{$\pm$0.01} & 0.90\tiny{$\pm$0.01} & 0.87\tiny{$\pm$0.01} & 0.87\tiny{$\pm$0.01} & 0.89\tiny{$\pm$0.01} \\
    \midrule
    \multirow{9}{*}{\mbox{local-$\alpha\beta$}}
    & GNAR($p, \boldsymbol{1}$) & 1.02\tiny{$\pm$0.01} & 0.93\tiny{$\pm$0.01} & 0.86\tiny{$\pm$0.02} & 0.86\tiny{$\pm$0.02} & 0.85\tiny{$\pm$0.02} \\
    & GNAR($p, \boldsymbol{2}$) & 0.97\tiny{$\pm$0.01} & 0.89\tiny{$\pm$0.01} & 0.83\tiny{$\pm$0.02} & 0.84\tiny{$\pm$0.02} & 0.85\tiny{$\pm$0.02} \\
    & GNAR($p, \boldsymbol{s}$) & 0.98\tiny{$\pm$0.01} & 0.90\tiny{$\pm$0.01} & 0.83\tiny{$\pm$0.02} & 0.84\tiny{$\pm$0.02} & 0.84\tiny{$\pm$0.02} \\
    & AvGNAR($\mathcal{P}_1,\mathcal{S}_1$) & 0.88\tiny{$\pm$0.01} & 0.87\tiny{$\pm$0.01} & 0.84\tiny{$\pm$0.01} & 0.85\tiny{$\pm$0.01} & 0.86\tiny{$\pm$0.01} \\
    & AvGNAR($\mathcal{P}_2,\mathcal{S}_1$) & 0.87\tiny{$\pm$0.01} & 0.88\tiny{$\pm$0.01} & 0.85\tiny{$\pm$0.01} & 0.86\tiny{$\pm$0.01} & 0.88\tiny{$\pm$0.01} \\
    & AvGNAR($\mathcal{P}_1,\mathcal{S}_2$) & 0.86\tiny{$\pm$0.01} & 0.86\tiny{$\pm$0.01} & 0.83\tiny{$\pm$0.02} & 0.85\tiny{$\pm$0.02} & 0.88\tiny{$\pm$0.07} \\
    & AvGNAR($\mathcal{P}_2,\mathcal{S}_2$) & 0.85\tiny{$\pm$0.01} & 0.87\tiny{$\pm$0.01} & 0.85\tiny{$\pm$0.02} & 0.86\tiny{$\pm$0.02} & 0.89\tiny{$\pm$0.07} \\
    & AvGNAR($\mathcal{P}_1,\mathcal{S}_3$) & 0.86\tiny{$\pm$0.01} & 0.86\tiny{$\pm$0.01} & 0.83\tiny{$\pm$0.01} & 0.84\tiny{$\pm$0.01} & 0.86\tiny{$\pm$0.02} \\
    & AvGNAR($\mathcal{P}_2,\mathcal{S}_3$) & 0.86\tiny{$\pm$0.01} & 0.87\tiny{$\pm$0.01} & 0.84\tiny{$\pm$0.01} & 0.85\tiny{$\pm$0.01} & 0.88\tiny{$\pm$0.02} \\
    \bottomrule
    \end{tabular}
    \caption{Relative to benchmark RMSEs for the same models as in Tables~\ref{tab:best_net} and \ref{tab:best_net_mavg}, with forecasts averaged across the $n=5$ best-performing networks. All RMSEs are reported relative to the AvAR($\mathcal{P}_2$) benchmark and averaged over 100 runs of the algorithm, with $\pm1$ standard deviation displayed in a smaller font.}
    \label{tab:averages5}
\end{table}

\subsection{Bank of England}

The Bank of England employs a forecasting platform consisting of a central model, COMPASS, which is a New Keynesian general equilibrium model, a suite of at least 40 other models, including ARMA and survey models for example, as well as judgemental insights from the Monetary Policy Committee~\cite{boe2013, boe2024}. Each forecasting round begins from the previous inflation rate forecasts and builds on these incrementally, combining these with new data as it becomes available. There are at least five meetings between the Monetary Policy Committee and the Bank staff, during which judgemental adjustments are made to the forecasts~\cite{boe2013, boe2024}. Moreover, to produce a single set of forecasts each quarter, the COMPASS algorithm is run hundreds of times in order to incorporate insights from the Monetary Policy Committee and explore different scenarios~\cite{boe2013, boe2024}.  In recent years, central bank forecasting accuracy has declined, particularly during the sharp rise in inflation following the COVID-19 pandemic, which both central banks and external forecasters struggled to predict~\cite{boe2024, central_banks_2024}. In fact, ~\citet{central_banks_2024} find that inflation forecasts produced by central banks at short horizons have generally been unbiased and quite efficient, but that this performance deteriorates for longer horizons. Further, \cite{boe2024} notes that ``the forecast errors made by the Bank and those made by external forecasters are barely distinguishable'', whereas the results below show that our errors, and forecasts, are demonstrably better.

We compare our forecasts with those of the Bank of England, published every quarter in the Bank of England's Monetary Policy Reports (see \cite{mpc_aug_2024} for the latest one). Until recently, the Bank of England has released inflation rate predictions for up to 3 months ahead; this horizon has now been extended to 6 months. Given that the Bank publishes forecasts on a quarterly basis, our comparisons are limited to the corresponding months.

Table~\ref{tab:boe} show the results for all models.
The Bank of England strongly outperforms our models for horizons of one, two and three months.
However, several GNAR models notably compete, and in many cases surpass, the Bank's forecast for horizons of four, five and six months, with the AvGNAR models emerging as the overall best performers. This is especially noteworthy given the sophistication
and expense of the Bank of England's forecasting procedure, which integrates multiple models, extensive data analysis, and expert judgment,
compared to the simplicity of RaGNAR. The Bank's approach involves complex econometric models and regular adjustments by the Monetary Policy Committee to account for the latest economic developments, making it a formidable benchmark. That RaGNAR can outperform such a comprehensive and refined forecasting system over longer horizons highlights the strength and potential of our approach, particularly in capturing trends and dynamics that may become more pronounced over extended periods. Moreover, since the Bank of England's forecasts are only available from the end of 2019 onwards, Table~\ref{tab:boe} also highlights how our GNAR models have outperformed the benchmarks significantly more in recent years. This is because RaGNAR leverages valuable information from disaggregated item series during periods of unstable inflation, whereas in calmer times, these series add little value to forecasts.

\begin{table}
    \centering
    \begin{tabular}{cl|llllll}
    \toprule
    \multicolumn{2}{c|}{} & \multicolumn{6}{c}{Forecast Horizon (months)}\\
    \multicolumn{2}{c|}{Forecaster}   & 1 & 2 & 3 & 4 & 5 & 6 \\
    \midrule
    & Bank of England &  0.17 & 0.32 & 0.48 & 0.85 & 0.93 & 0.92 \\ 
    & AvAR($\mathcal{P}_2$) & 0.72 & 0.88 & 0.98 & 1.03 & 1.06 & 1.00 \\
    \midrule
    \multirow{9}{*}{\rotatebox{90}{\mbox{global-$\alpha$}}}
    & GNAR($p,\boldsymbol{1}$) &  0.72\tiny{$\pm$0.04} & 0.90\tiny{$\pm$0.04} & 1.00\tiny{$\pm$0.06} & 0.99\tiny{$\pm$0.12} & 1.09\tiny{$\pm$0.13} & 1.07\tiny{$\pm$0.13} \\
    & GNAR($p,\boldsymbol{2}$) & 0.72\tiny{$\pm$0.03} & 0.89\tiny{$\pm$0.03} & 0.98\tiny{$\pm$0.04} & 1.48\tiny{$\pm$0.10} & 1.62\tiny{$\pm$0.11} & 1.75\tiny{$\pm$0.13} \\
    & GNAR($p,\boldsymbol{s}$) &  0.72\tiny{$\pm$0.03} & 0.90\tiny{$\pm$0.03} & 0.98\tiny{$\pm$0.04} & 1.48\tiny{$\pm$0.10} & 1.62\tiny{$\pm$0.11} & 1.75\tiny{$\pm$0.13} \\
    & AvGNAR($\mathcal{P}_1,\mathcal{S}_1$) & 0.63\tiny{$\pm$0.02} & 0.78\tiny{$\pm$0.02} & 0.91\tiny{$\pm$0.03} & 0.31\tiny{$\pm$0.06} & 0.53\tiny{$\pm$0.09} & 0.70\tiny{$\pm$0.11} \\
    & AvGNAR($\mathcal{P}_2,\mathcal{S}_1$) & 0.61\tiny{$\pm$0.02} & 0.77\tiny{$\pm$0.02} & 0.89\tiny{$\pm$0.03} & 0.30\tiny{$\pm$0.06} & 0.48\tiny{$\pm$0.09} & 0.66\tiny{$\pm$0.13} \\
    & AvGNAR($\mathcal{P}_1,\mathcal{S}_2$) & 0.61\tiny{$\pm$0.02} & 0.72\tiny{$\pm$0.02} & 0.79\tiny{$\pm$0.03} & 0.49\tiny{$\pm$0.08} & 0.53\tiny{$\pm$0.10} & 0.58\tiny{$\pm$0.13} \\
    & AvGNAR($\mathcal{P}_2,\mathcal{S}_2$) & 0.59\tiny{$\pm$0.02} & 0.72\tiny{$\pm$0.02} & 0.80\tiny{$\pm$0.03} & 0.51\tiny{$\pm$0.08} & 0.56\tiny{$\pm$0.11} & 0.63\tiny{$\pm$0.15} \\
    & AvGNAR($\mathcal{P}_1,\mathcal{S}_3$) & 0.61\tiny{$\pm$0.01} & 0.74\tiny{$\pm$0.02} & 0.84\tiny{$\pm$0.02} & 0.32\tiny{$\pm$0.05} & 0.43\tiny{$\pm$0.06} & 0.53\tiny{$\pm$0.08} \\
    & AvGNAR($\mathcal{P}_2,\mathcal{S}_3$) & 0.60\tiny{$\pm$0.02} & 0.74\tiny{$\pm$0.01} & 0.83\tiny{$\pm$0.02} & 0.35\tiny{$\pm$0.05} & 0.44\tiny{$\pm$0.07} & 0.56\tiny{$\pm$0.10} \\
    \midrule
    \multirow{9}{*}{\rotatebox{90}{\mbox{standard}}}
    & GNAR($p,\boldsymbol{1}$) & 0.74\tiny{$\pm$0.02} & 0.89\tiny{$\pm$0.03} & 0.99\tiny{$\pm$0.04} & 1.14\tiny{$\pm$0.21} & 1.26\tiny{$\pm$0.22} & 1.28\tiny{$\pm$0.26} \\
    & GNAR($p,\boldsymbol{2}$) & 0.72\tiny{$\pm$0.03} & 0.87\tiny{$\pm$0.04} & 0.94\tiny{$\pm$0.05} & 1.11\tiny{$\pm$0.21} & 1.26\tiny{$\pm$0.22} & 1.31\tiny{$\pm$0.24} \\
    & GNAR($p,\boldsymbol{s}$) & 0.72\tiny{$\pm$0.03} & 0.88\tiny{$\pm$0.04} & 0.94\tiny{$\pm$0.04} & 1.11\tiny{$\pm$0.21} & 1.26\tiny{$\pm$0.22} & 1.31\tiny{$\pm$0.24} \\
    & AvGNAR($\mathcal{P}_1,\mathcal{S}_1$) & 0.64\tiny{$\pm$0.01} & 0.78\tiny{$\pm$0.02} & 0.85\tiny{$\pm$0.02} & 0.47\tiny{$\pm$0.05} & 0.47\tiny{$\pm$0.06} & 0.43\tiny{$\pm$0.07} \\
    & AvGNAR($\mathcal{P}_2,\mathcal{S}_1$) & 0.64\tiny{$\pm$0.01} & 0.79\tiny{$\pm$0.02} & 0.86\tiny{$\pm$0.02} & 0.50\tiny{$\pm$0.05} & 0.55\tiny{$\pm$0.07} & 0.60\tiny{$\pm$0.08} \\
    & AvGNAR($\mathcal{P}_1,\mathcal{S}_2$) & 0.64\tiny{$\pm$0.01} & 0.77\tiny{$\pm$0.02} & 0.81\tiny{$\pm$0.03} & 0.56\tiny{$\pm$0.06} & 0.56\tiny{$\pm$0.07} & 0.51\tiny{$\pm$0.06} \\
    & AvGNAR($\mathcal{P}_2,\mathcal{S}_2$) & 0.62\tiny{$\pm$0.01} & 0.76\tiny{$\pm$0.02} & 0.81\tiny{$\pm$0.02} & 0.52\tiny{$\pm$0.05} & 0.50\tiny{$\pm$0.06} & 0.47\tiny{$\pm$0.06} \\
    & AvGNAR($\mathcal{P}_1,\mathcal{S}_3$) & 0.64\tiny{$\pm$0.01} & 0.77\tiny{$\pm$0.01} & 0.82\tiny{$\pm$0.02} & 0.49\tiny{$\pm$0.04} & 0.48\tiny{$\pm$0.04} & 0.42\tiny{$\pm$0.05} \\
    & AvGNAR($\mathcal{P}_2,\mathcal{S}_3$) & 0.63\tiny{$\pm$0.01} & 0.77\tiny{$\pm$0.01} & 0.83\tiny{$\pm$0.02} & 0.47\tiny{$\pm$0.03} & 0.48\tiny{$\pm$0.04} & 0.47\tiny{$\pm$0.05} \\
    \midrule
    \multirow{9}{*}{\rotatebox{90}{\mbox{local-$\alpha\beta$}}}
    & GNAR($p,\boldsymbol{1}$) &  0.71\tiny{$\pm$0.01} & 0.82\tiny{$\pm$0.02} & 0.89\tiny{$\pm$0.03} & 0.75\tiny{$\pm$0.13} & 0.80\tiny{$\pm$0.13} & 0.72\tiny{$\pm$0.14} \\
    & GNAR($p,\boldsymbol{2}$) & 0.67\tiny{$\pm$0.01} & 0.78\tiny{$\pm$0.02} & 0.82\tiny{$\pm$0.02} & 0.67\tiny{$\pm$0.07} & 0.77\tiny{$\pm$0.08} & 0.72\tiny{$\pm$0.09} \\
    & GNAR($p,\boldsymbol{s}$) & 0.67\tiny{$\pm$0.01} & 0.79\tiny{$\pm$0.02} & 0.83\tiny{$\pm$0.02} & 0.67\tiny{$\pm$0.07} & 0.77\tiny{$\pm$0.08} & 0.72\tiny{$\pm$0.09} \\
    & AvGNAR($\mathcal{P}_1,\mathcal{S}_1$) & 0.58\tiny{$\pm$0.01} & 0.73\tiny{$\pm$0.01} & 0.80\tiny{$\pm$0.02} & 0.54\tiny{$\pm$0.04} & 0.47\tiny{$\pm$0.04} & 0.57\tiny{$\pm$0.06} \\
    & AvGNAR($\mathcal{P}_2,\mathcal{S}_1$) & 0.55\tiny{$\pm$0.01} & 0.73\tiny{$\pm$0.01} & 0.79\tiny{$\pm$0.02} & 0.62\tiny{$\pm$0.04} & 0.53\tiny{$\pm$0.04} & 0.67\tiny{$\pm$0.05} \\
    & AvGNAR($\mathcal{P}_1,\mathcal{S}_2$) & 0.56\tiny{$\pm$0.01} & 0.69\tiny{$\pm$0.02} & 0.75\tiny{$\pm$0.03} & 0.57\tiny{$\pm$0.07} & 0.53\tiny{$\pm$0.08} & 0.60\tiny{$\pm$0.07} \\
    & AvGNAR($\mathcal{P}_2,\mathcal{S}_2$) & 0.54\tiny{$\pm$0.02} & 0.69\tiny{$\pm$0.02} & 0.75\tiny{$\pm$0.03} & 0.66\tiny{$\pm$0.07} & 0.61\tiny{$\pm$0.09} & 0.66\tiny{$\pm$0.09} \\
    & AvGNAR($\mathcal{P}_1,\mathcal{S}_3$) & 0.57\tiny{$\pm$0.01} & 0.71\tiny{$\pm$0.01} & 0.77\tiny{$\pm$0.02} & 0.54\tiny{$\pm$0.04} & 0.47\tiny{$\pm$0.05} & 0.55\tiny{$\pm$0.04} \\
    & AvGNAR($\mathcal{P}_2,\mathcal{S}_3$) & 0.54\tiny{$\pm$0.01} & 0.70\tiny{$\pm$0.01} & 0.76\tiny{$\pm$0.02} & 0.63\tiny{$\pm$0.04} & 0.55\tiny{$\pm$0.05} & 0.63\tiny{$\pm$0.05} \\
    \bottomrule
    \end{tabular}
    \caption{Comparison of GNAR and Bank of England inflation rate forecasts on available dates. GNAR forecasts are computed as the average of the $n=5$ best-performing networks. Model parameters $p$ and $\boldsymbol{s}$ are selected monthly using the BIC, while for AvGNAR processes, we define $\mathcal{P}_1 = \{1,13,25\}$, $\mathcal{P}_2 = \{2,13,25\}$, $\mathcal{S}_1 = \{\boldsymbol{1}\}$, $\mathcal{S}_2 = \{\boldsymbol{2}\}$, and $\mathcal{S}_3 = \{\boldsymbol{1}, \boldsymbol{2}\}$. RMSEs are averaged over 100 runs of the algorithm, with $\pm1$ standard deviation displayed in a smaller font.  All RMSEs displayed are \underline{absolute} and not relative to AvAR($\mathcal{P}_2$).}
    \label{tab:boe}
\end{table}

\subsection{Mean Absolute Percentage Error}

We now compare RaGNAR’s forecasts to the benchmark and the Bank of England using Mean Absolute Percentage Error (MAPE) instead of RMSE. In our setting, MAPE is defined as
\begin{equation}
    \operatorname{MAPE} = T^{-1}\sum_{t=1}^{T}\frac{|X_t - \hat{X}_t|}{\left|X_t\right| + 1},
\end{equation}
where $T$ is the number of predictions. The denominator includes a +1 adjustment to avoid division by zero when inflation is zero. MAPE values for different models against the benchmark are presented in Table~\ref{tab:mape_averages5}, while comparisons to the Bank of England’s forecasts are displayed in Table~\ref{tab:boe_mape}. 

The MAPEs in Table~\ref{tab:mape_averages5} follow a similar pattern to their RMSE counterparts in Table~\ref{tab:averages5}, though the margin of improvement is slightly lower in some cases. Nevertheless, as before, AvGNAR models consistently outperform the AvAR($\mathcal{P}_1$) benchmark by a wide margin in nearly all cases.

A similar trend is observed in Table~\ref{tab:boe_mape}. While the BIC-selected GNAR processes and local-$\alpha\beta$ models, which previously outperformed the Bank of England in terms of RMSE, no longer do, the global-$\alpha$ and standard AvGNAR forecasts remain remarkably strong. RaGNAR's strength lies in its ability to leverage disaggregated item series, which proves most effective during turbulent periods. During calmer periods, however, the Bank of England’s forecasts remain highly accurate, likely due to the role of judgment-based adjustments. Notably, several AvGNAR processes outperform the Bank of England at the 4- to 6-month horizon in terms of both RMSE and MAPE, underscoring RaGNAR’s strength as a forecasting model.

In particular, the \mbox{global-$\alpha$} and standard
AvGNAR($\mathcal{P}_1, \mathcal{S}_1$) and
AvGNAR($\mathcal{P}_2, \mathcal{S}_1$) models {\it always} improve
on the Bank's forecasts at horizons 4, 5 and 6. Taking \mbox{global-$\alpha$} and standard models together they (themselves)
have a median percentage improvement of 15.1\% with inter-quartile range of
$(9.4, 23.0)$ percent improvement. The global-$\alpha$ models are better
still, with a median percentage improvement of 19.2\% with inter-quartile
range of $(10.6, 38.4)$ percent improvement. These are impressive
improvements especially consider the simplicity,
efficiency and speed of our models.

\begin{figure}
    \centering
    \includegraphics[width=1\linewidth]{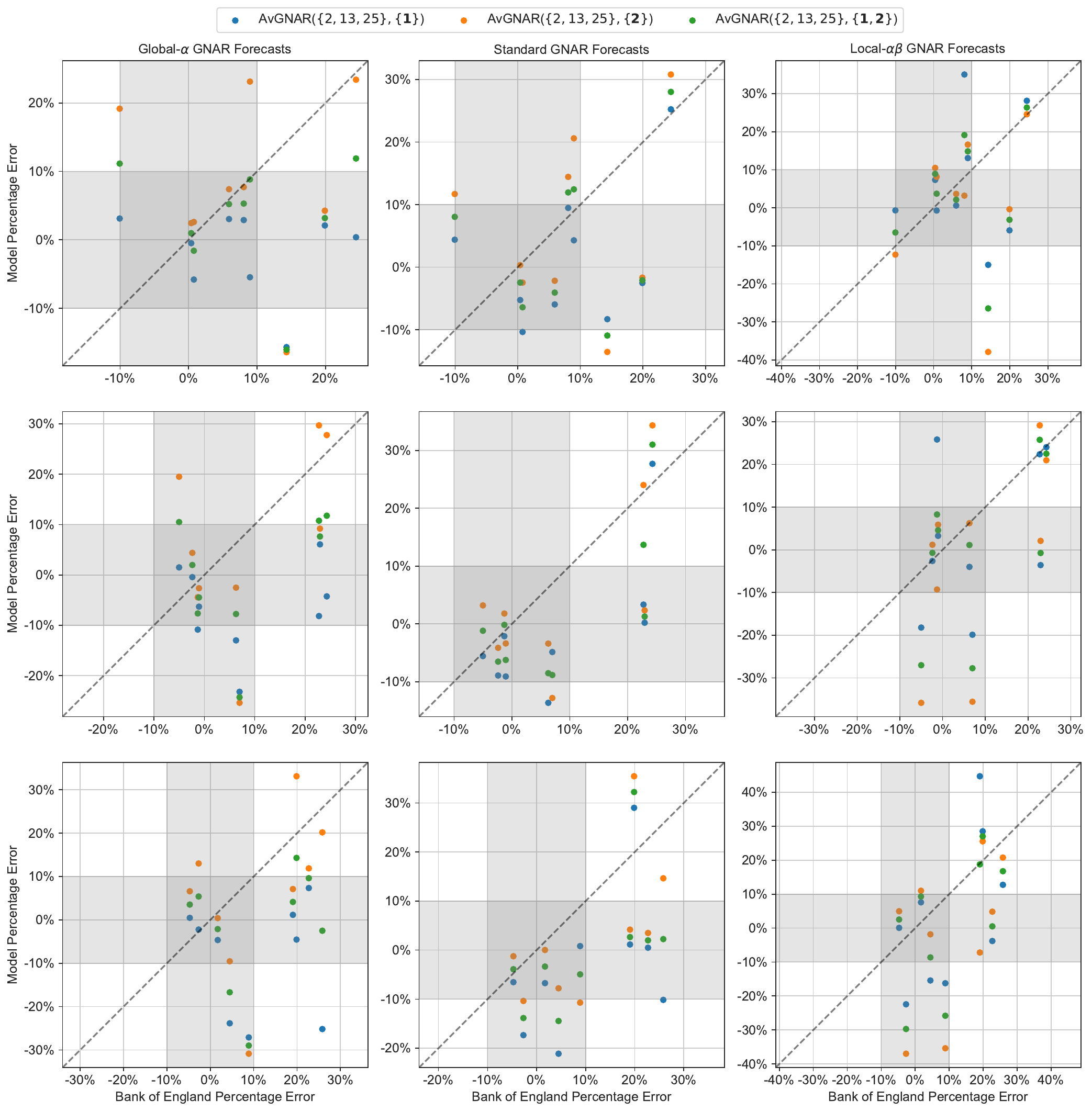}
    \caption{Scatter plots comparing the Bank of England’s forecast percentage errors to those of different RaGNAR models. From top to bottom, the plots correspond to forecast horizons of $h=4, 5,$ and $6$ months. Points along the line $y=x$ indicate cases where the Bank of England and RaGNAR models have the same error. Points outside the vertical shaded area represent Bank of England forecasts with more than 10\% error, while points outside the horizontal shaded area represent RaGNAR forecasts with more than 10\% error.}
    \label{fig:pct_scatter}
\end{figure}

Figure~\ref{fig:pct_scatter} compares the percentage errors of the Bank of England’s forecasts with those of our models across different forecast horizons ($h=4,5,$ and $6$ from top to bottom) and model classes, based on a single run of the RaGNAR algorithm. To aid visualisation, we include the $y=x$ line, where forecasting errors between our models and the Bank of England are identical, along with gray shaded regions representing errors below 10\%. More specifically, points outside the vertical shaded region indicate Bank of England forecasts with more than 10\% error, while points outside the horizontal region indicate the same for our models.

Examining the global-$\alpha$ and standard models, we find that many of our forecasts fall within the shaded region when the Bank of England’s do not, with the global AvGNAR($\mathcal{P}_1, \mathcal{S}_1$) model performing
{\it particularly well} --- only a single forecast lies outside the shaded region at the $h=4$ month horizon. For the local-$\alpha\beta$ processes, the number of forecasts outside the shaded region is comparable to that of the Bank of England’s forecasts.
These results concur with those in Table~\ref{tab:boe_mape}.
Interestingly, in all cases, there are numerous instances where RaGNAR forecasts inflation accurately while the Bank of England does not, and vice versa. Additionally, some forecasts show opposite error directions. For example, the cluster in the bottom right of the top-left plot shows that while the Bank of England had an error of around 15\%, all displayed RaGNAR forecasts had an error of approximately -15\%. This suggests that RaGNAR could add significant value to the Bank of England’s forecasts, particularly in areas where its predictions fall short,
as RaGNAR is forecasting differently.

\begin{table}
    \centering
    \begin{tabular}{cl|lllll}
    \toprule
        \multicolumn{2}{c|}{RaGNAR} &  \multicolumn{5}{c}{Forecast Horizon (months)} \\
    Class & \multicolumn{1}{c|}{Model} & 1 & 3 & 6 & 9 & 12 \\
    \midrule
    & AvAR$(\mathcal{P}_2)$  & 8.16 & 15.33 & 27.15 & 38.50 & 50.24 \\
    \midrule
    \multirow{9}{*}{\mbox{global-$\alpha$}}
    & GNAR($p,\boldsymbol{1}$) & 9.13\tiny{$\pm$0.12} & 17.50\tiny{$\pm$0.38} & 30.35\tiny{$\pm$0.53} & 40.97\tiny{$\pm$0.51} & 49.42\tiny{$\pm$0.44} \\
    & GNAR($p,\boldsymbol{2}$) & 8.72\tiny{$\pm$0.11} & 15.97\tiny{$\pm$0.35} & 28.86\tiny{$\pm$0.58} & 40.84\tiny{$\pm$0.68} & 51.50\tiny{$\pm$0.74} \\
    & GNAR($p,\boldsymbol{s}$) & 8.93\tiny{$\pm$0.10} & 16.94\tiny{$\pm$0.33} & 30.44\tiny{$\pm$0.48} & 42.52\tiny{$\pm$0.54} & 52.90\tiny{$\pm$0.63} \\
    & AvGNAR($\mathcal{P}_1,\mathcal{S}_1$) & 8.30\tiny{$\pm$0.07} & 15.94\tiny{$\pm$0.24} & 27.51\tiny{$\pm$0.40} & 38.72\tiny{$\pm$0.57} & 48.90\tiny{$\pm$0.63} \\
    & AvGNAR($\mathcal{P}_2,\mathcal{S}_1$) & 8.22\tiny{$\pm$0.07} & 15.52\tiny{$\pm$0.24} & 26.71\tiny{$\pm$0.43} & 38.05\tiny{$\pm$0.61} & 48.51\tiny{$\pm$0.70} \\
    & AvGNAR($\mathcal{P}_1,\mathcal{S}_2$) & 7.84\tiny{$\pm$0.09} & 14.01\tiny{$\pm$0.25} & 24.42\tiny{$\pm$0.43} & 34.78\tiny{$\pm$0.56} & 45.35\tiny{$\pm$0.64} \\
    & AvGNAR($\mathcal{P}_2,\mathcal{S}_2$) & 7.81\tiny{$\pm$0.10} & 13.96\tiny{$\pm$0.22} & 24.21\tiny{$\pm$0.40} & 34.67\tiny{$\pm$0.57} & 45.34\tiny{$\pm$0.69} \\
    & AvGNAR($\mathcal{P}_1,\mathcal{S}_3$) & 8.03\tiny{$\pm$0.06} & 14.80\tiny{$\pm$0.18} & 25.56\tiny{$\pm$0.29} & 36.30\tiny{$\pm$0.43} & 46.51\tiny{$\pm$0.49} \\
    & AvGNAR($\mathcal{P}_2,\mathcal{S}_3$) & 7.98\tiny{$\pm$0.06} & 14.58\tiny{$\pm$0.18} & 25.10\tiny{$\pm$0.30} & 35.92\tiny{$\pm$0.47} & 46.42\tiny{$\pm$0.57} \\
    \midrule
    \multirow{9}*{\parbox{1.5cm}{standard\\\mbox{(local-$\alpha$)}}}
    & GNAR($p,\boldsymbol{1}$) & 8.66\tiny{$\pm$0.07} & 14.96\tiny{$\pm$0.17} & 25.35\tiny{$\pm$0.55} & 36.28\tiny{$\pm$0.83} & 47.33\tiny{$\pm$1.19} \\
    & GNAR($p,\boldsymbol{2}$) & 8.42\tiny{$\pm$0.10} & 14.45\tiny{$\pm$0.32} & 25.32\tiny{$\pm$0.73} & 36.65\tiny{$\pm$0.99} & 48.23\tiny{$\pm$1.24} \\
    & GNAR($p,\boldsymbol{s}$) & 8.57\tiny{$\pm$0.08} & 14.76\tiny{$\pm$0.29} & 25.67\tiny{$\pm$0.66} & 37.16\tiny{$\pm$0.93} & 48.77\tiny{$\pm$1.19} \\
    & AvGNAR($\mathcal{P}_1,\mathcal{S}_1$) & 7.96\tiny{$\pm$0.05} & 14.29\tiny{$\pm$0.15} & 25.15\tiny{$\pm$0.29} & 36.51\tiny{$\pm$0.40} & 47.43\tiny{$\pm$0.47} \\
    & AvGNAR($\mathcal{P}_2,\mathcal{S}_1$) & 8.04\tiny{$\pm$0.06} & 14.52\tiny{$\pm$0.16} & 25.68\tiny{$\pm$0.29} & 37.19\tiny{$\pm$0.39} & 47.92\tiny{$\pm$0.48} \\
    & AvGNAR($\mathcal{P}_1,\mathcal{S}_2$) & 7.76\tiny{$\pm$0.09} & 13.69\tiny{$\pm$0.21} & 24.36\tiny{$\pm$0.45} & 35.37\tiny{$\pm$0.61} & 46.16\tiny{$\pm$0.72} \\
    & AvGNAR($\mathcal{P}_2,\mathcal{S}_2$) & 7.81\tiny{$\pm$0.09} & 13.78\tiny{$\pm$0.21} & 24.69\tiny{$\pm$0.45} & 35.91\tiny{$\pm$0.64} & 46.57\tiny{$\pm$1.24} \\
    & AvGNAR($\mathcal{P}_1,\mathcal{S}_3$) & 7.83\tiny{$\pm$0.05} & 13.90\tiny{$\pm$0.13} & 24.54\tiny{$\pm$0.27} & 35.69\tiny{$\pm$0.37} & 46.39\tiny{$\pm$0.46} \\
    & AvGNAR($\mathcal{P}_2,\mathcal{S}_3$) & 7.89\tiny{$\pm$0.06} & 14.06\tiny{$\pm$0.15} & 24.92\tiny{$\pm$0.27} & 36.26\tiny{$\pm$0.36} & 46.69\tiny{$\pm$0.61} \\
    \midrule
    \multirow{9}{*}{\mbox{local-$\alpha\beta$}}
    & GNAR($p,\boldsymbol{1}$) & 8.41\tiny{$\pm$0.09} & 13.97\tiny{$\pm$0.23} & 22.80\tiny{$\pm$0.46} & 32.21\tiny{$\pm$0.65} & 42.00\tiny{$\pm$0.99} \\
    & GNAR($p,\boldsymbol{2}$) & 8.04\tiny{$\pm$0.09} & 13.36\tiny{$\pm$0.23} & 22.86\tiny{$\pm$0.50} & 32.71\tiny{$\pm$0.83} & 43.09\tiny{$\pm$1.16} \\
    & GNAR($p,\boldsymbol{s}$) & 8.30\tiny{$\pm$0.10} & 13.59\tiny{$\pm$0.25} & 22.81\tiny{$\pm$0.49} & 32.21\tiny{$\pm$0.76} & 42.14\tiny{$\pm$1.07} \\
    & AvGNAR($\mathcal{P}_1,\mathcal{S}_1$) & 7.61\tiny{$\pm$0.09} & 13.89\tiny{$\pm$0.22} & 23.82\tiny{$\pm$0.49} & 33.93\tiny{$\pm$0.73} & 45.28\tiny{$\pm$0.87} \\
    & AvGNAR($\mathcal{P}_2,\mathcal{S}_1$) & 7.64\tiny{$\pm$0.10} & 14.12\tiny{$\pm$0.21} & 24.67\tiny{$\pm$0.49} & 35.22\tiny{$\pm$0.71} & 46.29\tiny{$\pm$0.86} \\
    & AvGNAR($\mathcal{P}_1,\mathcal{S}_2$) & 7.53\tiny{$\pm$0.12} & 13.42\tiny{$\pm$0.30} & 23.14\tiny{$\pm$0.65} & 33.00\tiny{$\pm$0.93} & 44.20\tiny{$\pm$1.38} \\
    & AvGNAR($\mathcal{P}_2,\mathcal{S}_2$) & 7.58\tiny{$\pm$0.13} & 13.63\tiny{$\pm$0.30} & 23.82\tiny{$\pm$0.64} & 34.15\tiny{$\pm$0.94} & 45.35\tiny{$\pm$1.49} \\
    & AvGNAR($\mathcal{P}_1,\mathcal{S}_3$) & 7.48\tiny{$\pm$0.08} & 13.50\tiny{$\pm$0.21} & 23.21\tiny{$\pm$0.45} & 33.04\tiny{$\pm$0.64} & 44.21\tiny{$\pm$0.83} \\
    & AvGNAR($\mathcal{P}_2,\mathcal{S}_3$) & 7.52\tiny{$\pm$0.08} & 13.72\tiny{$\pm$0.20} & 24.02\tiny{$\pm$0.45} & 34.32\tiny{$\pm$0.63} & 45.35\tiny{$\pm$0.83} \\
    \bottomrule
    \end{tabular}
    \caption{MAPEs for the same models as in Tables~\ref{tab:best_net} and \ref{tab:best_net_mavg}, with forecasts averaged across the $n=5$ best-performing networks. MAPEs are averaged over 100 runs of the algorithm, with $\pm1$ standard deviation displayed in a smaller font.  All MAPEs displayed are \underline{absolute} and not relative to AvAR($\mathcal{P}_2$). }
    \label{tab:mape_averages5}
\end{table}

\begin{table}
    \centering
    \begin{tabular}{cl|llllll}
    \toprule
    \multicolumn{2}{c|}{} & \multicolumn{6}{c}{Forecast Horizon (months)}\\
    \multicolumn{2}{c|}{Forecaster}   & 1 & 2 & 3 & 4 & 5 & 6 \\
    \midrule
    & Bank of England & 3.96 & 7.08 & 8.71 & 8.24 & 8.42 & 9.58 \\
    & AvAR($\mathcal{P}_2$)  & 12.26 & 18.70 & 15.25 & 14.44 & 14.20 & 14.83 \\
    \midrule
    \multirow{9}{*}{\rotatebox{90}{\mbox{global-$\alpha$}}}
    & GNAR($p,\boldsymbol{1}$) & 13.62\tiny{$\pm$0.37} & 20.43\tiny{$\pm$0.70} & 18.13\tiny{$\pm$0.76} & 15.45\tiny{$\pm$1.38} & 19.08\tiny{$\pm$1.82} & 21.22\tiny{$\pm$1.91} \\
    & GNAR($p,\boldsymbol{2}$) & 13.90\tiny{$\pm$0.30} & 19.61\tiny{$\pm$0.74} & 17.59\tiny{$\pm$0.71} & 22.03\tiny{$\pm$1.25} & 26.99\tiny{$\pm$1.66} & 31.24\tiny{$\pm$1.93} \\
    & GNAR($p,\boldsymbol{s}$) & 13.68\tiny{$\pm$0.30} & 20.54\tiny{$\pm$0.73} & 18.53\tiny{$\pm$0.79} & 22.03\tiny{$\pm$1.25} & 26.99\tiny{$\pm$1.66} & 31.24\tiny{$\pm$1.93} \\
    & AvGNAR($\mathcal{P}_1,\mathcal{S}_1$) & 11.54\tiny{$\pm$0.24} & 17.99\tiny{$\pm$0.47} & 16.03\tiny{$\pm$0.53} & 4.60\tiny{$\pm$0.75} & 6.96\tiny{$\pm$1.04} & 9.20\tiny{$\pm$1.27} \\
    & AvGNAR($\mathcal{P}_2,\mathcal{S}_1$) & 11.63\tiny{$\pm$0.23} & 17.61\tiny{$\pm$0.40} & 15.50\tiny{$\pm$0.44} & 4.48\tiny{$\pm$0.63} & 6.64\tiny{$\pm$1.05} & 8.78\tiny{$\pm$1.39} \\
    & AvGNAR($\mathcal{P}_1,\mathcal{S}_2$) & 11.96\tiny{$\pm$0.28} & 16.19\tiny{$\pm$0.50} & 13.77\tiny{$\pm$0.55} & 7.37\tiny{$\pm$1.10} & 8.20\tiny{$\pm$1.11} & 8.96\tiny{$\pm$1.34} \\
    & AvGNAR($\mathcal{P}_2,\mathcal{S}_2$) & 11.90\tiny{$\pm$0.31} & 16.20\tiny{$\pm$0.45} & 13.85\tiny{$\pm$0.49} & 7.45\tiny{$\pm$1.05} & 8.19\tiny{$\pm$1.03} & 9.13\tiny{$\pm$1.30} \\
    & AvGNAR($\mathcal{P}_1,\mathcal{S}_3$) & 11.74\tiny{$\pm$0.19} & 17.01\tiny{$\pm$0.37} & 14.73\tiny{$\pm$0.43} & 5.09\tiny{$\pm$0.73} & 6.74\tiny{$\pm$0.69} & 7.29\tiny{$\pm$0.86} \\
    & AvGNAR($\mathcal{P}_2,\mathcal{S}_3$) & 11.76\tiny{$\pm$0.20} & 16.83\tiny{$\pm$0.33} & 14.53\tiny{$\pm$0.35} & 5.32\tiny{$\pm$0.67} & 6.64\tiny{$\pm$0.72} & 7.48\tiny{$\pm$0.99} \\
    \midrule
    \multirow{9}{*}{\rotatebox{90}{\mbox{standard}}}
    & GNAR($p,\boldsymbol{1}$) & 14.04\tiny{$\pm$0.27} & 19.47\tiny{$\pm$0.42} & 16.81\tiny{$\pm$0.43} & 15.77\tiny{$\pm$2.73} & 19.48\tiny{$\pm$3.12} & 21.74\tiny{$\pm$3.80} \\
    & GNAR($p,\boldsymbol{2}$) & 13.69\tiny{$\pm$0.30} & 18.96\tiny{$\pm$0.64} & 16.25\tiny{$\pm$0.79} & 16.91\tiny{$\pm$3.04} & 21.14\tiny{$\pm$3.40} & 24.35\tiny{$\pm$3.93} \\
    & GNAR($p,\boldsymbol{s}$) & 13.76\tiny{$\pm$0.30} & 19.19\tiny{$\pm$0.64} & 16.64\tiny{$\pm$0.76} & 16.91\tiny{$\pm$3.04} & 21.14\tiny{$\pm$3.40} & 24.35\tiny{$\pm$3.93} \\
    & AvGNAR($\mathcal{P}_1,\mathcal{S}_1$) & 11.98\tiny{$\pm$0.17} & 16.85\tiny{$\pm$0.37} & 14.29\tiny{$\pm$0.43} & 7.35\tiny{$\pm$0.80} & 6.57\tiny{$\pm$0.88} & 7.09\tiny{$\pm$1.04} \\
    & AvGNAR($\mathcal{P}_2,\mathcal{S}_1$) & 11.76\tiny{$\pm$0.18} & 17.12\tiny{$\pm$0.37} & 14.55\tiny{$\pm$0.42} & 7.58\tiny{$\pm$0.80} & 7.33\tiny{$\pm$0.76} & 8.64\tiny{$\pm$1.01} \\
    & AvGNAR($\mathcal{P}_1,\mathcal{S}_2$) & 11.84\tiny{$\pm$0.25} & 16.57\tiny{$\pm$0.42} & 13.54\tiny{$\pm$0.53} & 8.54\tiny{$\pm$1.01} & 7.92\tiny{$\pm$1.00} & 7.80\tiny{$\pm$1.05} \\
    & AvGNAR($\mathcal{P}_2,\mathcal{S}_2$) & 11.59\tiny{$\pm$0.26} & 16.67\tiny{$\pm$0.40} & 13.62\tiny{$\pm$0.49} & 8.04\tiny{$\pm$0.86} & 7.33\tiny{$\pm$0.84} & 7.29\tiny{$\pm$0.87} \\
    & AvGNAR($\mathcal{P}_1,\mathcal{S}_3$) & 11.88\tiny{$\pm$0.15} & 16.65\tiny{$\pm$0.28} & 13.88\tiny{$\pm$0.37} & 7.73\tiny{$\pm$0.71} & 6.88\tiny{$\pm$0.65} & 6.70\tiny{$\pm$0.71} \\
    & AvGNAR($\mathcal{P}_2,\mathcal{S}_3$) & 11.61\tiny{$\pm$0.16} & 16.82\tiny{$\pm$0.29} & 14.06\tiny{$\pm$0.35} & 7.52\tiny{$\pm$0.57} & 6.91\tiny{$\pm$0.53} & 6.93\tiny{$\pm$0.64} \\
    \midrule
    \multirow{9}{*}{\rotatebox{90}{\mbox{local-$\alpha\beta$}}}
    & GNAR($p,\boldsymbol{1}$) & 13.91\tiny{$\pm$0.27} & 17.61\tiny{$\pm$0.48} & 14.72\tiny{$\pm$0.49} & 10.50\tiny{$\pm$1.51} & 13.32\tiny{$\pm$1.89} & 12.15\tiny{$\pm$1.83} \\
    & GNAR($p,\boldsymbol{2}$) & 13.26\tiny{$\pm$0.26} & 16.78\tiny{$\pm$0.50} & 13.70\tiny{$\pm$0.51} & 10.55\tiny{$\pm$1.18} & 13.79\tiny{$\pm$1.25} & 13.97\tiny{$\pm$2.08} \\
    & GNAR($p,\boldsymbol{s}$) & 13.53\tiny{$\pm$0.27} & 17.02\tiny{$\pm$0.51} & 13.96\tiny{$\pm$0.49} & 10.55\tiny{$\pm$1.18} & 13.79\tiny{$\pm$1.25} & 13.97\tiny{$\pm$2.08} \\
    & AvGNAR($\mathcal{P}_1,\mathcal{S}_1$) & 11.48\tiny{$\pm$0.28} & 17.49\tiny{$\pm$0.50} & 14.70\tiny{$\pm$0.48} & 8.92\tiny{$\pm$0.83} & 9.37\tiny{$\pm$0.95} & 11.63\tiny{$\pm$1.13} \\
    & AvGNAR($\mathcal{P}_2,\mathcal{S}_1$) & 11.20\tiny{$\pm$0.25} & 17.52\tiny{$\pm$0.52} & 14.73\tiny{$\pm$0.52} & 9.87\tiny{$\pm$0.71} & 10.38\tiny{$\pm$0.90} & 12.82\tiny{$\pm$1.09} \\
    & AvGNAR($\mathcal{P}_1,\mathcal{S}_2$) & 11.16\tiny{$\pm$0.40} & 16.09\tiny{$\pm$0.61} & 13.11\tiny{$\pm$0.57} & 9.54\tiny{$\pm$1.15} & 9.88\tiny{$\pm$1.46} & 11.63\tiny{$\pm$1.70} \\
    & AvGNAR($\mathcal{P}_2,\mathcal{S}_2$) & 10.89\tiny{$\pm$0.42} & 16.19\tiny{$\pm$0.61} & 13.13\tiny{$\pm$0.58} & 11.09\tiny{$\pm$1.22} & 11.35\tiny{$\pm$1.64} & 12.83\tiny{$\pm$1.82} \\
    & AvGNAR($\mathcal{P}_1,\mathcal{S}_3$) & 11.23\tiny{$\pm$0.23} & 16.77\tiny{$\pm$0.41} & 13.76\tiny{$\pm$0.41} & 9.10\tiny{$\pm$0.78} & 9.34\tiny{$\pm$0.92} & 11.05\tiny{$\pm$0.99} \\
    & AvGNAR($\mathcal{P}_2,\mathcal{S}_3$) & 10.95\tiny{$\pm$0.23} & 16.84\tiny{$\pm$0.41} & 13.74\tiny{$\pm$0.42} & 10.36\tiny{$\pm$0.79} & 10.52\tiny{$\pm$1.02} & 12.66\tiny{$\pm$1.13} \\
    \bottomrule
    \end{tabular}
    \caption{Same as Table~\ref{tab:boe}, but using MAPE instead of RMSE. MAPEs are averaged over 100 runs of the algorithm, with $\pm1$ standard deviation displayed in a smaller font.  All MAPEs displayed are \underline{absolute} and not relative to AvAR($\mathcal{P}_2$).}
    \label{tab:boe_mape}
\end{table}

\section{Important CPI Components for Forecasting} \label{sec:structure}

This final section analyses the structure of the best-performing undirected graphs to identify the most influential CPI components (nodes) and compares their time series to the inflation rate. To achieve this, we track how often each component/node is included in the \mbox{stage-1} neighbour set of the CPI node for the networks ranked as the best each month.

\begin{figure}
    \centering
    \resizebox{\textwidth}{!}{\includegraphics{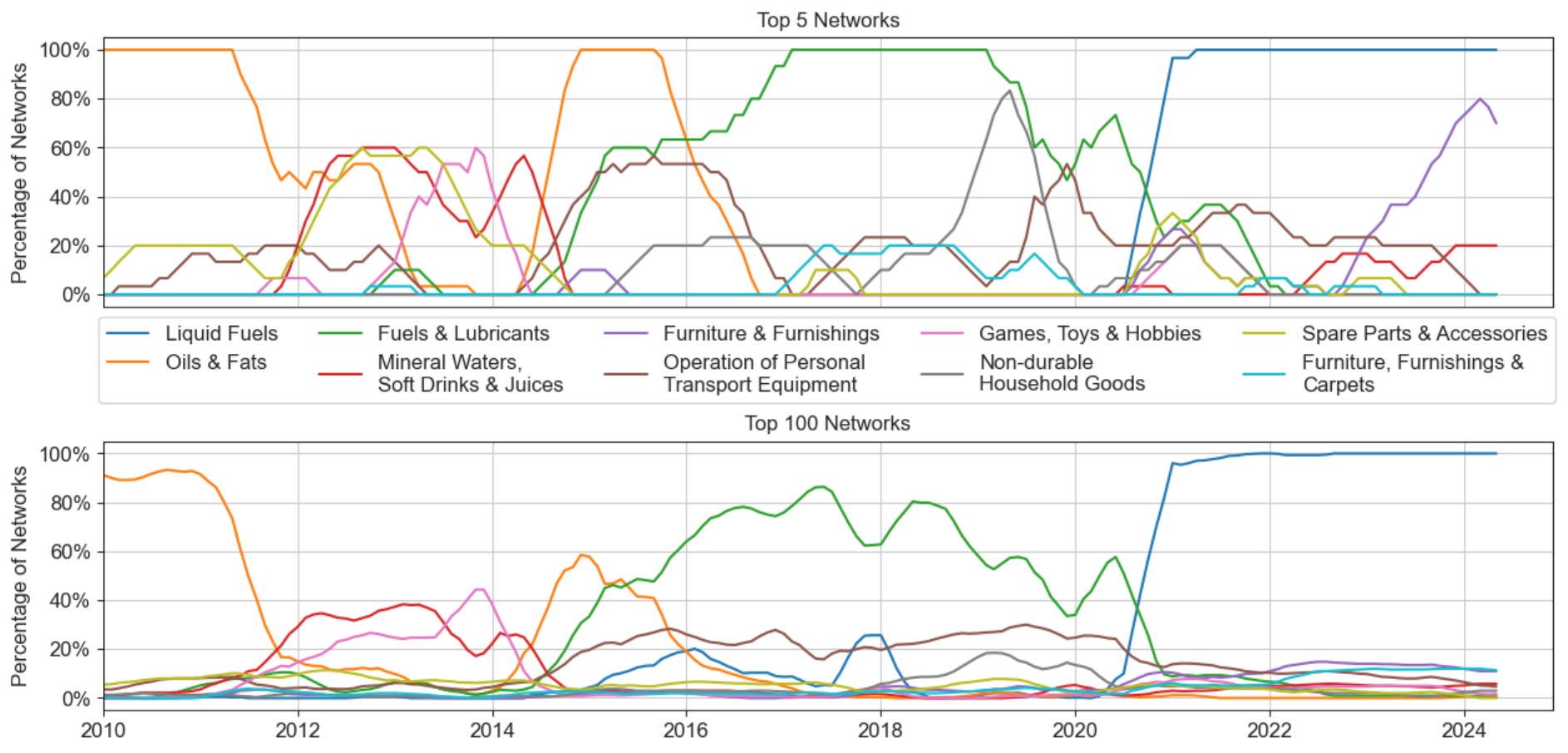}}
    \caption{Percentage of times each component appears in the CPI node's neighbour set for the best-performing networks, smoothed using a 6-month moving average. Only the components which appear most consistently are displayed.}
    \label{fig:popular_components}
\end{figure}

Figure~\ref{fig:popular_components} displays the percentage of times each component appeared in the CPI component’s neighbour set within the top-performing networks. Throughout most periods, a single component tends to dominate: oils and fats from 2010 to 2012 and 2015 to 2016, fuels and lubricants from 2016 to 2020, and liquid fuels from 2020 onwards. Notably, as one component phases out, another takes its place. Other disaggregated series appear less frequently in the CPI node’s neighbour set, as shown in Figure~\ref{fig:popular_components}, with the percentages for these components remaining relatively low. The exception is the period from 2012 to 2014, during which components cycled in and out of the neighbour set more frequently, when the best network selected by the algorithm was also changing more often. A similar pattern emerges between 2019 and 2020. These periods generally coincide with relatively stable inflation rates, meaning that there is less new information to be gained from the disaggregated item series.

\begin{figure}
    \centering
    \resizebox{\textwidth}{!}{\includegraphics{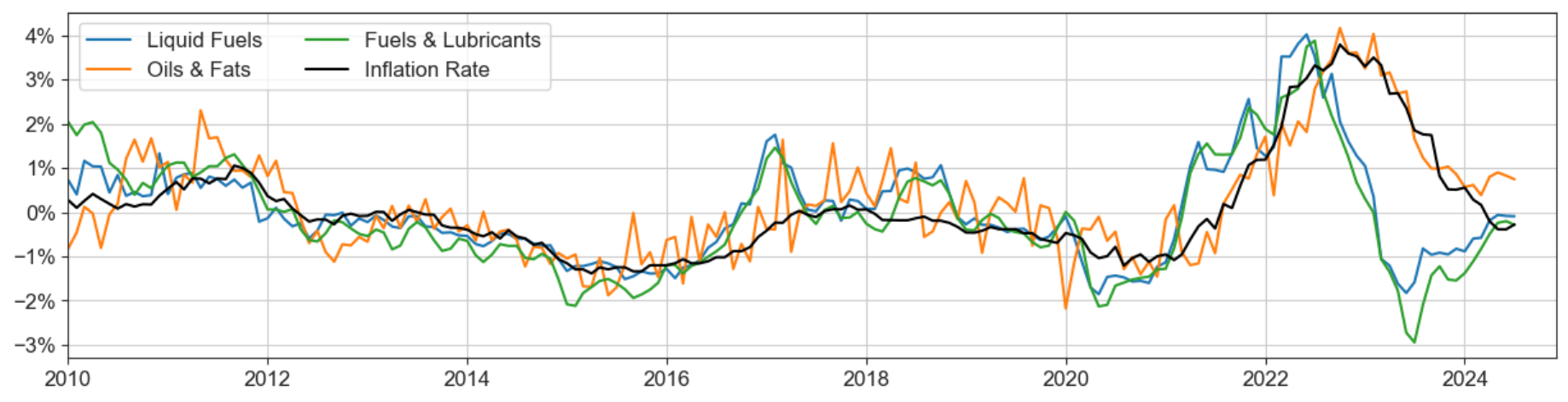}}
    \caption{Time series of the components that most frequently appear in the CPI node's neighbour set. For visualisation purposes, these series have been standardised by subtracting their means and dividing by their standard deviations.}
    \label{fig:some_comps}
\end{figure}

Figure~\ref{fig:some_comps} displays the time series for the most common disaggregated item series, according to Figure~\ref{fig:popular_components}. These series tend to anticipate movements in the inflation rate. For example, the year-on-year percentage change in oils \& fats precedes the inflation rise in 2011. Similarly, both liquid fuels and fuels \& lubricants predict the sharp increase in inflation to over 10\% in 2022, followed by a return to more stable levels at the start of 2024. Although these two series display similar behaviour in the figure, this in mainly due to the rescaling and, in reality, liquid fuels exhibit much larger percentage changes than fuels \& lubricants. In fact, during recent periods of significant inflationary change, networks that included liquid fuels in the stage-1 neighbour set tended to perform better. Both series also predicted the inflation rise in 2016, but since this increase was more moderate, graphs with an edge between CPI and fuels \& lubricants performed better.

\begin{figure}
    \centering
    \resizebox{\textwidth}{!}{\includegraphics{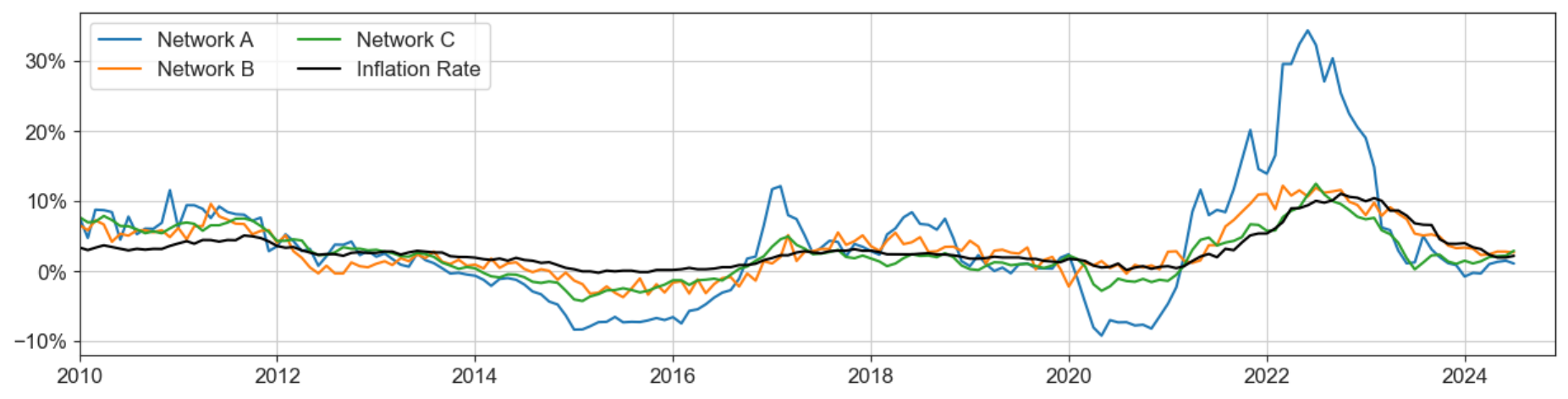}}
    \caption{Time series of the stage-1 neighbour set contributions for some of the best-performing networks. The colours correspond to those used in Figure~\ref{fig:some_comps}, where, for instance, Network A includes liquid fuels in the neighbour set of the CPI node, and similarly for the other networks.}
    \label{fig:ns_series}
\end{figure}

Figure~\ref{fig:ns_series} displays the neighbour set series for the corresponding best-performing networks, with colours matching those in Figure~\ref{fig:some_comps}. For instance, the blue neighbour series corresponds to networks where liquid fuels are in the CPI’s neighbour set. These plots offer a clearer view of the dynamics described above. As shown in equation~\eqref{eq:GNAR_unweighted}, the neighbour sums correspond to averages of multiple disaggregated series. This has the advantage of not only incorporating information from other CPI components, but also of stabilising the magnitudes of year-on-year percentage changes for certain series. Overall, GNAR processes tend to outperform traditional benchmark models, particularly during periods of inflation instability, as they are able to leverage information in a few key dissaggregate series in order to anticipate such movements.

\section{Conclusion and Future Work}
This article has applied the GNAR processes from~\cite{GNAR2,GNAR} to networks defined on the CPI component time series to predict inflation rates in the United Kingdom up to 12 months ahead. The networks were constructed according to the
Erd\H{o}s–R\'{e}nyi–Gilbert model, and then ranked based on the 1-step-ahead RMSE at the CPI node. The best graphs were then used to compute our out-of-sample forecasts. Our forecasts consistently outperformed traditional benchmark models, as well as those of the Bank of England at the four,
five and six month horizons. This is especially impressive given the simplicity of the GNAR models, which rely on only a (relatively) small number of parameters, particularly when compared to the numerous attempts to outperform the classic benchmarks using more complex machine learning techniques. The strength of our approach lies in its ability to identify a few key CPI components each month while leveraging the GNAR processes' capacity to estimate parameters from the broader network of time series.

Although RaGNAR methods do not outperform the Bank of England at forecast
horizons of 1, 2 and 3 months, we submit that RaGNAR could be incorporated
into the Bank's current suite of forecasting methods and, appropriately
weighted, might improve their forecasts
at those shorter horizons. It is
impossible to be precise, but we estimate that a 20\% improvement in
current forecasting performance might be achievable by incorporating RaGNAR
techniques.

This article was primarily concerned with networks consisting of unweighted edges. A natural extension would be to explore random networks with weighted edges, using methods such as those of ~\citet{wrn}, for example. Weighted networks introduce significantly greater flexibility as neighbour stage coefficients can vary between nodes, making the model more akin to a VAR. However, the computational complexity increases due to the added variability in connection strengths. Therefore, extending this work to weighted networks would require careful handling of these issues.

There are several other interesting avenues for future work.  One potential direction is to explore alternative methods for constructing graphs on the CPI component time series, such as using genetic algorithms—an optimisation technique inspired by natural selection in biological evolution. Starting with a set of random networks, a fitness measure such as the RMSE could be used to iteratively combine the most promising graphs. Another approach could involve applying boosting to an ensemble of graphs, where GNAR models are iteratively fitted to the residuals of previous ones. Node pruning is another promising avenue, involving the removal of nodes that frequently appear in the worst-performing networks, followed by generating new random graphs on the reduced set. As nodes are dropped, edge probabilities would need to be adjusted in order to maintain network connectivity. As demonstrated in our work, the disaggregated CPI components contain valuable information that can improve inflation rate forecasts. Accurate inflation forecasting is crucial for effective economic policy-making, investment planning, and financial stability, making further research in this area highly beneficial to the broader economy.

\section{Declaration of competing interests}
The authors declare that they have no known competing financial interests or personal relationships that could have appeared to influence the work reported in this paper.

\section{Data and Code Availability}

The data used in this article can be found at \url{https://www.ons.gov.uk/economy/inflationandpriceindices/datasets/consumerpriceindices}, while the code is available on GitHub at \url{https://github.com/henrypalasciano/Forecasting-Inflation-with-RaGNAR.git}.

\section{Acknowledgments}
We would like to thank Fergal Shortall for directing us to relevant literature on the Bank of England forecasting system and explaining some aspects in more detail.
Nason gratefully acknowledges support from
EPSRC NeST Programme Grant EP/X002195/1.
Palasciano gratefully acknowledges support of the EPSRC studentship.

\bibliographystyle{guy3}

\bibliography{main}

\begin{thebibliography}{72}
\providecommand{\natexlab}[1]{#1}

\bibitem[{Almosova and Andresen(2023)}]{almosova2023nonlinear}
Almosova, A. and Andresen, N. (2023) Nonlinear inflation forecasting with recurrent neural networks, \emph{Journal of Forecasting}, \textbf{42}, 240--259.

\bibitem[{Ang et~al.(2007)Ang, Bekaert, and Wei}]{ang2007macro}
Ang, A., Bekaert, G., and Wei, M. (2007) Do macro variables, asset markets, or surveys forecast inflation better?, \emph{Journal of Monetary Economics}, \textbf{54}, 1163--1212.

\bibitem[{Ansari et~al.(2024)Ansari, Stella, Turkmen, Zhang, Mercado, Shen, Shchur, Rangapuram, Arango, Kapoor, Zschiegner, Maddix, Wang, Mahoney, Torkkola, Bohlke-Schneider, and Wang}]{chronos}
Ansari, A.~F., Stella, L., Turkmen, C., Zhang, X., Mercado, P., Shen, H., Shchur, O., Rangapuram, S.~S., Arango, S.~P., Kapoor, S., Zschiegner, J., Maddix, D.~C., Wang, H., Mahoney, M.~W., Torkkola, A., K.~Wilson, Bohlke-Schneider, M., and Wang, Y. (2024) Chronos: Learning the language of time series, \emph{Transactions on Machine Learning Research}.

\bibitem[{Aparicio and Bertolotto(2020)}]{aparicio2020forecasting}
Aparicio, D. and Bertolotto, M.~I. (2020) Forecasting inflation with online prices, \emph{International Journal of Forecasting}, \textbf{36}, 232--247.

\bibitem[{Argiri et~al.(2024)Argiri, Hall, Momtsia, Papadopoulou, Skotida, Tavlas, and Wang}]{central_banks_2024}
Argiri, E., Hall, S.~G., Momtsia, A., Papadopoulou, D.~M., Skotida, I., Tavlas, G.~S., and Wang, Y. (2024) {An evaluation of the inflation forecasting performance of the European Central Bank, the Federal Reserve, and the Bank of England}, \emph{Forecasting in Turbulent Times}, \textbf{43}.

\bibitem[{Atkeson and Ohanian(2001)}]{atkeson2001phillips}
Atkeson, A. and Ohanian, L.~E. (2001) Are {Phillips} curves useful for forecasting inflation?, \emph{Federal Reserve bank of Minneapolis Quarterly Review}, \textbf{25}, 2--11.

\bibitem[{Barkan et~al.(2023)Barkan, Benchimol, Caspi, Cohen, Hammer, and Koenigstein}]{barkan2023forecasting}
Barkan, O., Benchimol, J., Caspi, I., Cohen, E., Hammer, A., and Koenigstein, N. (2023) Forecasting {CPI} inflation components with hierarchical recurrent neural networks, \emph{International Journal of Forecasting}, \textbf{39}, 1145--1162.

\bibitem[{Bates and Granger(1969)}]{model_averaging_1}
Bates, J.~M. and Granger, C. W.~J. (1969) The combination of forecasts, \emph{OR}, \textbf{20}, 451--468.

\bibitem[{Batini and Haldane(1999)}]{batini1999monetary}
Batini, N. and Haldane, A. (1999) Monetary policy rules and inflation forecasts, \emph{Bank of England. Quarterly Bulletin}, \textbf{39}, 60.

\bibitem[{Bernanke(2024)}]{boe2024}
Bernanke, B.~S. (2024) Forecasting for monetary policy making and communication at the {Bank of England}: a review, Technical report, Bank of England, {12 April 2024}, \url{https://www.bankofengland.co.uk/independent-evaluation-office/forecasting-for-monetary-policy-making-and-communication-at-the-bank-of-england-a-review/forecasting-for-monetary-policy-making-and-communication-at-the-bank-of-england-a-review}.

\bibitem[{Bernanke and Woodford(1997)}]{bernanke1997inflation}
Bernanke, B.~S. and Woodford, M. (1997) Inflation forecasts and monetary policy, Technical Report 6157, National Bureau of Economic Research, Cambridge, Massachusetts, USA.

\bibitem[{Binder and Sekkel(2024)}]{central_banks}
Binder, C.~C. and Sekkel, R. (2024) Central bank forecasting: A survey, \emph{Journal of Economic Surveys}, \textbf{38}, 342--364.

\bibitem[{Boaretto and Medeiros(2023)}]{boaretto2023forecasting}
Boaretto, G. and Medeiros, M.~C. (2023) Forecasting inflation using disaggregates and machine learning, \emph{arXiv preprint arXiv:2308.11173}.

\bibitem[{Boero et~al.(2008)Boero, Smith, and Wallis}]{boe_survey}
Boero, G., Smith, J., and Wallis, K.~F. (2008) Uncertainty and disagreement in economic prediction: {The Bank of England Survey of External Forecasters}, \emph{The Economic Journal}, \textbf{118}, 1107--1127.

\bibitem[{Brockwell and Davis(1991)}]{brd}
Brockwell, P.~J. and Davis, R.~A. (1991) \emph{Time Series: Theory and Methods}, Springer, New York.

\bibitem[{Bullmore and Sporns(2009)}]{bullmore2009complex}
Bullmore, E. and Sporns, O. (2009) Complex brain networks: graph theoretical analysis of structural and functional systems, \emph{Nature Reviews Neuroscience}, \textbf{10}, 186--198.

\bibitem[{Burgess et~al.(2013)Burgess, Fernandez-Corugedo, Groth, Harrison, Monti, Theodoridis, and Waldron}]{boe2013}
Burgess, S., Fernandez-Corugedo, E., Groth, C., Harrison, R., Monti, F., Theodoridis, K., and Waldron, M. (2013) The {Bank of England's} forecasting platform: {COMPASS, MAPS, EASE} and the suite of models, Technical Report 471, Bank of England, {17 May 2013}, \url{https://www.bankofengland.co.uk/working-paper/2013/the-boes-forecasting-platform-compass-maps-ease-and-the-suite-of-models}.

\bibitem[{Chu et~al.(2018)Chu, Huynh, Jacho-Ch{\'a}vez, and Kryvtsov}]{chu2018evolution}
Chu, B., Huynh, K., Jacho-Ch{\'a}vez, D., and Kryvtsov, O. (2018) On the evolution of the {United Kingdom} price distributions, \emph{The Annals of Applied Statistics}, \textbf{12}, 2618 -- 2646.

\bibitem[{Clements(2024)}]{clements2024}
Clements, M.~P. (2024) Do professional forecasters believe in the phillips curve?, \emph{International Journal of Forecasting}, \textbf{40}, 1238--1254.

\bibitem[{Clements et~al.(2023)Clements, Rich, and Tracy}]{clements2023}
Clements, M.~P., Rich, R.~W., and Tracy, J.~S. (2023) Chapter 3 - surveys of professionals, in R.~Bachmann, G.~Topa, and W.~{van der Klaauw}, eds., \emph{Handbook of Economic Expectations}, pp. 71--106, Academic Press.

\bibitem[{{Consumer Prices Team}(2024)}]{cpi_manual}
{Consumer Prices Team} (2024) Consumer prices indices technical manual, 2019, \emph{Office for National Statistics (ONS)}.

\bibitem[{Domit et~al.(2019)Domit, Monti, and Sokol}]{uk_var}
Domit, S., Monti, F., and Sokol, A. (2019) Forecasting the {UK} economy with a medium-scale {Bayesian} {VAR}, \emph{International Journal of Forecasting}, \textbf{35}, 1669--1678.

\bibitem[{Eugster and Uhl(2024)}]{eugster2024forecasting}
Eugster, P. and Uhl, M.~W. (2024) Forecasting inflation using sentiment, \emph{Economics Letters}, \textbf{236}, 111575.

\bibitem[{Faust and Wright(2013)}]{inflationforecasting2013}
Faust, J. and Wright, J.~H. (2013) Forecasting inflation, in G.~Elliott and A.~Timmermann, eds., \emph{Handbook of Economic Forecasting}, volume~2, pp. 2--56, Elsevier.

\bibitem[{Garlaschelli(2009)}]{wrn}
Garlaschelli, D. (2009) The weighted random graph model, \emph{arXiv preprint arXiv:0902.0897v2}.

\bibitem[{Giannellis et~al.(2024)Giannellis, Hall, Kouretas, and Tavlas}]{long_ar2}
Giannellis, N., Hall, S.~G., Kouretas, G.~P., and Tavlas, G.~S. (2024) Forecasting in turbulent times, \emph{Forecasting in Turbulent Times}, \textbf{43}.

\bibitem[{Gilbert(1959)}]{random_graphs}
Gilbert, E.~N. (1959) Random graphs, \emph{The Annals of Mathematical Statistics}, \textbf{30}, 1141--1144.

\bibitem[{Hall et~al.(2024)Hall, Tavlas, Wang, and Gefang}]{long_ar1}
Hall, S.~G., Tavlas, G.~S., Wang, Y., and Gefang, D. (2024) Inflation forecasting with rolling windows: An appraisal, \emph{Forecasting in Turbulent Times}, \textbf{43}.

\bibitem[{Heckerman(2008)}]{heckerman2008tutorial}
Heckerman, D. (2008) A tutorial on learning with {B}ayesian networks, in D.~E. Holmes and L.~C. Jain, eds., \emph{Innovations in Bayesian Networks: Theory and Applications}, pp. 33--82, Springer Berlin Heidelberg, Berlin, Heidelberg.

\bibitem[{Hendry and Hubrich(2011)}]{hendry2011combining}
Hendry, D.~F. and Hubrich, K. (2011) Combining disaggregate forecasts or combining disaggregate information to forecast an aggregate, \emph{Journal of Business \& Economic Statistics}, \textbf{29}, 216--227.

\bibitem[{Hubrich(2005)}]{hubrich2005forecasting}
Hubrich, K. (2005) Forecasting euro area inflation: Does aggregating forecasts by {HICP} component improve forecast accuracy?, \emph{International Journal of Forecasting}, \textbf{21}, 119--136.

\bibitem[{Joseph et~al.(2024)Joseph, Potjagailo, Chakraborty, and Kapetanios}]{uk_2024}
Joseph, A., Potjagailo, G., Chakraborty, C., and Kapetanios, G. (2024) Forecasting {UK} inflation bottom up, \emph{International Journal of Forecasting}, \textbf{40}, 1521--1538.

\bibitem[{Kang et~al.(2021)Kang, Ganguly, and Kolaczyk}]{kang2021dynamic}
Kang, X., Ganguly, A., and Kolaczyk, E.~D. (2021) Dynamic networks with multi-scale temporal structure, \emph{Sankhya A}, \textbf{84}, 218--260.

\bibitem[{Kapetanios et~al.(2008)Kapetanios, Labhard, and Price}]{boe_avg}
Kapetanios, G., Labhard, V., and Price, S. (2008) Forecasting using {Bayesian} and information-theoretic model averaging: An application to {UK} inflation, \emph{Journal of Business \& Economic Statistics}, \textbf{26}, 33--41.

\bibitem[{Knight et~al.(2020)Knight, Leeming, Nason, and Nunes}]{GNAR}
Knight, M., Leeming, K., Nason, G.~P., and Nunes, M. (2020) {Generalized Network Autoregressive Processes} and the {GNAR} package, \emph{Journal of Statistical Software}, \textbf{96}, 1–36.

\bibitem[{Knight et~al.(2016)Knight, Nunes, and Nason}]{GNARorig}
Knight, M.~I., Nunes, M.~A., and Nason, G.~P. (2016) {Modelling, Detrending and Decorrelation of Network Time Series}, \emph{arXiv preprint arXiv:1603.032221}.

\bibitem[{Kolaczyk(2009)}]{Kolaczyk09}
Kolaczyk, E.~D. (2009) \emph{Statistical Analysis of Network Data: Methods and Models}, Springer.

\bibitem[{Kolaczyk and Cs\'{a}rdi(2020)}]{KolaczykCsardi20}
Kolaczyk, E.~D. and Cs\'{a}rdi, G. (2020) \emph{Statistical Analysis of Network Data with R (Use R!)}, Springer.

\bibitem[{Koop and Korobilis(2012)}]{inf_avg_1}
Koop, G. and Korobilis, D. (2012) Forecasting inflation using dynamic model averaging, \emph{International Economic Review}, \textbf{53}, 867--886.

\bibitem[{Koski and Noble(2009)}]{koski2011bayesian}
Koski, T. and Noble, J. (2009) \emph{Bayesian Networks: an Introduction}, Wiley.

\bibitem[{Leeming(2019)}]{GNAR2}
Leeming, K. (2019) \emph{New Methods in Time Series Analysis: Univariate Testing and Network Autoregression Modelling}, Ph.D. thesis, University of Bristol.

\bibitem[{Lorenzo-Vald{\'e}s(2024)}]{GNAR_sm}
Lorenzo-Vald{\'e}s, A. (2024) Network analysis of the {Mexican} stock market, \emph{Investigaci{\'o}n Econ{\'o}mica}, \textbf{83}, 55--78.

\bibitem[{Lucchese et~al.(2023)Lucchese, Pakkanen, and Veraart}]{lucchese2023estimation}
Lucchese, L., Pakkanen, M.~S., and Veraart, A. E.~D. (2023) Estimation and inference for multivariate continuous-time autoregressive processes, \emph{arXiv preprint arXiv:2307.13020}.

\bibitem[{Makridakis et~al.(2018)Makridakis, Spiliotis, and Assimakopoulos}]{makridakis2018statistical}
Makridakis, S., Spiliotis, E., and Assimakopoulos, V. (2018) Statistical and machine learning forecasting methods: Concerns and ways forward, \emph{PLOS ONE}, \textbf{13}, e0194889.

\bibitem[{Marcellino et~al.(2006)Marcellino, Stock, and Watson}]{marcellino2006comparison}
Marcellino, M., Stock, J.~H., and Watson, M.~W. (2006) {A comparison of direct and iterated multistep {AR} methods for forecasting macroeconomic time series}, \emph{Journal of Econometrics}, \textbf{135}, 499--526.

\bibitem[{Medeiros et~al.(2021)Medeiros, Vasconcelos, Veiga, and Zilberman}]{medeiros2021forecasting}
Medeiros, M.~C., Vasconcelos, G. F.~R., Veiga, {\'A}., and Zilberman, E. (2021) Forecasting inflation in a data-rich environment: the benefits of machine learning methods, \emph{Journal of Business \& Economic Statistics}, \textbf{39}, 98--119.

\bibitem[{Mohammad and Khochiani(2022)}]{GNAR_GDP}
Mohammad, S. and Khochiani, R. (2022) Comparison of classical and dynamic network models efficiency in the application of generalized network autoregressive models, \emph{Journal of Economic Research}, \textbf{21}, 171--194.

\bibitem[{{Monetary Policy Committee}(2024)}]{mpc_aug_2024}
{Monetary Policy Committee} (2024) {Monetary Policy Report}, Technical report, Bank of England, {A}ugust, \url{https://www.bankofengland.co.uk/-/media/boe/files/monetary-policy-report/2024/august/monetary-policy-report-august-2024.pdf}.

\bibitem[{Nason and Wei(2022)}]{GNAR_covid}
Nason, G.~P. and Wei, J.~L. (2022) {Quantifying the economic response to {COVID-19} mitigations and death rates via forecasting purchasing managers' indices using generalised network autoregressive models with exogenous variables (with discussion)}, \emph{Journal of the Royal Statistical Society Series A: Statistics in Society}, \textbf{185}, 1778--1792.

\bibitem[{Nason et~al.(2023)Nason, Salnikov, and Cortina-Borja}]{GNAR_covid2}
Nason, G.~P., Salnikov, D., and Cortina-Borja, M. (2023) {New tools for network time series with an application to {COVID-19} hospitalisations}, \emph{arXiv preprint arXiv:2312.00530}.

\bibitem[{Nason et~al.(2024)Nason, Salnikov, and Cortina-Borja}]{us_politics}
Nason, G.~P., Salnikov, D., and Cortina-Borja, M. (2024) {Modelling clusters in network time series with an application to presidential elections in the {USA}}, \emph{arXiv preprint arXiv:2401.09381}.

\bibitem[{{Office for National Statistics}(2024)}]{cpi_data}
{Office for National Statistics} (2024) Consumer price inflation time series, {22 May 2024}, \url{https://www.ons.gov.uk/economy/inflationandpriceindices/datasets/consumerpriceindices}.

\bibitem[{Peach et~al.(2020)Peach, Arnaudon, and Barahona}]{peach2019semi}
Peach, R., Arnaudon, A., and Barahona, M. (2020) Semi-supervised classification on graphs using explicit diffusion dynamics, \emph{Foundations of Data Science}, \textbf{2}, 19--33.

\bibitem[{Peach et~al.(2021)Peach, Arnaudon, Schmidt, Palasciano, Bernier, Jelfs, Yaliraki, and Barahona}]{peach2021hcga}
Peach, R.~L., Arnaudon, A., Schmidt, J.~A., Palasciano, H.~A., Bernier, N.~R., Jelfs, K.~E., Yaliraki, S.~N., and Barahona, M. (2021) {HCGA}: Highly comparative graph analysis for network phenotyping, \emph{Patterns}, \textbf{2}.

\bibitem[{Pearl(2009{\natexlab{a}})}]{pearl2009causal}
Pearl, J. (2009{\natexlab{a}}) {Causal inference in statistics: An overview}, \emph{Statistics Surveys}, \textbf{3}, 96--146.

\bibitem[{Pearl(2009{\natexlab{b}})}]{pearl2009causality}
Pearl, J. (2009{\natexlab{b}}) \emph{Causality}, Cambridge University Press.

\bibitem[{Raffel et~al.(2019)Raffel, Shazeer, Roberts, Lee, Narang, Matena, Zhou, Li, and Liu}]{t5}
Raffel, C., Shazeer, N., Roberts, A., Lee, K., Narang, S., Matena, M., Zhou, Y., Li, W., and Liu, P. (2019) Exploring the limits of transfer learning with a unified text-to-text transformer, \emph{arXiv preprint arXiv:1910.10683v4}.

\bibitem[{Ralph et~al.(2015)Ralph, O'Neill, and Winton}]{Ralph15}
Ralph, J., O'Neill, R., and Winton, J. (2015) \emph{A Practical Introduction to Index Numbers}, Wiley.

\bibitem[{Scarselli et~al.(2008)Scarselli, Gori, Tsoi, Hagenbuchner, and Monfardini}]{scarselli2008graph}
Scarselli, F., Gori, M., Tsoi, A.~C., Hagenbuchner, M., and Monfardini, G. (2008) The graph neural network model, \emph{IEEE Transactions on Neural Networks}, \textbf{20}, 61--80.

\bibitem[{Stock and Watson(1999)}]{stock1999forecasting}
Stock, J.~H. and Watson, M.~W. (1999) Forecasting inflation, \emph{Journal of Monetary Economics}, \textbf{44}, 293--335.

\bibitem[{Stock and Watson(2003)}]{stock2003forecasting}
Stock, J.~H. and Watson, M.~W. (2003) Forecasting output and inflation: The role of asset prices, \emph{Journal of Economic Literature}, \textbf{41}, 788--829.

\bibitem[{Stock and Watson(2004)}]{stock_avg}
Stock, J.~H. and Watson, M.~W. (2004) Combination forecasts of output growth in a seven-country data set, \emph{Journal of Forecasting}, \textbf{23}, 405--430.

\bibitem[{Stock and Watson(2007)}]{stock2007has}
Stock, J.~H. and Watson, M.~W. (2007) Why has {US} inflation become harder to forecast?, \emph{Journal of Money, Credit and Banking}, \textbf{39}, 3--33.

\bibitem[{Stock and Watson(2009)}]{stock2009phillips}
Stock, J.~H. and Watson, M.~W. (2009) Phillips curve inflation forecasts, Technical Report 14322, National Bureau of Economic Research, Cambridge, Massachusetts, USA.

\bibitem[{Stock and Watson(2020)}]{stock2020slack}
Stock, J.~H. and Watson, M.~W. (2020) Slack and cyclically sensitive inflation, \emph{Journal of Money, Credit and Banking}, \textbf{52}, 393--428.

\bibitem[{Tucker and Gooding(2017)}]{cpi_bg}
Tucker, J. and Gooding, P. (2017) Consumer price indices, a brief guide: 2017, \emph{Office for National Statistics (ONS)}.

\bibitem[{Wright(2009)}]{wright_avg}
Wright, J.~H. (2009) Forecasting {US} inflation by {Bayesian} model averaging, \emph{Journal of Forecasting}, \textbf{28}, 131--144.

\bibitem[{Wu et~al.(2021)Wu, Pan, Chen, Long, Zhang, and Philip}]{wu2020comprehensive}
Wu, Z., Pan, S., Chen, F., Long, G., Zhang, C., and Philip, S.~Y. (2021) A comprehensive survey on graph neural networks, \emph{IEEE Transactions on Neural Networks and Learning Systems}, \textbf{32}, 4--24.

\bibitem[{Yao et~al.(2021)Yao, Chu, Li, Li, Gao, and Zhang}]{yao2021survey}
Yao, L., Chu, Z., Li, S., Li, Y., Gao, J., and Zhang, A. (2021) A survey on causal inference, \emph{ACM Transactions on Knowledge Discovery from Data}, \textbf{15}, 1--46.

\bibitem[{Zhou et~al.(2020{\natexlab{a}})Zhou, Cui, Hu, Zhang, Yang, Liu, Wang, Li, and Sun}]{zhou2020graph}
Zhou, J., Cui, G., Hu, S., Zhang, Z., Yang, C., Liu, Z., Wang, L., Li, C., and Sun, M. (2020{\natexlab{a}}) Graph neural networks: A review of methods and applications, \emph{AI Open}, \textbf{1}, 57--81.

\bibitem[{Zhou et~al.(2020{\natexlab{b}})Zhou, Chen, Zhang, Hu, Qiao, Yu, Yap, Pan, Zhang, and Shen}]{zhou2020toolbox}
Zhou, Z., Chen, X., Zhang, Y., Hu, D., Qiao, L., Yu, R., Yap, P., Pan, G., Zhang, H., and Shen, D. (2020{\natexlab{b}}) A toolbox for brain network construction and classification {(BrainNetClass)}, \emph{Human Brain Mapping}, \textbf{41}, 2808--2826.

\bibitem[{Zhu et~al.(2017)Zhu, Pan, Li, Liu, and Wang}]{zhu2017network}
Zhu, X., Pan, R., Li, G., Liu, Y., and Wang, H. (2017) Network vector autoregression, \emph{The Annals of Statistics}, \textbf{45}, 1096 -- 1123.

\end{thebibliography}

\section*{Appendix A - Proofs}

\subsection*{Proof of Proposition~\ref{prop:dist}}
Let
\begin{equation}
    \Lambda_{r-1}(N)=\left\{\boldsymbol{n}=(n_1,\ldots,n_{r-1})\in \mathbb{N}^{r-1} : \sum_{s=0}^{r-1} n_s \leq N\right\}
\end{equation}
be the set of all possible combinations of neighbour set sizes up to stage $r-1$. Applying the law of total probability and Bayes rule, we obtain
\begin{align}
     \nonumber \mathbb{P}\left(\bigl|\mathcal{N}_i^{(r)}\bigr| = n_r\right)
    &= \sum_{\boldsymbol{n}\in\Lambda_{r-1}(N)} \mathbb{P}\left(\bigl|\mathcal{N}_i^{(r)}\bigr| = n_r \Big | \bigl|\mathcal{N}_i^{(r-1)}\bigr| = n_{r-1}, \ldots, \bigl|\mathcal{N}_i^{(1)}\bigr| = n_1 \right)\\[-3mm]
    & \hspace{5cm}\mathbb{P}\left( \bigl|\mathcal{N}_i^{(r-1)}\bigr| = n_{r-1}, \ldots, \bigl|\mathcal{N}_i^{(1)}\bigr| = n_1 \right)\\
    \nonumber
    &= \sum_{\boldsymbol{n}\in\Lambda_{r-1}(N)} \mathbb{P}\left(\bigl|\mathcal{N}_i^{(r)}\bigr| = n_r \Big | \bigl|\mathcal{N}_i^{(r-1)}\bigr| = n_{r-1}, \ldots, \bigl|\mathcal{N}_i^{(1)}\bigr| = n_1 \right)\\[-3mm]
    \nonumber
    & \hspace{2.5cm} \mathbb{P}\left(\bigl|\mathcal{N}_i^{(r-1)}\bigr| = n_{r-1} \Big | \bigl|\mathcal{N}_i^{(r-2)}\bigr| = n_{r-2}, \ldots, \bigl|\mathcal{N}_i^{(1)}\bigr| = n_1 \right)\ldots\\
    &\hspace{3.5cm} \ldots\mathbb{P}\left(\bigl|\mathcal{N}_i^{(2)}\bigr| = n_2 \Big |\bigl|\mathcal{N}_i^{(1)}\bigr| = n_1 \right)\mathbb{P}\left(\bigl|\mathcal{N}_i^{(1)}\bigr| = n_1 \right)\\
    &= \sum_{\boldsymbol{n}\in\Lambda_{r-1}(N)} \prod_{s=1}^r\mathbb{P}\left(\bigl|\mathcal{N}_i^{(s)}\bigr| = n_s \Big | \bigl|\mathcal{N}_i^{(s-1)}\bigr| = n_{s-1}, \ldots, \bigl|\mathcal{N}_i^{(1)}\bigr| = n_1 \right)\\
    &= \sum_{\boldsymbol{n}\in\Lambda_{r-1}(N)} \prod_{s=1}^r\mathbb{P}\left(Z_{n_1:n_{s-1}}(N) = n_s \right),
\end{align}
where
\begin{equation}
    Z_{n_1:n_{s-1}}(N) = \bigl|\mathcal{N}_i^{(s)}\bigr| \Big| \bigl|\mathcal{N}_i^{(s-1)}\bigr| = n_{s-1}, \ldots, \bigl|\mathcal{N}_i^{(1)}\bigr| = n_1
\end{equation}
is a random variable that takes values in the set of all possible sizes of the neighbour set at stage $s$, conditioned on the sizes of the neighbour sets from stages $1$ through $s-1$.
Now note that $\bigl|\mathcal{N}_i^{(1)}\bigr|\sim\text{Bin}(N-1, \pi)$, since there are a total of $n-1$ nodes to choose from, and each of these is contained in $\mathcal{N}_i^{(1)}$ with probability $\pi$. More generally, given that $\bigl|\mathcal{N}_i^{(1)}\bigr|=n_1,\ldots, \bigl|\mathcal{N}_i^{(s-1)}\bigr|=n_{s-1}$, the remaining number of nodes to choose from is $N - (1 + n_1 + \cdots + n_{r-1})$. Letting $j$ denote any of the remaining nodes, the probability that this is included in $\mathcal{N}_i^{(s)}$ is given by
\begin{equation}
    \mathbb{P}\left(j\in \mathcal{N}_i^{(s)} \Big | \bigl|\mathcal{N}_i^{(s-1)}\bigr| = n_{s-1} \right)  = 1 - \mathbb{P}\left(j\notin \mathcal{N}_i^{(s)} \Big | \bigl|\mathcal{N}_i^{(s-1)}\bigr| = n_{s-1}\right) = 1 - (1 - \pi)^{n_{s-1}},
\end{equation}
since the probability that no edge exists between node $j$ and node $i\in \mathcal{N}_i^{(s-1)}$ is $1-\pi$, and $j\notin \mathcal{N}_i^{(s)}$ if and only if $j\centernot\leftrightsquigarrow i$ for all $i\in \mathcal{N}_i^{(s-1)}$. Hence, 
\begin{equation}
    Z_{n_1:n_{s-1}}(N)\sim  \text{Bin}\left(N - \sum_{k=0}^{s-1} n_k,  1 - (1 - \pi)^{n_{s-1}}\right),
\end{equation}
where $n_0$ = 1, completing the proof. \hfill\qed

\subsection*{Proof of Proposition~\ref{prop:element}}
To begin with, by the law of total probability,
\begin{align}
    \pi_r 
    & = \mathbb{P}\left(j\in\mathcal{N}_i^{(r)}\right) = \mathbb{P}\left(j\in\mathcal{N}_i^{(r)}\Big| j\notin\bigcup_{s=1}^{r-1}\mathcal{N}_i^{(s)}\right)\mathbb{P}\left(j\notin\bigcup_{s=1}^{r-1}\mathcal{N}_i^{(s)}\right),
\end{align}
since the probability of $j\in\mathcal{N}_i^{(r)}$ conditioned on $j$ being a member of any $\mathcal{N}_i^{(1)},\ldots,\mathcal{N}_i^{(r-1)}$ is zero. Now 
\begin{align}
    \mathbb{P}\left(j\notin\bigcup_{s=1}^{r-1}\mathcal{N}_i^{(s)}\right) &= 1 - \mathbb{P}\left(j\in\bigcup_{s=1}^{r-1}\mathcal{N}_i^{(s)}\right) = 1- \mathbb{P}\left(\bigcup_{s=1}^{r-1}\left\{j\in\mathcal{N}_i^{(s)}\right\}\right)  \\
    &= 1 - \sum_{s=1}^{r-1}\mathbb{P}\left(j\in\mathcal{N}_i^{(s)}\right) = 1 - \sum_{s=1}^{r-1}\pi_s,
\end{align}
which follows from the fact that these events are disjoint, since $j$ can only be in a single neighbour set at a time. Conditioning on the size of the previous stage neighbour set, we have
\begin{align}
    \nonumber
    \mathbb{P}\left(j\in\mathcal{N}_i^{(r)}\Big| j\notin\bigcup_{s=1}^{r-1}\mathcal{N}_i^{(s)}\right) & = \sum_{n=0}^{N-r}\mathbb{P}\left(j\in\mathcal{N}_i^{(r)}\Big| \bigl|\mathcal{N}_i^{(r-1)}\bigr| = n, j\notin\bigcup_{s=1}^{r-1}\mathcal{N}_i^{(s)}\right) \\
    &\hspace{3.5cm}\mathbb{P}\left(\bigl|\mathcal{N}_i^{(r-1)}\bigr| = n \Big|  j\notin\bigcup_{s=1}^{r-1}\mathcal{N}_i^{(s)}\right)\\
    \nonumber
    & = \sum_{n=0}^{N-r}\left\{1-\mathbb{P}\left(j\notin\mathcal{N}_i^{(r)}\Big| \bigl|\mathcal{N}_i^{(r-1)}\bigr| = n, j\notin\bigcup_{s=1}^{r-1}\mathcal{N}_i^{(s)}\right)\right\} \\
    &\hspace{3.5cm}\mathbb{P}\left(\bigl|\mathcal{N}_i^{(r-1)}\bigr| = n \Big|  j\notin\bigcup_{s=1}^{r-1}\mathcal{N}_i^{(s)}\right)\\
    & = \sum_{n=0}^{N-r}\left\{1 - (1-\pi)^n\right\}\mathbb{P}\left(\bigl|\mathcal{N}_i^{(r-1)}\bigr| = n \Big|  j\notin\bigcup_{s=1}^{r-1}\mathcal{N}_i^{(s)}\right)\\
    & = 1 - \sum_{n=0}^{N-r}(1-\pi)^n\mathbb{P}\left(\bigl|\mathcal{N}_i^{(r-1)}\bigr| = n \Big|  j\notin\bigcup_{s=1}^{r-1}\mathcal{N}_i^{(s)}\right),
\end{align}
as 
\begin{equation}
     \sum_{n=0}^{N-r}\mathbb{P}\left(\bigl|\mathcal{N}_i^{(r-1)}\bigr| = n \Big|  j\notin\bigcup_{s=1}^{r-1}\mathcal{N}_i^{(s)}\right) = 1.
\end{equation}
Moreover, using Proposition~\ref{prop:dist},
\begin{equation}
     \mathbb{P}\left(\bigl|\mathcal{N}_i^{(r-1)}\bigr| = n \Big|  j\notin\bigcup_{s=1}^{r-1}\mathcal{N}_i^{(s)}\right) = \sum_{\boldsymbol{n}\in\Lambda_{r-2}(N-1)} \prod_{s=1}^{r-1}\mathbb{P}\left(Z_{n_1:n_{s-1}}(N-1) = n \right),
\end{equation}
since conditioning on the event $j\notin\bigcup_{s=1}^{r-1}\mathcal{N}_i^{(s)}$ is equivalent to constructing the neighbour sets from stage $1$ through stage $r-1$ with one fewer node available to choose from. Hence,
\begin{align}\nonumber
    \pi_r &= \mathbb{P}\left(j\in\mathcal{N}_i^{(r)}\right) \\
    &= \left(1 - \sum_{s=1}^{r-1} \pi_s\right)\left\{1 - \sum_{n=0}^{N - r}(1 - \pi)^n\sum_{\boldsymbol{n}\in\Lambda_{r-2}(N-1)} \prod_{s=1}^{r-1}\mathbb{P}\left(Z_{n_1:n_{s-1}}(N-1) = n \right)\right\},
\end{align}
as required. \hfill\qed

\section*{Appendix B - Model Averaging Extended}
\begin{figure}
    \centering
    \includegraphics[width=1\linewidth]{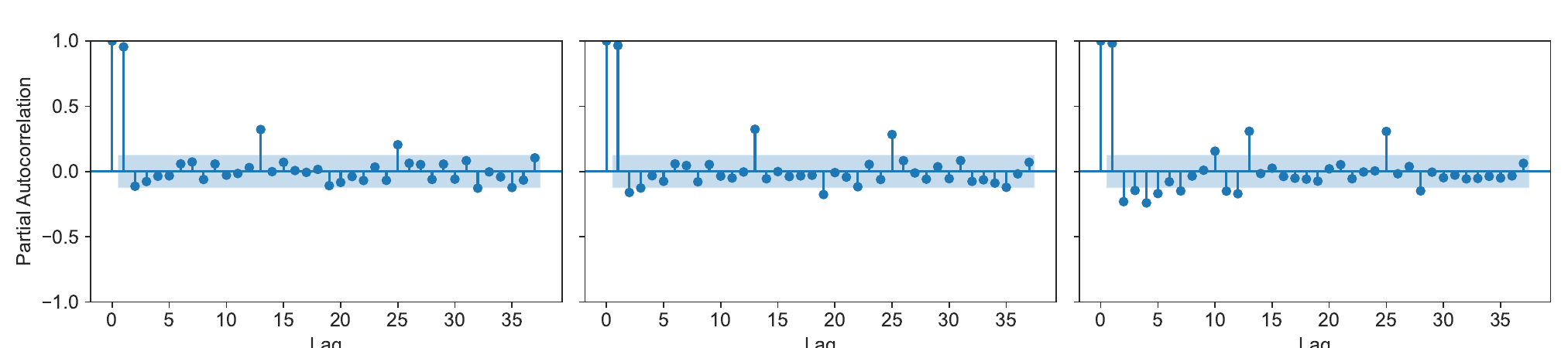}
    \caption{Partial autocorrelations for the inflation rate time series across different time periods. From left to right, data from: 1\textsuperscript{st} January 1995 to 31\textsuperscript{st} December 2014, 1\textsuperscript{st} January 2000 to 31\textsuperscript{st} December 2019 and 1\textsuperscript{st} January 2005 to 31\textsuperscript{st} December 2024.}
    \label{fig:pacfs}
\end{figure}

This section outlines our rationale for selecting model orders to average across. Figure~\ref{fig:pacfs} shows the partial autocorrelation function (PACF) computed over different 20-year windows of the inflation rate time series. We observe three notable spikes at lags 1, 13, and 25, suggesting these lags hold important predictive information. As a result, we average across models of these orders. Additionally, lag 2 appears to carry some useful information. Given the widespread use of AR(2) models in the literature, see for example~\citet{uk_2024}, we sometimes include AR(2) instead of AR(1) in our averages. We use the same orders when averaging the predictions of our GNAR processes.

\section*{Appendix C - Absolute RMSEs}

This section presents the results from the main article using \textbf{absolute} RMSEs rather than relative to our AR($\mathcal{P}_2$) benchmark, allowing for clearer comparisons in other sections of the Appendix.

\begin{table}[h]
    \centering
    \begin{tabular}{cl|lllll}
    \toprule
        \multicolumn{2}{c|}{GNAR} &  \multicolumn{5}{c}{Forecast Horizon (months)} \\
     \multicolumn{2}{c|}{Model} & 1 & 3 & 6 & 9 & 12 \\
    \midrule
    \multirow{9}{*}{\mbox{global-$\alpha$}}
    & GNAR($p, \boldsymbol{1}$) & 0.40\tiny{$\pm$0.02} & 0.79\tiny{$\pm$0.05} & 1.37\tiny{$\pm$0.08} & 1.93\tiny{$\pm$0.07} & 2.38\tiny{$\pm$0.06} \\
    & GNAR($p, \boldsymbol{2}$) & 0.39\tiny{$\pm$0.01} & 0.78\tiny{$\pm$0.04} & 1.38\tiny{$\pm$0.08} & 2.02\tiny{$\pm$0.10} & 2.55\tiny{$\pm$0.13} \\
    & GNAR($p, \boldsymbol{s}$) & 0.40\tiny{$\pm$0.01} & 0.79\tiny{$\pm$0.05} & 1.40\tiny{$\pm$0.08} & 2.03\tiny{$\pm$0.10} & 2.55\tiny{$\pm$0.12} \\
    & AvGNAR($\mathcal{P}_1,\mathcal{S}_1$) & 0.36\tiny{$\pm$0.01} & 0.72\tiny{$\pm$0.02} & 1.23\tiny{$\pm$0.03} & 1.80\tiny{$\pm$0.04} & 2.30\tiny{$\pm$0.04} \\
    & AvGNAR($\mathcal{P}_2,\mathcal{S}_1$) & 0.36\tiny{$\pm$0.01} & 0.71\tiny{$\pm$0.02} & 1.20\tiny{$\pm$0.03} & 1.76\tiny{$\pm$0.04} & 2.27\tiny{$\pm$0.04} \\
    & AvGNAR($\mathcal{P}_1,\mathcal{S}_2$) & 0.35\tiny{$\pm$0.01} & 0.68\tiny{$\pm$0.04} & 1.15\tiny{$\pm$0.05} & 1.71\tiny{$\pm$0.06} & 2.23\tiny{$\pm$0.06} \\
    & AvGNAR($\mathcal{P}_2,\mathcal{S}_2$) & 0.35\tiny{$\pm$0.02} & 0.69\tiny{$\pm$0.06} & 1.16\tiny{$\pm$0.08} & 1.73\tiny{$\pm$0.11} & 2.25\tiny{$\pm$0.11} \\
    & AvGNAR($\mathcal{P}_1,\mathcal{S}_3$) & 0.35\tiny{$\pm$0.01} & 0.69\tiny{$\pm$0.02} & 1.17\tiny{$\pm$0.03} & 1.73\tiny{$\pm$0.03} & 2.24\tiny{$\pm$0.03} \\
    & AvGNAR($\mathcal{P}_2,\mathcal{S}_3$) & 0.35\tiny{$\pm$0.01} & 0.68\tiny{$\pm$0.02} & 1.16\tiny{$\pm$0.03} & 1.72\tiny{$\pm$0.05} & 2.23\tiny{$\pm$0.05} \\
    \midrule
    \multirow{9}*{\parbox{1.5cm}{standard\\\mbox{(local-$\alpha$)}}}
    & GNAR($p, \boldsymbol{1}$) & 0.40\tiny{$\pm$0.01} & 0.76\tiny{$\pm$0.04} & 1.30\tiny{$\pm$0.08} & 1.92\tiny{$\pm$0.12} & 2.45\tiny{$\pm$0.16} \\
    & GNAR($p, \boldsymbol{2}$) & 0.39\tiny{$\pm$0.01} & 0.74\tiny{$\pm$0.04} & 1.28\tiny{$\pm$0.09} & 1.89\tiny{$\pm$0.13} & 2.45\tiny{$\pm$0.17} \\
    & GNAR($p, \boldsymbol{s}$) & 0.39\tiny{$\pm$0.01} & 0.75\tiny{$\pm$0.04} & 1.29\tiny{$\pm$0.08} & 1.90\tiny{$\pm$0.13} & 2.46\tiny{$\pm$0.17} \\
    & AvGNAR($\mathcal{P}_1,\mathcal{S}_1$) & 0.36\tiny{$\pm$0.01} & 0.68\tiny{$\pm$0.02} & 1.17\tiny{$\pm$0.03} & 1.72\tiny{$\pm$0.03} & 2.24\tiny{$\pm$0.04} \\
    & AvGNAR($\mathcal{P}_2,\mathcal{S}_1$) & 0.36\tiny{$\pm$0.01} & 0.68\tiny{$\pm$0.02} & 1.16\tiny{$\pm$0.03} & 1.72\tiny{$\pm$0.03} & 2.23\tiny{$\pm$0.03} \\
    & AvGNAR($\mathcal{P}_1,\mathcal{S}_2$) & 0.35\tiny{$\pm$0.01} & 0.67\tiny{$\pm$0.02} & 1.17\tiny{$\pm$0.03} & 1.73\tiny{$\pm$0.04} & 2.25\tiny{$\pm$0.05} \\
    & AvGNAR($\mathcal{P}_2,\mathcal{S}_2$) & 0.36\tiny{$\pm$0.03} & 0.68\tiny{$\pm$0.03} & 1.17\tiny{$\pm$0.03} & 1.74\tiny{$\pm$0.15} & 2.34\tiny{$\pm$0.79} \\
    & AvGNAR($\mathcal{P}_1,\mathcal{S}_3$) & 0.35\tiny{$\pm$0.01} & 0.67\tiny{$\pm$0.01} & 1.15\tiny{$\pm$0.02} & 1.71\tiny{$\pm$0.03} & 2.23\tiny{$\pm$0.03} \\
    & AvGNAR($\mathcal{P}_2,\mathcal{S}_3$) & 0.35\tiny{$\pm$0.01} & 0.67\tiny{$\pm$0.02} & 1.15\tiny{$\pm$0.02} & 1.71\tiny{$\pm$0.05} & 2.26\tiny{$\pm$0.32} \\
    \midrule
    \multirow{9}{*}{\mbox{local-$\alpha\beta$}}
    & GNAR($p, \boldsymbol{1}$) & 0.39\tiny{$\pm$0.01} & 0.71\tiny{$\pm$0.02} & 1.16\tiny{$\pm$0.04} & 1.71\tiny{$\pm$0.06} & 2.18\tiny{$\pm$0.09} \\
    & GNAR($p, \boldsymbol{2}$) & 0.37\tiny{$\pm$0.01} & 0.69\tiny{$\pm$0.02} & 1.15\tiny{$\pm$0.05} & 1.70\tiny{$\pm$0.08} & 2.20\tiny{$\pm$0.11} \\
    & GNAR($p, \boldsymbol{s}$) & 0.38\tiny{$\pm$0.01} & 0.69\tiny{$\pm$0.02} & 1.14\tiny{$\pm$0.05} & 1.69\tiny{$\pm$0.08} & 2.18\tiny{$\pm$0.11} \\
    & AvGNAR($\mathcal{P}_1,\mathcal{S}_1$) & 0.34\tiny{$\pm$0.01} & 0.66\tiny{$\pm$0.02} & 1.13\tiny{$\pm$0.03} & 1.68\tiny{$\pm$0.05} & 2.20\tiny{$\pm$0.07} \\
    & AvGNAR($\mathcal{P}_2,\mathcal{S}_1$) & 0.33\tiny{$\pm$0.01} & 0.67\tiny{$\pm$0.01} & 1.15\tiny{$\pm$0.03} & 1.70\tiny{$\pm$0.04} & 2.24\tiny{$\pm$0.06} \\
    & AvGNAR($\mathcal{P}_1,\mathcal{S}_2$) & 0.34\tiny{$\pm$0.01} & 0.67\tiny{$\pm$0.02} & 1.15\tiny{$\pm$0.04} & 1.73\tiny{$\pm$0.15} & 2.42\tiny{$\pm$1.34} \\
    & AvGNAR($\mathcal{P}_2,\mathcal{S}_2$) & 0.34\tiny{$\pm$0.01} & 0.67\tiny{$\pm$0.02} & 1.18\tiny{$\pm$0.04} & 1.76\tiny{$\pm$0.15} & 2.47\tiny{$\pm$1.34} \\
    & AvGNAR($\mathcal{P}_1,\mathcal{S}_3$) & 0.33\tiny{$\pm$0.01} & 0.65\tiny{$\pm$0.01} & 1.12\tiny{$\pm$0.03} & 1.66\tiny{$\pm$0.06} & 2.25\tiny{$\pm$0.59} \\
    & AvGNAR($\mathcal{P}_2,\mathcal{S}_3$) & 0.33\tiny{$\pm$0.01} & 0.66\tiny{$\pm$0.01} & 1.14\tiny{$\pm$0.03} & 1.70\tiny{$\pm$0.05} & 2.30\tiny{$\pm$0.59} \\
    \bottomrule
    \end{tabular}
    \caption{Absolute RMSEs corresponding to Tables~\ref{tab:best_net} and \ref{tab:best_net_mavg}.}
    \label{tab:abs_rmse_1}
\end{table}

\begin{table}[h]
    \centering
    \begin{tabular}{cl|lllll}
    \toprule
        \multicolumn{2}{c|}{GNAR} &  \multicolumn{5}{c}{Forecast Horizon (months)} \\
     \multicolumn{2}{c|}{Model} & 1 & 3 & 6 & 9 & 12 \\
    \midrule
    \multirow{9}{*}{\mbox{global-$\alpha$}}
    & GNAR($p, \boldsymbol{1}$) & 0.39\tiny{$\pm$0.01} & 0.76\tiny{$\pm$0.03} & 1.32\tiny{$\pm$0.04} & 1.88\tiny{$\pm$0.03} & 2.33\tiny{$\pm$0.03} \\
    & GNAR($p, \boldsymbol{2}$) & 0.38\tiny{$\pm$0.01} & 0.75\tiny{$\pm$0.02} & 1.34\tiny{$\pm$0.03} & 1.98\tiny{$\pm$0.04} & 2.52\tiny{$\pm$0.05} \\
    & GNAR($p, \boldsymbol{s}$) & 0.39\tiny{$\pm$0.01} & 0.76\tiny{$\pm$0.02} & 1.36\tiny{$\pm$0.03} & 2.00\tiny{$\pm$0.04} & 2.53\tiny{$\pm$0.05} \\
    & AvGNAR($\mathcal{P}_1,\mathcal{S}_1$) & 0.36\tiny{$\pm$0.00} & 0.71\tiny{$\pm$0.01} & 1.21\tiny{$\pm$0.02} & 1.78\tiny{$\pm$0.02} & 2.29\tiny{$\pm$0.02} \\
    & AvGNAR($\mathcal{P}_2,\mathcal{S}_1$) & 0.35\tiny{$\pm$0.00} & 0.70\tiny{$\pm$0.01} & 1.18\tiny{$\pm$0.02} & 1.75\tiny{$\pm$0.02} & 2.26\tiny{$\pm$0.02} \\
    & AvGNAR($\mathcal{P}_1,\mathcal{S}_2$) & 0.34\tiny{$\pm$0.01} & 0.66\tiny{$\pm$0.01} & 1.12\tiny{$\pm$0.02} & 1.68\tiny{$\pm$0.02} & 2.20\tiny{$\pm$0.02} \\
    & AvGNAR($\mathcal{P}_2,\mathcal{S}_2$) & 0.33\tiny{$\pm$0.01} & 0.66\tiny{$\pm$0.01} & 1.12\tiny{$\pm$0.02} & 1.68\tiny{$\pm$0.03} & 2.21\tiny{$\pm$0.03} \\
    & AvGNAR($\mathcal{P}_1,\mathcal{S}_3$) & 0.34\tiny{$\pm$0.00} & 0.67\tiny{$\pm$0.01} & 1.15\tiny{$\pm$0.01} & 1.71\tiny{$\pm$0.02} & 2.23\tiny{$\pm$0.02} \\
    & AvGNAR($\mathcal{P}_2,\mathcal{S}_3$) & 0.34\tiny{$\pm$0.00} & 0.67\tiny{$\pm$0.01} & 1.14\tiny{$\pm$0.01} & 1.70\tiny{$\pm$0.02} & 2.22\tiny{$\pm$0.02} \\
    \midrule
    \multirow{9}*{\parbox{1.5cm}{standard\\\mbox{(local-$\alpha$)}}}
    & GNAR($p, \boldsymbol{1}$) & 0.39\tiny{$\pm$0.01} & 0.74\tiny{$\pm$0.02} & 1.26\tiny{$\pm$0.04} & 1.88\tiny{$\pm$0.07} & 2.42\tiny{$\pm$0.10} \\
    & GNAR($p, \boldsymbol{2}$) & 0.38\tiny{$\pm$0.01} & 0.71\tiny{$\pm$0.02} & 1.25\tiny{$\pm$0.05} & 1.86\tiny{$\pm$0.07} & 2.42\tiny{$\pm$0.08} \\
    & GNAR($p, \boldsymbol{s}$) & 0.38\tiny{$\pm$0.01} & 0.72\tiny{$\pm$0.02} & 1.26\tiny{$\pm$0.05} & 1.87\tiny{$\pm$0.07} & 2.43\tiny{$\pm$0.08} \\
    & AvGNAR($\mathcal{P}_1,\mathcal{S}_1$) & 0.36\tiny{$\pm$0.00} & 0.68\tiny{$\pm$0.01} & 1.15\tiny{$\pm$0.01} & 1.71\tiny{$\pm$0.02} & 2.22\tiny{$\pm$0.03} \\
    & AvGNAR($\mathcal{P}_2,\mathcal{S}_1$) & 0.36\tiny{$\pm$0.01} & 0.68\tiny{$\pm$0.01} & 1.16\tiny{$\pm$0.01} & 1.71\tiny{$\pm$0.02} & 2.22\tiny{$\pm$0.02} \\
    & AvGNAR($\mathcal{P}_1,\mathcal{S}_2$) & 0.35\tiny{$\pm$0.01} & 0.66\tiny{$\pm$0.01} & 1.15\tiny{$\pm$0.01} & 1.71\tiny{$\pm$0.02} & 2.24\tiny{$\pm$0.02} \\
    & AvGNAR($\mathcal{P}_2,\mathcal{S}_2$) & 0.35\tiny{$\pm$0.01} & 0.66\tiny{$\pm$0.01} & 1.15\tiny{$\pm$0.01} & 1.71\tiny{$\pm$0.02} & 2.25\tiny{$\pm$0.08} \\
    & AvGNAR($\mathcal{P}_1,\mathcal{S}_3$) & 0.35\tiny{$\pm$0.00} & 0.67\tiny{$\pm$0.01} & 1.15\tiny{$\pm$0.01} & 1.70\tiny{$\pm$0.01} & 2.22\tiny{$\pm$0.02} \\
    & AvGNAR($\mathcal{P}_2,\mathcal{S}_3$) & 0.35\tiny{$\pm$0.00} & 0.67\tiny{$\pm$0.01} & 1.15\tiny{$\pm$0.01} & 1.70\tiny{$\pm$0.01} & 2.22\tiny{$\pm$0.03} \\
    \midrule
    \multirow{9}{*}{\mbox{local-$\alpha\beta$}}
    & GNAR($p, \boldsymbol{1}$) & 0.38\tiny{$\pm$0.00} & 0.69\tiny{$\pm$0.01} & 1.13\tiny{$\pm$0.03} & 1.68\tiny{$\pm$0.04} & 2.14\tiny{$\pm$0.06} \\
    & GNAR($p, \boldsymbol{2}$) & 0.36\tiny{$\pm$0.00} & 0.66\tiny{$\pm$0.01} & 1.09\tiny{$\pm$0.02} & 1.63\tiny{$\pm$0.04} & 2.13\tiny{$\pm$0.05} \\
    & GNAR($p, \boldsymbol{s}$) & 0.37\tiny{$\pm$0.00} & 0.66\tiny{$\pm$0.01} & 1.09\tiny{$\pm$0.02} & 1.63\tiny{$\pm$0.04} & 2.12\tiny{$\pm$0.05} \\
    & AvGNAR($\mathcal{P}_1,\mathcal{S}_1$) & 0.33\tiny{$\pm$0.00} & 0.64\tiny{$\pm$0.01} & 1.10\tiny{$\pm$0.02} & 1.65\tiny{$\pm$0.02} & 2.17\tiny{$\pm$0.03} \\
    & AvGNAR($\mathcal{P}_2,\mathcal{S}_1$) & 0.32\tiny{$\pm$0.00} & 0.65\tiny{$\pm$0.01} & 1.12\tiny{$\pm$0.02} & 1.67\tiny{$\pm$0.02} & 2.20\tiny{$\pm$0.03} \\
    & AvGNAR($\mathcal{P}_1,\mathcal{S}_2$) & 0.32\tiny{$\pm$0.00} & 0.63\tiny{$\pm$0.01} & 1.10\tiny{$\pm$0.02} & 1.65\tiny{$\pm$0.03} & 2.20\tiny{$\pm$0.17} \\
    & AvGNAR($\mathcal{P}_2,\mathcal{S}_2$) & 0.32\tiny{$\pm$0.00} & 0.64\tiny{$\pm$0.01} & 1.12\tiny{$\pm$0.02} & 1.68\tiny{$\pm$0.03} & 2.25\tiny{$\pm$0.17} \\
    & AvGNAR($\mathcal{P}_1,\mathcal{S}_3$) & 0.32\tiny{$\pm$0.00} & 0.63\tiny{$\pm$0.01} & 1.09\tiny{$\pm$0.01} & 1.64\tiny{$\pm$0.02} & 2.17\tiny{$\pm$0.06} \\
    & AvGNAR($\mathcal{P}_2,\mathcal{S}_3$) & 0.32\tiny{$\pm$0.00} & 0.64\tiny{$\pm$0.01} & 1.11\tiny{$\pm$0.01} & 1.66\tiny{$\pm$0.02} & 2.20\tiny{$\pm$0.06} \\
    \bottomrule
    \end{tabular}
    \caption{Absolute RMSEs corresponding to Table~\ref{tab:averages5}.}
    \label{tab:abs_rmse_5}
\end{table}

\section*{Appendix D - Fixed Lag Comparisons}
In this section we compare fixed order AR and GNAR models. Table~\ref{tab:ar_processes} displays the RMSEs for AR processes of different orders, Table~\ref{tab:fixed_order} present those for the corresponding GNAR models, for which forecasts were computed by averaging the predictions from the $n=5$ best networks each month.

\begin{table}
    \centering
    \begin{tabular}{c|lllll}
    \toprule
    Model & 1 & 3 & 6 & 9 & 12 \\
    \midrule
    AR(1) & 0.43 & 0.91 & 1.66 & 2.45 & 3.17 \\
    AR(2) & 0.43 & 0.87 & 1.59 & 2.32 & 2.96 \\
    AR(12) & 0.40 & 0.75 & 1.33 & 1.95 & 2.46 \\
    AR(13) & 0.38 & 0.78 & 1.38 & 2.02 & 2.58 \\
    AR(25) & 0.37 & 0.73 & 1.32 & 2.02 & 2.66 \\
    \bottomrule
    \end{tabular}
    \caption{RMSEs for AR model CPI inflation forecasts from 2010 onwards.}
    \label{tab:ar_processes}
\end{table}

\begin{table}
    \centering
    \begin{tabular}{cl|lllll}
    \toprule
        \multicolumn{2}{c|}{GNAR} &  \multicolumn{5}{c}{Forecast Horizon (months)} \\
    Class & Model & 1 & 3 & 6 & 9 & 12 \\
    \midrule
    \multirow{10}{*}{\mbox{global-$\alpha$}}
    & GNAR(1,\textbf{1}) & 0.39\tiny{$\pm$0.01} & 0.76\tiny{$\pm$0.03} & 1.32\tiny{$\pm$0.04} & 1.88\tiny{$\pm$0.03} & 2.33\tiny{$\pm$0.03} \\
    & GNAR(1,\textbf{2}) & 0.38\tiny{$\pm$0.01} & 0.75\tiny{$\pm$0.02} & 1.34\tiny{$\pm$0.03} & 1.98\tiny{$\pm$0.04} & 2.52\tiny{$\pm$0.05} \\
    & GNAR(2,\textbf{1}) & 0.38\tiny{$\pm$0.01} & 0.73\tiny{$\pm$0.02} & 1.27\tiny{$\pm$0.03} & 1.88\tiny{$\pm$0.03} & 2.42\tiny{$\pm$0.04} \\
    & GNAR(2,\textbf{2}) & 0.37\tiny{$\pm$0.01} & 0.75\tiny{$\pm$0.02} & 1.36\tiny{$\pm$0.04} & 2.04\tiny{$\pm$0.07} & 2.65\tiny{$\pm$0.09} \\
    & GNAR(12,\textbf{1}) & 0.40\tiny{$\pm$0.01} & 0.80\tiny{$\pm$0.02} & 1.35\tiny{$\pm$0.04} & 2.01\tiny{$\pm$0.07} & 2.58\tiny{$\pm$0.10} \\
    & GNAR(12,\textbf{2}) & 0.37\tiny{$\pm$0.01} & 0.71\tiny{$\pm$0.02} & 1.17\tiny{$\pm$0.04} & 1.73\tiny{$\pm$0.06} & 2.24\tiny{$\pm$0.08} \\
    & GNAR(13,\textbf{1}) & 0.37\tiny{$\pm$0.01} & 0.75\tiny{$\pm$0.02} & 1.27\tiny{$\pm$0.04} & 1.87\tiny{$\pm$0.05} & 2.41\tiny{$\pm$0.05} \\
    & GNAR(13,\textbf{2}) & 0.35\tiny{$\pm$0.01} & 0.71\tiny{$\pm$0.03} & 1.18\tiny{$\pm$0.04} & 1.77\tiny{$\pm$0.06} & 2.32\tiny{$\pm$0.06} \\
    & GNAR(25,\textbf{1}) & 0.37\tiny{$\pm$0.01} & 0.77\tiny{$\pm$0.03} & 1.32\tiny{$\pm$0.04} & 1.99\tiny{$\pm$0.07} & 2.58\tiny{$\pm$0.08} \\
    & GNAR(25,\textbf{2}) & 0.35\tiny{$\pm$0.01} & 0.70\tiny{$\pm$0.03} & 1.17\tiny{$\pm$0.05} & 1.78\tiny{$\pm$0.07} & 2.33\tiny{$\pm$0.07} \\
    \midrule
      \multirow{10}*{\parbox{1.5cm}{standard\\\mbox{(local-$\alpha$)}}}
    & GNAR(1,\textbf{1}) & 0.39\tiny{$\pm$0.01} & 0.74\tiny{$\pm$0.02} & 1.26\tiny{$\pm$0.04} & 1.88\tiny{$\pm$0.07} & 2.42\tiny{$\pm$0.10} \\
    & GNAR(1,\textbf{2}) & 0.38\tiny{$\pm$0.01} & 0.71\tiny{$\pm$0.02} & 1.25\tiny{$\pm$0.05} & 1.86\tiny{$\pm$0.07} & 2.42\tiny{$\pm$0.08} \\
    & GNAR(2,\textbf{1}) & 0.38\tiny{$\pm$0.01} & 0.70\tiny{$\pm$0.01} & 1.17\tiny{$\pm$0.02} & 1.73\tiny{$\pm$0.03} & 2.21\tiny{$\pm$0.04} \\
    & GNAR(2,\textbf{2}) & 0.37\tiny{$\pm$0.01} & 0.70\tiny{$\pm$0.01} & 1.23\tiny{$\pm$0.04} & 1.85\tiny{$\pm$0.07} & 2.44\tiny{$\pm$0.42} \\
    & GNAR(12,\textbf{1}) & 0.39\tiny{$\pm$0.01} & 0.73\tiny{$\pm$0.02} & 1.23\tiny{$\pm$0.03} & 1.81\tiny{$\pm$0.04} & 2.27\tiny{$\pm$0.05} \\
    & GNAR(12,\textbf{2}) & 0.38\tiny{$\pm$0.01} & 0.68\tiny{$\pm$0.02} & 1.16\tiny{$\pm$0.03} & 1.71\tiny{$\pm$0.03} & 2.16\tiny{$\pm$0.04} \\
    & GNAR(13,\textbf{1}) & 0.38\tiny{$\pm$0.01} & 0.76\tiny{$\pm$0.02} & 1.28\tiny{$\pm$0.02} & 1.87\tiny{$\pm$0.03} & 2.41\tiny{$\pm$0.03} \\
    & GNAR(13,\textbf{2}) & 0.37\tiny{$\pm$0.01} & 0.72\tiny{$\pm$0.01} & 1.25\tiny{$\pm$0.02} & 1.84\tiny{$\pm$0.03} & 2.38\tiny{$\pm$0.03} \\
    & GNAR(25,\textbf{1}) & 0.38\tiny{$\pm$0.01} & 0.73\tiny{$\pm$0.02} & 1.24\tiny{$\pm$0.03} & 1.85\tiny{$\pm$0.04} & 2.41\tiny{$\pm$0.05} \\
    & GNAR(25,\textbf{2}) & 0.37\tiny{$\pm$0.01} & 0.71\tiny{$\pm$0.02} & 1.23\tiny{$\pm$0.03} & 1.81\tiny{$\pm$0.04} & 2.36\tiny{$\pm$0.06} \\
    \midrule
    \multirow{10}{*}{\mbox{local-$\alpha\beta$}}
    & GNAR(1,\textbf{1}) & 0.38\tiny{$\pm$0.00} & 0.68\tiny{$\pm$0.01} & 1.13\tiny{$\pm$0.03} & 1.68\tiny{$\pm$0.04} & 2.15\tiny{$\pm$0.06} \\
    & GNAR(1,\textbf{2}) & 0.36\tiny{$\pm$0.00} & 0.66\tiny{$\pm$0.01} & 1.09\tiny{$\pm$0.02} & 1.63\tiny{$\pm$0.04} & 2.13\tiny{$\pm$0.05} \\
    & GNAR(2,\textbf{1}) & 0.35\tiny{$\pm$0.00} & 0.70\tiny{$\pm$0.01} & 1.19\tiny{$\pm$0.02} & 1.77\tiny{$\pm$0.03} & 2.26\tiny{$\pm$0.04} \\
    & GNAR(2,\textbf{2}) & 0.34\tiny{$\pm$0.00} & 0.68\tiny{$\pm$0.01} & 1.17\tiny{$\pm$0.03} & 1.77\tiny{$\pm$0.04} & 2.31\tiny{$\pm$0.12} \\
    & GNAR(12,\textbf{1}) & 0.36\tiny{$\pm$0.01} & 0.71\tiny{$\pm$0.01} & 1.19\tiny{$\pm$0.02} & 1.75\tiny{$\pm$0.03} & 2.22\tiny{$\pm$0.04} \\
    & GNAR(12,\textbf{2}) & 0.34\tiny{$\pm$0.01} & 0.68\tiny{$\pm$0.02} & 1.17\tiny{$\pm$0.03} & 1.70\tiny{$\pm$0.04} & 2.24\tiny{$\pm$0.56} \\
    & GNAR(13,\textbf{1}) & 0.35\tiny{$\pm$0.00} & 0.72\tiny{$\pm$0.01} & 1.24\tiny{$\pm$0.02} & 1.82\tiny{$\pm$0.03} & 2.38\tiny{$\pm$0.05} \\
    & GNAR(13,\textbf{2}) & 0.34\tiny{$\pm$0.01} & 0.70\tiny{$\pm$0.02} & 1.22\tiny{$\pm$0.03} & 1.78\tiny{$\pm$0.05} & 2.33\tiny{$\pm$0.07} \\
    & GNAR(25,\textbf{1}) & 0.39\tiny{$\pm$0.02} & 0.75\tiny{$\pm$0.04} & 1.27\tiny{$\pm$0.05} & 1.88\tiny{$\pm$0.07} & 2.45\tiny{$\pm$0.08} \\
    & GNAR(25,\textbf{2}) & 0.41\tiny{$\pm$0.05} & 0.77\tiny{$\pm$0.06} & 1.29\tiny{$\pm$0.06} & 1.89\tiny{$\pm$0.08} & 2.45\tiny{$\pm$0.10} \\
    \bottomrule
    \end{tabular}
    \caption{Absolute RMSEs for inflation rate forecasts computed by averaging predictions from GNAR models fitted to the $n=5$ best-performing networks each month. RMSEs are averaged over 100 runs of the algorithm, with $\pm1$ standard deviation displayed in a smaller font. }
    \label{tab:fixed_order}
\end{table}

Overall, our GNAR models outperform all AR processes, regardless of the order $p$. Unlike AR models, where performance generally improves with lag order, low-order GNAR processes often perform competitively or even better than their higher-order counterparts, particularly at longer horizons. In fact, local-$\alpha\beta$ GNAR($1,\boldsymbol{1}$) and GNAR($1,\boldsymbol{2}$) processes significantly outperform all higher-order models at longer horizons. This suggests that disaggregated item series contain valuable predictive information, as GNAR model performance seems to be less dependent on order selection compared to AR models.

\section*{Appendix E - Directed Graphs}
For comparison, we also consider an algorithm which constructs a directed network based on the one-month ahead RMSE at all nodes. To do so, we select the best graph for each node separately, and then construct a single directed network by placing edges from the stage one
neighbours of a given node to that node itself. Formally, let $\mathcal{G}_t^i$ be the network with the lowest RMSE at node $i$ at time $t$ and let $\mathcal{G}_t$ denote the directed graph we are constructing. Then $j \rightsquigarrow i$ in $\mathcal{G}_t$ only if $j\in\mathcal{N}(i)$ in $\mathcal{G}_t^i$, for all CPI components $i$. By using directed edges, each node is only directly affected by its own stage one neighbours. We only fit a few GNAR models for comparison purposes.
The results, summarised in Table~\ref{tab:directed}, closely resemble those obtained for the undirected networks discussed in Section~\ref{sec:best_graph}. This similarity suggests that constraining the network's structure offers little to no benefit, especially given its significantly higher computational complexity. In fact, allowing the network structure to remain random may help prevent overfitting.

\begin{table}[h]
    \centering
    \begin{tabular}{cc|lllll}
    \toprule
      \multicolumn{2}{c|}{GNAR} &  \multicolumn{5}{c}{Forecast Horizon (months)} \\
    Class & Model & 1 & 3 & 6 & 9 & 12 \\
    \midrule
         \multirow{2}{*}{\mbox{global-$\alpha$}}
    & GNAR(1,\textbf{1}) & 0.40\tiny{$\pm$0.01} & 0.78\tiny{$\pm$0.04} & 1.34\tiny{$\pm$0.06} & 1.92\tiny{$\pm$0.06} & 2.39\tiny{$\pm$0.06} \\
    & GNAR(2,\textbf{1}) & 0.39\tiny{$\pm$0.01} & 0.76\tiny{$\pm$0.04} & 1.33\tiny{$\pm$0.07} & 1.97\tiny{$\pm$0.10} & 2.54\tiny{$\pm$0.13} \\
    \midrule
    \multirow{2}*{\parbox{1.5cm}{standard\\\mbox{(local-$\alpha$)}}}
    & GNAR(1,\textbf{1}) & 0.40\tiny{$\pm$0.02} & 0.77\tiny{$\pm$0.04} & 1.32\tiny{$\pm$0.08} & 1.97\tiny{$\pm$0.11} & 2.53\tiny{$\pm$0.16} \\
    & GNAR(2,\textbf{1}) & 0.38\tiny{$\pm$0.01} & 0.70\tiny{$\pm$0.02} & 1.19\tiny{$\pm$0.05} & 1.78\tiny{$\pm$0.08} & 2.31\tiny{$\pm$0.10} \\
    \midrule
        \multirow{2}{*}{\mbox{local-$\alpha\beta$}}
    & GNAR(1,\textbf{1}) & 0.39\tiny{$\pm$0.01} & 0.71\tiny{$\pm$0.02} & 1.17\tiny{$\pm$0.05} & 1.74\tiny{$\pm$0.07} & 2.24\tiny{$\pm$0.09} \\
    & GNAR(2,\textbf{1}) & 0.35\tiny{$\pm$0.01} & 0.70\tiny{$\pm$0.02} & 1.22\tiny{$\pm$0.04} & 1.84\tiny{$\pm$0.07} & 2.38\tiny{$\pm$0.09} \\
    \bottomrule
    \end{tabular}
    \caption{Absolute RMSEs for inflation rate forecasts computed by fitting a GNAR process to a directed network constructed each month.
    RMSEs are averaged over 100 runs of the algorithm, with $\pm1$ standard deviation displayed in a smaller font.}
    \label{tab:directed}
\end{table}

\section*{Appendix F - Selecting Networks on a Quarterly Basis}
In this section we select the best networks and fit GNAR models once every three months. For brevity, we only examine the forecasts from GNAR models chosen according to the BIC. The results of averaging the forecasts of $n=5$ best graphs selected each quarter are displayed in Table~\ref{tab:quarterly}. Switching from monthly to quarterly yields slightly weaker forecasts on average, although may be advantageous from a computational perspective.

\begin{table}[h]
    \centering
    \begin{tabular}{cc|lllll}
    \toprule
      \multicolumn{2}{c|}{GNAR} &  \multicolumn{5}{c}{Forecast Horizon (months)} \\
    Class & Model & 1 & 3 & 6 & 9 & 12 \\
    \midrule
    \multirow{3}{*}{\mbox{global-$\alpha$}}
    & GNAR($p, \boldsymbol{1}$) & 0.39\tiny{$\pm$0.01} & 0.77\tiny{$\pm$0.03} & 1.36\tiny{$\pm$0.04} & 1.92\tiny{$\pm$0.03} & 2.36\tiny{$\pm$0.03} \\
    & GNAR($p, \boldsymbol{2}$) & 0.38\tiny{$\pm$0.01} & 0.76\tiny{$\pm$0.02} & 1.38\tiny{$\pm$0.04} & 2.02\tiny{$\pm$0.03} & 2.56\tiny{$\pm$0.05} \\
    & GNAR($p, \boldsymbol{s}$) & 0.39\tiny{$\pm$0.01} & 0.78\tiny{$\pm$0.02} & 1.38\tiny{$\pm$0.03} & 2.03\tiny{$\pm$0.04} & 2.55\tiny{$\pm$0.05} \\
    \midrule
    \multirow{3}*{\parbox{1.5cm}{standard\\\mbox{(local-$\alpha$)}}}
    & GNAR($p, \boldsymbol{1}$) & 0.40\tiny{$\pm$0.01} & 0.76\tiny{$\pm$0.02} & 1.31\tiny{$\pm$0.04} & 1.92\tiny{$\pm$0.07} & 2.45\tiny{$\pm$0.10} \\
    & GNAR($p, \boldsymbol{2}$) & 0.38\tiny{$\pm$0.01} & 0.71\tiny{$\pm$0.02} & 1.27\tiny{$\pm$0.05} & 1.87\tiny{$\pm$0.07} & 2.43\tiny{$\pm$0.08} \\
    & GNAR($p, \boldsymbol{s}$) & 0.38\tiny{$\pm$0.01} & 0.73\tiny{$\pm$0.02} & 1.29\tiny{$\pm$0.05} & 1.89\tiny{$\pm$0.07} & 2.44\tiny{$\pm$0.08} \\
    \midrule
    \multirow{3}{*}{\mbox{local-$\alpha\beta$}}
    & GNAR($p, \boldsymbol{1}$) & 0.39\tiny{$\pm$0.00} & 0.68\tiny{$\pm$0.01} & 1.18\tiny{$\pm$0.03} & 1.74\tiny{$\pm$0.04} & 2.18\tiny{$\pm$0.06} \\
    & GNAR($p, \boldsymbol{2}$) & 0.37\tiny{$\pm$0.00} & 0.67\tiny{$\pm$0.01} & 1.13\tiny{$\pm$0.03} & 1.67\tiny{$\pm$0.04} & 2.15\tiny{$\pm$0.05} \\
    & GNAR($p, \boldsymbol{s}$) & 0.37\tiny{$\pm$0.00} & 0.66\tiny{$\pm$0.01} & 1.13\tiny{$\pm$0.02} & 1.67\tiny{$\pm$0.04} & 2.16\tiny{$\pm$0.05} \\
    \bottomrule
    \end{tabular}
    \caption{Absolute RMSEs for inflation rate forecasts computed by averaging predictions from GNAR models fitted to the $n=5$ best-performing networks every three months. RMSEs are averaged over 100 runs of the algorithm, with $\pm1$ standard deviation displayed in a smaller font.}
    \label{tab:quarterly}
\end{table}

\end{document}